\documentclass[prd,reprint]{revtex4-1}
\usepackage{amsmath}
\usepackage{amsfonts}
\usepackage{graphicx}
\usepackage{booktabs}
\usepackage{ctable}
\usepackage{tabulary}
\usepackage{graphicx}
\usepackage{epsfig}
\usepackage{fancyhdr}
\usepackage{eucal}
\usepackage{mathbbol}
\usepackage{pstricks}
\usepackage{color}
\usepackage{bm} 
\usepackage{multirow}
\usepackage{array} 
\usepackage{nccmath} 
\usepackage{ctable} 
\usepackage{nicefrac} 
\usepackage{bigstrut}
\usepackage{rotate}
\usepackage{textpos}
\usepackage[normalem]{ulem}

\def\hyperfine#1{\Delta M_{#1\uS}}
\def\epemwidth#1{\Gamma_{{\epem}\to{#1^3 \rm S_1}}}
\newcommand{\sixj}[6]{ \left\{\begin{matrix} {#1} & {#2} & {#3} \\    {#4} & {#5} & {#6} \end{matrix}   \right\} }
\newcommand{\ninej}[9]{ \left\{\begin{matrix}  {#1} & {#2} & {#3} \\    {#4} & {#5} & {#6}\\    {#7} & {#8} & {#9}  \end{matrix}   \right\} }

\def\pmat#1{\begin{pmatrix}#1\end{pmatrix}} 

\def\>{\big>}
\def\<{\big<}
\def\|{\big\vert}
\def\£{\big\Vert}

\def\etal{\textit{et~al}}

\newcommand{\rf}[1]{(\ref{#1})}

\def\etab{\end{tabular}}  
  
\def\bit{\begin{itemize}}
\def\eit{\end{itemize}}

\def\bml{\begin{multline}}
\def\eml{\end{multline}}
\def\be{\begin{equation}} 
\def\ee{\end{equation}}  
\def\bea{\begin{eqnarray}}  
\def\eea{\end{eqnarray}} 
\def\bmu{\begin{multline}}
\def\emu{\end{multline}}
\def\bal{\begin{array}{l}} 	
\def\eal{\end{array}}

\def\h{\hat}

\def\ot{\times}
\def\half{\frac{1}{2}}

\def\MeV{\textrm{ MeV}}
\def\GeV{\textrm{ GeV}}

\newcommand{\ve}[1]{\mathbf{#1}} 	
\newcommand{\vc}[1]{\boldsymbol{#1}} 	
\newcommand{\vh}[1]{\mathbf{\hat{#1}}}	
\newcommand{\vb}[1]{\mathbf{\overline{#1}}} 	
\newcommand{\vehat}[1]{\mathbf{\hat{#1}}} 

\newcommand{\p}{\ve{p}}

\def\letterS{S}
\def\letterP{P}
\def\letterD{D}
\def\letterF{F}
\def\greek#1{\def\letter{#1}
\ifx\letter\letterS\Sigma
\else
	\ifx\letter\letterP\Pi
	\else
		\ifx\letter\letterD\Delta
		\else
			\ifx\letter\letterF\Phi
			\else XXX
			\fi
		\fi
	\fi
\fi
}

\def\q{\overline q}
\def\Q{\overline Q}

\def\qq{q\overline q}
\def\QQ{Q\overline Q}
\def\qQ{q\overline Q}
\def\Qq{Q\overline q}

\def\MeV{\textrm{ MeV}}
\def\GeV{\textrm{ GeV}}

\renewcommand{\u}[1]{\rm{#1}}
\def\cn#1#2#3{^{#1}{\u{#2}}_{#3}}     
\def\han#1#2#3{^{#1}\widehat{\u {#2}}_{#3}}     
\def\an{\cn} 
\def\sl#1#2{^{#1}{\u{#2}}}    
\def\slj#1#2#3{^{#1}{\u{#2}}_{#3}}    
\newcommand{\uS}{\rm{S}}
\newcommand{\uP}{\rm{P}}
\newcommand{\uD}{\rm{D}}
\newcommand{\uF}{\rm{F}}
\newcommand{\uG}{\rm{G}}

\def\epem{e^+e^-}

\def\nf#1/#2{\dfrac{#1}{#2}}		
\def\sf#1/#2{\sqrt{\nf{#1}/{#2}}}	

\def\nfr#1,#2{#1/#2}

\def\xn#1#2#3(#4){\an#1#2#3~(\u#4)}
\def\yn#1#2#3(#4){\han#1#2#3~(\u#4)}

\def\o{\overline}

\def\spato{\mathbf{ O}}
\def\spatom{O^{-m}}
\def\spino{\bm{\chi}}
\def\spinom{\chi^{m}}

\def\half{\nf 1/2}

\def\phrac{\phantom{\dfrac{\sqrt 0}{ 0}}}
\def\bsw{\begin{sideways}}
\def\esw{\end{sideways}}

\def\p{_{\phantom{0}}}
\def\Pp{\phantom{S_0}}

\def\qnset#1{\!\bigg[\!\begin{smallmatrix}#1\end{smallmatrix}\!\bigg]}

\def\spread[#1,#2,#3]{\begin{tabularx}{0.2\linewidth}{XXX}$#1$&$#2$ & $#3$ \end{tabularx}}

\begin{document}

\title{Angular momentum coefficients for meson strong decay and unquenched quark models}

\author{T.~J.~Burns }
\email{{\tt timothy.burns@durham.ac.uk}\href{mailto:timothy.burns@durham.ac.uk}}
\affiliation{Department of Mathematical Sciences, Durham University, Durham DH1 3LE, United Kingdom}
\begin{abstract}

In most meson strong decay and unquenched (coupled-channel) quark models, the pair-creation operator is a scalar product of vectors in the spin and spatial degrees of freedom. While differing in the spatial part, most models have the same spin part, which creates a $q\q$ pair coupled to spin triplet, with the spins of the initial quarks as spectators. This is a basic assumption of the $\an 3P0$ model, and is well-known to arise also in the flux tube model, starting from the strong coupling expansion of lattice QCD. In this article the same structure is shown to emerge in the Cornell model, in the dominant contributions of a more general microscopic decay model, and in the pseudoscalar-meson emission model. A solution is obtained for arbitrary matrix elements in these ``non-flip, triplet'' models, expressed as a weighted sum over spatial matrix elements. The coefficients in the expansion, which  involve the spin degrees of freedom and the associated angular momentum algebra, are model-independent. 
Tables of the angular momentum coefficients are presented which can be used in future calculations, avoiding tedious Clebsch-Gordan sums. The symmetry and orthogonality properties of the coefficients are discussed, as well as their application to transitions involving hybrid mesons and states of mixed spin.  New selection rules are derived, and existing ones generalised. The coefficients lead to model-independent relations among decay amplitudes and widths which can be tested in experiment and lattice QCD. They can also be used to explain how mass shifts in the unquenched quark model do not spoil successful predictions of the ordinary (quenched) quark model.

\pacs{12.39.Jh,12.39.Pn,12.40.Yx,14.40.Pq}
\end{abstract}


\maketitle

\thispagestyle{fancy}


\section{Introduction}

Meson strong decay widths are dominated by channels allowed by the OZI (Okubo-Zweig-Iizuka) rule, in which the initial $Q\Q$ and a created $q\q$ pair recombine into final states $Q\q$ and $q\Q$. Quark models have been applied extensively to such transitions, including the phenomenological $\an 3P0$ model, developed in the 1970s and still widely used today, and flux tube models more closely connected to QCD.

The same transition also plays a role below threshold, leading to a virtual meson-meson component in the physical wavefunction, and a mass shift with respect to the bare $Q\Q$ mass. ``Unquenched'' quark models, of which the Cornell model is the prototypical example, take account of this effect. Recently such models have been applied to the $X$, $Y$, and $Z$ mesons in the charmonium and bottomonium mass regions, in an attempt to explain their unusual masses and decay properties.

The matrix element for the transition is typically calculated as the overlap of the non-relativistic wavefunctions of the mesons, assuming some operator describing the created $q\q$ pair. The overall structure of most operators is the same: a scalar product of vectors acting separately on the spin and spatial degrees of freedom. While the models differ in the spatial degrees of freedom, usually the spin part involves a $q$ and $\q$ coupled to a triplet, with the projections of the spins of $Q$ and $\Q$ unchanged. Matrix elements in such models have  a common angular-momentum dependence, obtained by recoupling the intrinsic spins of the initial and created quark-antiquark pairs to match those of the final states, and from recoupling the orbital and total angular momenta. 

The purpose of this paper is to study this angular-momentum dependence. A general expression is obtained for arbitrary matrix elements in terms of a sum over coefficients $\xi$, which are common to all models and contain the angular-momentum dependence described above, multiplying model-dependent spatial matrix elements. The general properties of the coefficients are studied, and tables of values are presented.

The coefficients can be used in future strong decay and unquenched quark model calculations. Typically such calculations begin with sums over Clebsch-Gordan coefficients which, as well as being tedious, obscure the underlying patterns associated with the angular momentum algebra. The summations over $\xi$ coefficients are straightforward by comparison, resulting in simpler expressions for decay amplitudes, widths, mass shifts and mixing amplitudes. 

The coefficients have other applications, which will be introduced in this paper and discussed in future work. There are new selection rules, as well as model-independent relations among decay amplitudes and widths, which can be tested in experiment and lattice QCD. They can be used for model-independent predictions for mass shifts in the unquenched quark model.

The paper begins (Section \ref{models}) by demonstrating the common operator structure of $\an 3P0$, flux tube, microscopic and pseudoscalar-meson emission models. 

The bulk of the paper (Section \ref{coefficients}) is concerned with a formalism which can be applied to these ``non-flip, triplet'' models. An expression is obtained for arbitrary matrix elements in terms of the coefficients $\xi$. A general formula for the coefficients is derived, and their orthogonality and symmetry properties are discussed. Zeroes in the coefficients, which are selection rules, are classified.

Transitions involving heavy-light mesons with mixed quark spin require special treatment (Section~\ref{spin-mixed}). An expression is obtained for the $\xi$ coefficient corresponding to arbitrary singlet-triplet mixing.

The coefficients involving transitions to a pair of S-wave mesons (Section \ref{swavepairs}) are the most widely useful. A compact expression for these coefficients is written down, along with a table of values for arbitrary initial states.

For transitions involving hybrid mesons (Section~\ref{hybrids}) the same $\xi$ coefficients can be used in the flux tube model and, for the lightest hybrids, in the constituent gluon model.

An introduction to some applications of the $\xi$s is then given (Section \ref{applications}), including selection rules, relations among decay amplitudes and widths, and some general results in the unquenched quark model. 

The conclusion (Section \ref{conclusions}) is followed by appendices which contain a general expression for the spatial matrix element (Appendix \ref{spatial:sect}), and tables of $\xi$ coefficients (Appendix \ref{tables}): these can be used as a starting point for future calculations.  

This paper discusses in more detail the results introduced in the workshop proceedings \cite{Burns:2013xoa}.

\section{Models}
\label{models}

In this section various approaches to modelling the coupling of a meson to a meson-meson pair are introduced: $\an 3P0$ models, flux tube models, microscopic models, and pseudoscalar-meson emission models. Their pair-creation operators have the same overall structure, and while they differ in the spatial part, most are identical in the quark spin degrees of freedom. Everything else in this paper follows from this common structure.

\subsection{Non-flip, triplet models}

The aim is to show that most models involve an operator of the form
\be
\spino\cdot\spato=\sum_m (-)^m \spinom\spatom,\label{operator}
\ee
where $\spino$ describes the creation of a spin triplet $\qq$ pair, $\spato$ contains the spatial dependence, and $m$ refers to the spherical components. The spin part creates a $q$ and $\q$ with projections $s$ and $\o s$, 
\be
\spinom=\sum_{s\o s}\|s\o s\>\spinom_{s\o s},
\label{spinomeqn}
\ee
with an amplitude given by a Clebsch-Gordan coefficient 
\be
\spinom_{s\o s}=\<\tfrac{1}{2} s, \tfrac{1}{2} \o s \| 1{m}\>.
\label{cgspin}
\ee
Collecting the spherical components into a vector, equation \rf{spinomeqn} can be expressed
\be
\spino=\sum_{s\o s}\|s\o s\>\spino_{s\o s}.
\label{spinoeqn}
\ee
An equivalent approach involves the Pauli matrix $\vc\sigma$ in place of $\spino$, and this is understood to be sandwiched between spinors, leading to the analogous expression
\be
\vc \sigma=\sum_{s\o s}\|s\o s\>\chi_s^\dag\vc \sigma\chi_{-\o s},
\label{sigmasandwich}
\ee
where the $\chi$s are two-component Pauli spinors, not to be confused with the $\chi$s appearing above. If the spinors are chosen
\begin{align}
\chi_{\uparrow}&=\pmat{1\\0},
&
\chi_{\downarrow}&=\pmat{0\\ 1},
\\
\chi_{\o\uparrow}&=\pmat{-1\\0},
&
\chi_{\o\downarrow}&=\pmat{0\\1},
\end{align}
then $\spino_{s\o s}$ is related to the above matrix element by
\be
\spino_{s\o s}=\frac{1}{\sqrt 2}
\chi_s^\dag
\vc \sigma
\chi_{-\o s},
\label{chisigma}
\ee
so that
\be
\spino=\frac{1}{\sqrt 2}\vc\sigma.
\ee

There are several different choices in the literature for the normalisation of the pair-creation operator. Some authors use a scalar product defined as above, while others form a scalar product by means of a Clebsch-Gordan coefficient, which introduces a factor of $1/\sqrt 3$. In some cases the operator involves an explicit colour singlet wavefunction, which gives a common factor of 1/3 for all transitions due to the overlap of colour wavefunctions; some authors compensate this with a factor of 3 in the operator itself. 

These differences have no relevance to this paper, which only requires that the operator is  proportional to $\spino\cdot\spato$. Models of this type will be called ``non-flip triplet'' models, because the created $q\q$ pair is spin triplet, and the initial $Q\Q$ spins (but not necessarily their spatial degrees of freedom) are spectators.

In Section \ref{coefficients} a general expression for the matrix element of $\spino\cdot\spato$ is obtained as a sum over the matrix elements of the spatial part $\spato$ (which varies from model to model) weighted by angular momentum coefficients (common to all non-flip, triplet models). For practical calculations the different normalisations discussed above can be absorbed into the definition of $\spato$.

\subsection{$\an 3P0$ models}
\label{3P0model}

The assumption of early pair-creation models is that the $\qq$ pair,  since it has the $0^{++}$  quantum numbers of the vacuum, is in a $\an 3P0$ state. The operator is therefore proportional to the scalar product \rf{operator} above. Micu \cite{Micu:1968mk} applied this model to the decays of scalar, axial and tensor mesons, treating the spatial part of the decay matrix element as a free parameter.

Models which treat the spatial degrees of freedom explicitly involve the creation of a $q$ and $\q$ with momenta $\ve k$ and $\ve{\o k}$, with a certain spatial amplitude $\spato(\ve k, \ve{\o k})$, 
\be
\spato=\int d^3\ve k\int d^3\ve {\o k}\|\ve k, \ve{\o k} \> \spato(\ve k,\ve {\o k}).\label{doublespat}
\ee
In practice $\spato(\ve k,\ve {\o k})$ contains a momentum-conserving delta function, leading to
\be
\spato=\int d^3\ve k\|\ve k, -\ve k \> \spato(\ve k).\label{singlespat}
\ee
Different versions of the model have different spatial amplitudes $\spato(\ve k)$, but each involves the same integral.

The approach of Le Yaouanc \etal.~\cite{LeYaouanc:1972ae} assumes that the $\qq$ pair is created with equal amplitude everywhere in space, and the $\an 3P0$ wavefunction for the pair leads to an operator of the form
\begin{widetext}
\be
H=\gamma\sum_m \<1m,1-m\|00\>\int d^3\ve k\int d^3\ve {\o k}\sum_{s\o s}
\delta^3(\ve {\o k}+\ve{k})
\mathcal{Y}_1^{-m}(\ve k-\ve{\o k})
\spinom_{s\o s}
\|s\o s\>
\|\ve k \ve{\o k} \>,
\ee
\end{widetext}
where the pair creation strength $\gamma$ is fit to data, and the solid harmonic is the $-m$th component of the vector
\be
\mathcal{Y}_1^{-m}(\ve k-\ve{\o k})=\sqrt\frac{3}{4\pi}(\ve k-\ve{\o k})^{-m}.
\ee
Carrying out the trivial integral and expressing the sum over $m$ as a dot product, the operator reduces to 
\be
H=-\frac{\gamma}{\sqrt{\pi}}\int d^3\ve k\sum_{s\o s}
\|s\o s\>
\|\ve k ,-\ve{k} \>
\spino_{s\o s}\cdot \ve k.
\ee
Recognising the vector $\spino$ of equation \rf{spinoeqn}, the amplitude can be written
\be
H=\spino\cdot\spato,
\ee
where $\spato$ is defined by \rf{singlespat} above, with amplitude
\be
\spato(\ve k)=-\frac{\gamma}{\sqrt{\pi}}\ve k. \label{basicspat}
\ee
The matrix elements of $\spino\cdot\spato$ can be computed using explicit meson radial wavefunctions, which are commonly taken to be harmonic oscillators.

Ackleh \etal.~\cite{Ackleh:1996yt} formulated the $\an 3P0$ model in terms of quark fields, with operator
\be
 H=g\int d^3\ve x\overline \psi(\ve x) \psi(\ve x).
\ee
Expanding in terms of creation and annihilation operators gives
\be
H
=
g\sum_{s\o s}
\int d^3\ve k
\frac{m}{E_k}
a_{s\ve k}^\dag
b_{\o s -\ve {k}}^\dag
 \o u_{s\ve k}
 v_{\o s -\ve{k}}
\ee
where $m$ is the mass of the created quark and antiquark. In the non-relativistic limit, the spinors reduce to 
\be
  \o u_{s\ve k} v_{\o s -\ve{k}}
=
-\frac{1}{m}
\chi_s^\dag
\vc \sigma
\chi_{-\o s}
\cdot \ve k,
  \ee
which can be written in terms of the operator $\spino_{s\o s}$ using equation \rf{chisigma}. Making the identification
\be
a_{s\ve k}^\dag
b_{\o s -\ve {k}}^\dag
=\|s\o s\> \|\ve k, -\ve {k} \>
\ee
gives
\be
H=-
\frac{g\sqrt 2}{m}\sum_{s\o s}
\int d^3\ve k
\|s\o s\> \|\ve k, -\ve k\>
\spino_{s\o s}\cdot \ve k .
\ee
Again the operator has the form $\spino\cdot\spato$, where $\spino$ and $\spato$ are given by equations \rf{spinoeqn} and \rf{singlespat} respectively, and the spatial pair-creation amplitude is (apart from a numerical factor) the same as \rf{basicspat},
\be
 \spato(\ve k)=-\frac{g\sqrt 2}{m}\ve k.
\ee

The authors compared their results to a $\an 3P0$ model whose spatial operator is normalised, in the notation of this paper,
\be
\spato(\ve k)=-2\sqrt 2\gamma\ve k \label{acklehspatial}.
\ee
Comparison with the above gives $\gamma=g/2m$, as in ref.~\cite{Ackleh:1996yt}.

Using harmonic oscillator wavefunctions the $\an 3P0$ model has been applied widely to the decays of light mesons \cite{Blundell:1995ev,Blundell:1995au,Page:1995rh,Barnes:1996ff,Close:1997dj,Barnes:2002mu,Li:2008mza,Li:2008xy,Bing:2013fva,He:2013ttg}, charmonia \cite{LeYaouanc:1977ux,Barnes:2003vb,Barnes:2005pb,Liu:2009fe,Yang:2009fj,Limphirat:2013jga,Chen:2014wva}, charmed mesons~\cite{Godfrey:1986wj,Close:2005se,Lu:2006ry,Zhang:2006yj,Close:2006gr,Segovia:2008zz,Segovia:2009zz,Sun:2009tg,Li:2009qu,Li:2010vx,Sun:2010pg,Yuan:2012ej,Sun:2013qca,Yu:2014dda} and bottom mesons \cite{Luo:2009wu,Sun:2014wea}. Formulae for transitions between arbitrary mesons described by harmonic oscillator wavefunctions are derived in references \cite{Burns07production,Burns:2006rz}.

Chaichian and K\"ogerler \cite{Chaichian:1978th} and Ono \cite{Ono:1980js} used the $\an 3P0$ model with more realistic meson wavefunctions for a Coulomb plus linear potential. Segovia \etal. \cite{Segovia:2012cd} considered the dependence of $\spato(\ve k)$ on the reduced mass of the initial $Q\Q$ pair. Ribeiro \cite{Ribeiro:1981fk} and van Beveren \cite{vanBeveren:1983im,vanBeveren:1982sj,vanBeveren:1998qg} derived algebraic expressions for the matrix elements in which the pair-creation amplitude $\spato(\ve k)$ is itself a harmonic oscillator function. 

The $\an 3P0$ model has also been applied to baryon decays, for example in references \cite{LeYaouanc:1973xz,LeYaouanc:1974mr,Capstick:1992th, Capstick:1993kb}. Roberts and Silvestre-Brac \cite{Roberts:1992js} generalised the $\an 3P0$ model for decays of hadrons with arbitrary numbers of quarks: mesons, baryons and multiquark states. These results were applied to the decays of light mesons with a modified spatial operator $\spato(\ve k)$ \cite{Roberts:1997kq,BonnazSilvestre-Brac99discussion}. 

The $\an 3P0$ operator has been used in the unquenched quark model for charmonia and bottomonia \cite{Ono:1983rd,vanBeveren:1982qb,Heikkila:1983wd,Tornqvist85quarkonium,Kalashnikova:1993tv,Kalashnikova:2005ui,Barnes:2007xu,Li:2009ad,Ortega:2010qq,Liu:2011yp,Coito:2012ka,Ferretti:2012zz,Ferretti:2013faa,Ferretti:2013vua,Ortega:2012rs,Zhou:2013ada,Ferretti:2014xqa}, as well as charmed mesons \cite{vanBeveren:2004ve,Coito:2011qn}, bottom mesons~\cite{Vijande:2007ke}, and baryons \cite{Bijker:2009up,Santopinto:2008zz,Santopinto:2010zza,Bijker:2012zza}.

All of the models described above are characterised by the operator $\spino\cdot\spato$; the models differ only in the spatial part $\spato$. On the other hand there are some variants of the $\an 3P0$ model which do not have this structure. The ``corrected'' $\an 3P0$ model \cite{daSilva:2008rp,DeQuadros:2010ai} differs  by including corrections associated with the bound state nature of the mesons. The operator creates a $\an 3P0$ $q\q$ pair, but since the bound state corrections are sensitive to the spin degrees of freedom of the initial and final states, the angular-momentum dependence of the matrix elements differs from that of other $\an 3P0$ models. The ordinary $\an 3P0$ model is recovered in the appropriate limit, and for most channels the corrections turn out to be small.

More recently, Fuda \cite{Fuda:2012xd} has shown that there is nothing inherently non-relativistic about the $\an 3P0$ operator. Nevertheless the matrix elements in the relativistic quark model of that paper, such as $\rho\to\pi\pi$, have a more complicated angular momentum algebra than the non-relativistic models discussed here.

\subsection{Flux tube models}
\label{fluxtubemodels}

\begin{figure*}
 \vspace{1cm}
 \rput(0.1,0.8){$\ve {\o X}$}
 \rput(2.1,0.8){$\ve x$}
 \rput(5.9,0.8){$\ve X$}
 \rput(8.5,-0.2){$\ve {\o X}$}
 \rput(10.8,2.3){$\ve x$}
 \rput(14.4,-0.2){$\ve X$}
 \rput(12.0,-0.2){$\ve r$}
 \rput(11.2,1.3){$\vc \rho_\perp$}
  \rput(2.5,3.0){$\an 3S1$ flux tube model}
  \rput(11,3.0){Isgur-Paton flux tube model}
  \includegraphics[width=0.85\linewidth]{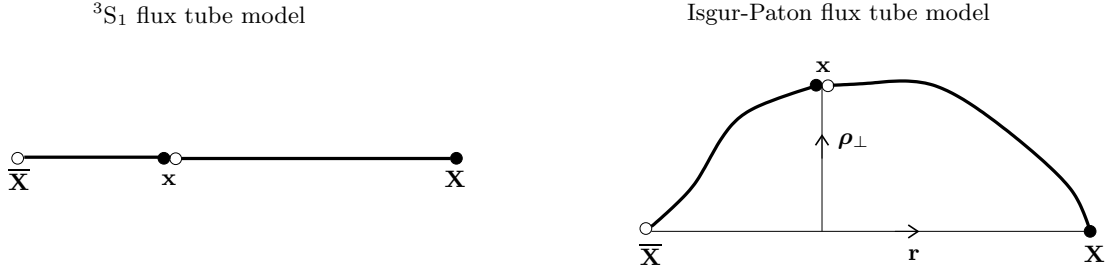}
\vspace{0.5cm}
 \caption{The geometry of quark-pair creation in the $\an 3S1$ and Isgur-Paton flux tube models. The initial $Q$ and $\Q$ are at $\ve X$ and $\ve {\o X}$ respectively, and the created $q\q$ pair is at $\ve x$.}
 \label{geometry}
\end{figure*}

In flux tube models the $q\q$ pair is created not out of the vacuum, but from the breaking of the gluonic flux tube. Again it turns out that the pair-creation operator has the form $\spino\cdot\spato$. 

The spatial part $\spato$ is usually formulated in position space, and creates
the $q$ and $\q$ at a point,
\be
\spato=\int d^3\ve x \|\ve x \ve {x} \> \spato(\ve x).
\label{posop}
\ee

Alcock \etal.~\cite{Alcock:1983gb} assume that the flux tube is straight, so that $q\q$ pair-creation is constrained to the $Q\Q$ axis, as shown in Fig. \ref{geometry}. The amplitude is proportional to the chromoelectric flux vector, which is aligned along the direction $\vh r$, where $\ve r=\ve X-\ve{\o X}$ is the vector connecting the initial $Q$ and $\Q$ at $\ve X$ and $\ve{\o X}$ respectively. As the flux tube is straight, the spatial amplitude is zero everywhere except along the $Q\Q$ axis where, up to an overall constant,
\be
\spato(\ve x)=\vh r. 
\ee
To form a scalar, the field vector is contracted with another vector describing a $q\q$ pair with spin triplet, and so the operator can be written in the form $\spino\cdot\spato$. The created $q\q$ pair has $\an 3S1$ quantum numbers.

The flux tube model of Isgur and Paton \cite{Isgur:1983wj,Isgur:1984bm} is derived from the strong coupling expansion of lattice QCD. Pair creation is due to a term which destroys a piece of flux connecting neighbouring lattice sites at $\ve x$ and $\ve x +a \vh n$ (where $a$ is the lattice spacing), replacing it with a $\q$ and $q$ at the ends of the broken flux links~\cite{Kokoski:1985is}. The matrix element involves the overlaps of the degrees of freedom not only of the quarks, but also the flux tubes.  

The flux tubes have transverse oscillations so that, unlike in the previous model, pair creation can occur away from the $Q\Q$ axis, as shown in Fig. \ref{geometry}. Within the adiabatic approximation the flux tube degrees of freedom are integrated out first, leaving a term $\gamma(\ve x)$, which is the overlap of the flux tubes as a function of the position $\ve x$ at which the $q\q$ pair is created \cite{Dowrick:1986ub}. (It is a function of the coordinate $\ve x$ only, because eventually one goes to the continuum limit so that pair creation occurs at a point.) This leaves an operator of the form
\be
\int d^3 \ve x
\gamma(\ve x)
\psi^\dag(\ve x +a \vh n) \vc \alpha  \cdot \vh n \psi(\ve x).
\ee
In terms of creation operators, the fields expand as
\begin{widetext}
\be
\psi^\dag(\ve x +a \vh n) \vc \alpha  \cdot \vh n \psi(\ve x)
=
\frac{1}{(2\pi)^3}
\sum_{s\o s}
\int d^3\ve k
\int d^3\ve{\o k}
\frac{m}{\sqrt{E_k E_{\o k}}}
a_{s\ve k}^\dag
b_{\o s\ve {\o k}}^\dag
 e^{-i\ve k\cdot (\ve x +a \vh n)}
 \o u_{s\ve k}
 \vc\gamma\cdot\vh n
 e^{-i\ve{\o k}\cdot\ve x}
 v_{\o s\ve{\o k}}.
 \label{operatorexpansion}
\ee

\end{widetext}
In the non-relativistic limit, the matrix element of the Dirac gamma matrix between quark and antiquark spinors $\o u$ and $ v$ is
\be
 \o u_{s\ve k}
 \vc\gamma
 v_{\o s\ve{\o k}}
=
\chi_s^\dag
\vc \sigma
\chi_{-\o s},
\ee
and one recognises the vector $\spino_{s\o s}$ of equation \rf{chisigma}. Writing the creation operators in terms of the state vectors used before,
\be
a_{s\ve k}^\dag
b_{\o s\ve {\o k}}^\dag
=\|s\o s\> \|\ve k \ve {\o k} \>,
\ee
and incorporating the definition \rf{spinoeqn}, gives
\be
\sum_{s\o s}
a_{s\ve k}^\dag
b_{\o s\ve {\o k}}^\dag
 \o u_{s\ve k}
 \vc\gamma
 v_{\o s\ve{\o k}}
=
 \|\ve k \ve {\o k} \>\sqrt{2}\spino,
\ee
The operator \rf{operatorexpansion} can therefore be written in the form $\spino\cdot\spato$, with the spatial part (in the non-relativistic limit) given by 
\begin{multline}
 \spato
=
\frac{\sqrt 2}{(2\pi)^3}
\vh n
\int d^3\ve x
\gamma(\ve x)
\\
\times
\int d^3\ve k
\int d^3\ve {\o k}
\| \ve k \ve{\o k} \>
e^{-i(\ve k+\ve{\o k})\cdot \ve x-ia\ve{k}\cdot\vh n}.
\end{multline}
Comparison with equation \rf{doublespat} leads to the definition of the amplitude in momentum space
\be
\spato(\ve k, \ve{\o k})
=
\frac{\sqrt 2}{(2\pi)^3}
\vh n
e^{-ia\ve{ k}\cdot\vh n}
\int d^3\ve x
\gamma(\ve x)
e^{-i(\ve k+\ve{\o k})\cdot \ve x}
\ee
Going to the continuum limit, comparison with equation \rf{posop} likewise gives the position-space amplitude,
\be
\spato (\ve x)=\sqrt{2}\ve{\h n}\gamma(\ve x).
\ee
For a straight flux tube $\vh n=\vh r$ and $\gamma (\ve x)$ is non-zero only along the $Q\Q$ axis, which recovers the $\an 3S1$ model described above.

Kokoski and Isgur \cite{Kokoski:1985is} argue that the flux tube is not straight, due to zero point oscillations of its Fourier modes. Pair creation occurs with equal amplitude on each of any of the six neighbouring lattice sites characterised by the vector $\vh n$. In the expansion
\be
\spato(\ve k, \ve{\o k})
\approx
\frac{\sqrt 2}{(2\pi)^3}
\vh n
(1-ia\vh n\cdot\ve{k})
\int d^3\ve x
\gamma(\ve x)
e^{-i(\ve k+\ve{\o k})\cdot \ve x}
\ee 
the first term vanishes after summing over nearest neighbours, while the second term survives,
\be
\sum_{\vh n}\vh n (\vh n\cdot \ve{ k})=2\ve{ k}.
\ee
The pair created by this operator has $\an 3P0$ quantum numbers. Indeed if $\gamma(\ve x)$ is taken as a constant the resulting spatial operator is identical (up to a numerical factor) to that of the old $\an 3P0$ model, equation \rf{basicspat}. In practice $\gamma(\ve x)$ is not constant, but localises pair creation to a ``cigar-shaped'' region surrounding the initial $Q\Q$ axis, approximately Gaussian in the distance $\rho_\perp$ away from the axis (shown in Fig. \ref{geometry}),
\be
\gamma(\ve x)\sim e^{-fb\rho_\perp^2},
\ee
where $b$ is the string tension and $f$ is approximately constant \cite{Dowrick:1986ub}. The model gives similar predictions to the $\an 3P0$ model, which mimics the above localisation in the overlap of meson wavefunctions \cite{Kokoski:1985is}. 

Kumano and Pandharipande \cite{Kumano:1988ga} found that both $\an 3S1$ and $\an 3P0$ models fit light meson decay data, although the latter requires stronger final state interactions. With a detailed treatment of final state interactions, Geiger and Swanson \cite{Geiger:1994kr} found a strong preference the $\an 3P0$ flux tube model. From the point of view of this paper, both are non-flip, triplet models, differing only in the spatial part $\spato$. 

The flux tube model has also been applied to light meson decays by other authors \cite{Blundell:1995ev,Blundell:1995au,Li:2008xy}, as well as to baryon decays \cite{Stassart:1990zt,Stancu:1988gb}, and to meson properties in the unquenched quark model \cite{Geiger:1989yc,Geiger:1991ab,Geiger:1991qe,GeigerIsgur91reconciling,GeigerIsgur93when}.

The model has been applied to the decays of hybrid mesons, whose matrix elements are normalised against those of conventional mesons.  The dynamics of the gluonic excitation enter in the form of a modified flux tube overlap~\cite{Dowrick:1986ub},
\be
\gamma(\ve x)\sim 
\vh e^{\vh r}_\Lambda\cdot\vc\rho_\perp
e^{-fb\rho_\perp^2},
\ee
where $\vh e^{\vh r}_\Lambda$ is a  spherical unit vector orthogonal to $\vh r$ and $\Lambda=\pm 1$ is the projection of the flux tube angular momentum along $\vh r$. The overlap for transitions with hybrids in the final state has a similar form \cite{Burns:2006rz,Burns07production}.

This additional term in the flux tube overlap leads to different dynamics for hybrid meson decays, in particular to a selection rule forbidding their decay to an identical pair of S-wave mesons \cite{Isgur:1985vy, Close:1994hc}. The modification is associated only with the spatial part $\spato$, however, and the overall structure is still $\spino\cdot\spato$. References~\cite{Burns:2006wz,Burns:2007hk,Burns07production} showed that the ratio of two hybrid meson decay amplitudes computed on the lattice \cite{McNeile:2006bz} is consistent with the flux tube model, and that this ratio is intrinsically connected to the assumed $\spino\cdot\spato$ structure; this observation will be discussed in Section~\ref{latticesection}.

This implementation of the flux tube model has been applied by several authors to the decays of hybrids with light quarks \cite{Isgur:1985vy,Close:1994hc,Bing:2013fva}.

In the above model, the pair-creation amplitude has a node along the $Q\Q$ axis. Page \etal. \cite{Page:1998gz,Swanson:1997wy} proposed an alternative model for hybrid decay in which pair creation is constrained to lie along the $Q\Q$ axis, with amplitude
\be
\spato(\ve x)
\sim
\sum_{\mu\Lambda}
\vh e_{\Lambda}^{\vh r}
\cos(\xi\pi)
(\alpha_{\mu\Lambda}-\alpha_{\mu\Lambda}^\dag).
\ee
Here $\alpha^\dag$ and $\alpha$ are creation and annihilation operators for phonons (gluonic excitations) with mode number $\mu$ and polarisation $\Lambda$, and
$
\xi={|\ve x - \ve {\o X}|}/{|\ve X - \ve {\o X}|}
$
measures how far along the $Q\Q$ axis pair creation takes place. The operator has the same $\spino\cdot\spato$ structure as the others described above, and so can be treated on the same footing from the point of view of this paper.

(In the constituent gluon model for hybrids, decay is triggered by the creation of a $\qq$ pair from the vector gluon. The angular-momentum dependence in such models does not, in general, correspond to the other models described in this paper, with a notable exception; this is discussed in Section \ref{hybrids}.)

The approach of Dosch and Gromes \cite{Dosch:1986dp} is similar in spirit to the above flux tube models, in that it starts from the strong coupling expansion of QCD and works within the adiabatic approximation. They obtain a general expression for meson decay widths, which involves an operator with the familiar $\spino\cdot\spato$ structure. The difference compared to other models is that the pair-creation amplitude is a function of the mass of the $q\q$ pair, but within the factorisation approach of this paper, this would be absorbed into the definition of $\spato$.

\subsection{Microscopic models}
\label{microscopicmodels}
\begin{figure*}
\vspace{1cm}
 \rput(3.5,2.5){$Q\o Q$ potential}
  \rput(12,2.5){$q\o q$ pair creation}
  \includegraphics[width=0.85\linewidth]{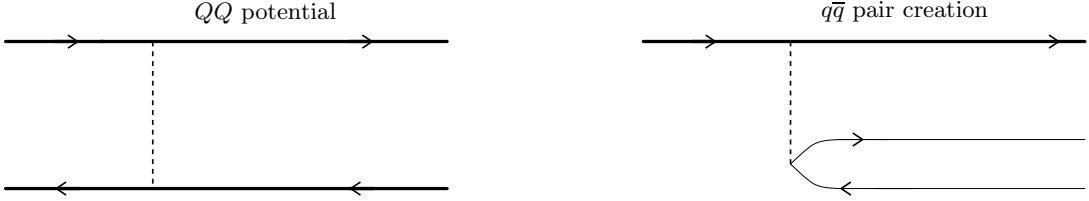}
\vspace{0.5cm}
 \caption{The operator \rf{microoperator} is responsible for both the quark-antiquark potential and pair-creation.}
\label{microfig}
\end{figure*}

In microscopic models pair creation arises from the same interaction which controls the hadron spectrum, so that   masses and decays are determined by the same parameters, unlike in the approaches outlined above, for which the pair-creation strength is fit to data. These models have been formulated in several ways, differing in the assumed Lorentz structure of the potential. As will be shown below, some of these lead uniquely to an operator of the familiar form $\spino\cdot\spato$. Others include additional operators with a different structure.

The term responsible for both hadron masses and pair creation has the form
\be
H=
\int d^3\ve x
\int d^3\ve y
\o \psi(\ve x)\Gamma \psi(\ve x)
V(|\ve x-\ve y|)
\o \psi(\ve y)\Gamma \psi(\ve y),
\label{microoperator}
\ee
where $V$ is the potential and the $\Gamma$ matrices are chosen according to the assumed Lorentz structure. The currents are colour octet, although this will have no role in what follows and so the corresponding labels are suppressed.  

 
This approach was formulated in the Cornell model~\cite{Eichten:1975ag,Eichten:1978tg,Eichten:1979ms}, using a linear confining potential $V$ which is the time-like component of a Lorentz vector ($\Gamma=\gamma_0$). Using this operator in the context of the coupled-channel equations, the model was applied to the spectra and decays of charmonia states. More recently this model has been applied to charmonia near threshold~\cite{Eichten:2004uh,Eichten:2005ga,Yang:2010am}, and to the $D_s(2317)$ \cite{Hwang:2004cd}.

Ackleh \etal.~\cite{Ackleh:1996yt} developed a similar model, but in their approach the linear confining potential is a Lorentz scalar ($\Gamma=I$), and they also include the time-like ($\Gamma=\gamma_0$) and transverse ($\Gamma=\vc \gamma$) contributions from the one-gluon exchange potential. Li \etal.~\cite{Li:2011qb} use the same model but incorporate the effect of a running gluon mass in the Fourier transforms of the one-gluon exchange potential. Segovia \etal. \cite{Segovia:2013kg} use a screened confinement potential which is a mixture of Lorentz scalar and vector.

For present purposes, the question is whether or not the operator \rf{microoperator} can be written in the familiar $\spino\cdot\spato $ form, and the answer depends on the Lorentz structure. For illustration, consider the contribution due to scattering off the initial quark (as opposed to antiquark). Using the earlier notation for the position coordinates (Fig. \ref{geometry}), the scattering and pair-creation currents are taken at $\ve X$ and $\ve x$ respectively. Following references \cite{Ackleh:1996yt,Segovia:2013kg}, expanding the quark fields in terms of creation and annihilation operators gives (suppressing the overall normalisation)
\begin{widetext}
\begin{multline}
H=
\frac{1}{(2\pi)^6}
\int d^3\ve X
\int d^3\ve x
V(|\ve X -\ve x|)
\int d^3\ve K
\int d^3\ve K'
\int d^3\ve k
\int d^3\ve {\o k}
\\
\times
\frac {Mm}{\sqrt{E_KE_{K'}E_kE_{\o k}}}
\sum_{SS's\o s}
a_{S'\ve K'}^\dag
a_{S\ve K}
a_{s\ve k}^\dag
b_{\o s\ve{\o k}}^\dag
e^{i(\ve K-\ve K')\cdot \ve X -i(\ve k+\ve{\o k})\cdot\ve x}
\o u_{S'\ve K'}\Gamma u_{S\ve K}
\o u_{s\ve k}\Gamma v_{\o s \ve{\o k}},
\end{multline}
where $M$ is the mass of the initial quark, and $m$ is the mass of both the created quark and antiquark. As before, $s$ and $\o s$ are the projections of the spins of the created quark and antiquark, and $\ve k$ and $\ve{\o k}$ are their momenta. Likewise $S$ and $\ve K$ are for the initial quark, while $S'$ and $\ve K'$ are the corresponding quantities after scattering. 
Identifying the creation and annihilation operators with the state vectors used earlier, in the non-relativistic limit the operator is
\be
H 
=
\sum_{SS's\o s}
\int d^3\ve K
\int d^3\ve K'
\int d^3\ve k
\int d^3\ve {\o k}
\|S's\o s\> 
\|\ve K' \ve k \ve{\o k}\>
\delta^3(\ve K-\ve K'-\ve k-\ve{\o k})
\widetilde V(\ve k+\ve{\o k})
\o u_{S'\ve K'}\Gamma u_{S\ve K}
\o u_{s\ve k}\Gamma v_{\o s \ve{\o k}}
\<\ve K\|\<S\|,
\ee
\end{widetext}
where $\widetilde V$ is the Fourier transform of the potential
\be
\widetilde V(\ve Q)=
\frac{1}{(2\pi)^3}
\int d^3\ve Y
e^{i\ve Q\cdot \ve Y}V(Y).
\ee

The spin structure of this amplitude is determined by the matrix elements of the $\Gamma$ matrices, and the three possibilities will now be discussed in turn. First consider the case $\Gamma=I$, denoted $sKs$ (for scalar-kernel-scalar) in ref. \cite{Ackleh:1996yt}. The non-relativistic limit of the Dirac bilinears is
\be
\o u_{S'\ve K'} u_{S\ve K}
\o u_{s\ve k} v_{\o s \ve{\o k}}
=
\frac{1}{2m}
\delta_{SS'}
\chi_s^\dag\vc\sigma\chi_{-\o s}
\cdot
(\ve{\o k}-\ve k).
\ee
The initial quark spins are unchanged by the interaction, and the $\vc \sigma$ matrix corresponds to the spin triplet wavefunction $\spino_{s\o s}$ for the created $\qq$ pair as encountered earlier, equation \rf{chisigma}. The summation of spin projections
\be
\sum_{SS's\o s}
\|S's\o s\> 
\delta_{SS'}
\spino_{s\o s}
\<S\|
=
\sum_{s\o s}
\|s\o s\>
\spino_{s\o s}
\ee
is nothing but the operator $\spino$ defined in equation \rf{spinoeqn}. The full operator can be written in the form $\spino\cdot\spato$,
where the spatial part is now an integral over four momenta
\begin{multline}
\spato
=
\int d^3\ve K
\int d^3\ve K'
\int d^3\ve k
\int d^3\ve {\o k}
\\
\times
\|\ve K' \ve k \ve{\o k}\>\spato(\ve K, \ve K',\ve k,\ve{\o k})
\<\ve K\|,
\label{quadruplespat}
\end{multline}
of the amplitude
\begin{multline}
\spato(\ve K, \ve K',\ve k,\ve{\o k})
\\
=
\frac{\sqrt 2}{2m}
(\ve{\o k}-\ve k)
\delta^3(\ve K-\ve K'-\ve k-\ve{\o k})
\widetilde V(\ve k+\ve{\o k}).
\label{sKs}
\end{multline}
In the special case $V(Y)=1$, corresponding to pair-creation with equal amplitude everywhere in space, the potential is
\be
\widetilde V(\ve k +\ve{\o k})=\delta^3(\ve k + \ve{\o k}),
\ee
and after doing two delta function integrals and removing a complete set of states, equation \rf{quadruplespat} reduces to equation \rf{singlespat}, and the corresponding spatial amplitude is identical (up to a numerical factor) with that of the $\an 3P0$ model discussed earlier \cite{Ackleh:1996yt}.

The second case $\Gamma=\gamma_0$ is known as the $j^0Kj^0$ term. In the model of Ackleh \etal. this is one of several contributions, and is due to the time-like component of the one-gluon exchange potential. In the Cornell model it is the sole contribution, and is due to the confining potential, which is assumed to be the time-like component of a Lorentz vector. The non-relativistic reduction is the same as for the $sKs$ term apart from a relative sign between the quark and antiquark momenta,
\be
\o u_{S'\ve K'}\gamma_0 u_{S\ve K}
\o u_{s\ve k}\gamma_0 v_{\o s \ve{\o k}}
=
\frac{1}{2m}
\delta_{SS'}
\chi_s^\dag\vc\sigma\chi_{-\o s}
\cdot
(\ve{\o k}+\ve k)
\ee
so that again the amplitude can be written in the familiar form, with the spatial part the same as equation \rf{sKs} but with the appropriate change of sign. 

The third case $\Gamma=\vc\gamma$, known as $j^TKj^T$, is the transverse contribution from the one-gluon exchange potential in the model of Ackleh \etal. The $i$ and $j$ components of the scattering and pair-creation bilinears are combined as
\be
(\delta^{ij}-\widehat Q^i\widehat Q^j)
\o u_{S'\ve K'}\gamma_i u_{S\ve K}
\o u_{s\ve k}\gamma_j v_{\o s \ve{\o k}},
\ee
where $\ve Q=\ve K'-\ve K= \ve k+\ve{\o k}$ is the momentum transfer, and their non-relativistic reductions are
\begin{widetext}
\be
\o u_{S'\ve K'}\gamma_i u_{S\ve K}
\o u_{s\ve k}\gamma_j v_{\o s \ve{\o k}}
=
\frac{1}{2m}
\left[
\delta_{SS'}
(\ve K + \ve K')_i
+ 
\left(
i\chi_{S'}^\dag\vc\sigma\chi_{S}
\times
(\ve K - \ve K')
\right)_i
\right]
\chi_s^\dag\sigma_j\chi_{-\o s}.
\ee
\end{widetext}
The $q\q$ pair is again created in spin triplet, however there is an additional term which has the effect of flipping the spin of the initial quark. This leads to different spin overlaps, and ultimately a different angular-momentum dependence compared to non-flip, triplet models.

The model of Ackleh \etal. involves all three types of operators, and the contribution from the $j^TKj^T$ term implies that, in general, its matrix elements differ from non-flip, triplet models. Nevertheless it is worth noting that empirically, at least for light meson decays, the $sKs$ contributions are often dominant~\cite{Ackleh:1996yt,Li:2011qb}, in which case the operator is approximately non-flip, triplet. 

The approach of Jaronski and Robson \cite{Jaronski:1986gd} is closely related to the Cornell model. However their potential $V$ includes not only the linear confinement term, but also the short-range spin-dependent terms. The latter generate terms in the operator which do not have the $\spino\cdot\spato$ structure.

\subsection{Pseudoscalar-meson emission models}
\label{pseudoscalar}

In contrast to the models described previously, the pseudoscalar-meson emission model does not explicitly assume the creation of a $\qq$ pair. Instead one treats the emitted meson as a pseudoscalar field with the appropriate flavour quantum numbers, interacting with either of the initial quark or antiquark. Suppressing the overall normalisation as well as flavour degrees of freedom, the operator in such models has the general form
\be
H=\vc\sigma\cdot \spato (\ve q),\label{pseudoop}
\ee
where spatial part $\spato(\ve q)$ is a function of the momentum $\ve q$ of the emitted pseudoscalar meson. It involves a momentum transfer for the scattered (anti)quark,
\be
\spato (\ve q)
=
\int
d^3\ve K
\| \ve K+\ve q \>
\spato(\ve K, \ve q)
\< \ve K \|.
\ee
Keeping only the leading order term, the simplest form for the spatial amplitude is~\cite{Becchi:1966zz,VanRoyen:1967nq,Faiman:1968js,PhysRev.164.1803}
\be
\spato(\ve K, \ve q)=\ve q.
\ee
Inclusion of the recoil term leads to the following form~\cite{Mitra:1967zz,Koniuk:1979vy,Godfrey:1985xj}
\be
\spato(\ve K, \ve q)=\ve q -\frac{E_0}{M_Q}\ve K,
\ee
where $E_0$ is the energy of the emitted meson, valid in the limit that the initial and final (anti)quark are degenerate, with mass $M_Q$. Keeping an additional term in the expansion \cite{Close:2003af}, and allowing for $SU(3)$ breaking in quark masses \cite{Zhong:2008kd,Zhong:2009sk,Zhong:2010vq}, gives
\be
\spato(\ve K, \ve q)=\ve q\left(1+\frac{E_0}{E'+M'}\right) -\frac{E_0}{2\mu}\ve K,
\ee
where $M'$ and $E'$ are the mass and outgoing energy of the final state meson and $\mu=M_QM_Q'/(M_Q+M_Q')$ is the reduced mass of the initial and final (anti) quark masses $M_Q$ and $M_Q'$. This is the operator which arises in the chiral quark model for the special case of pseudoscalar-meson emission  \cite{Zhong:2008kd,Zhong:2009sk,Zhong:2010vq}. The model has recently been applied to charmed-meson decays~\cite{Godfrey:2013aaa}. A discussion of the different parametrisations can be found in ref. \cite{LeYaouanc:1988fx}.

The overall structure of the operator \rf{pseudoop}, as a scalar product of two vectors corresponding separately to spin and spatial degrees of freedom, is suggestive of the familiar form discussed previously. However the $\vc \sigma$ operator appearing here is not the same as that described previously. In equation \rf{sigmasandwich}, $\vc \sigma$ is taken between the spinors of the created quark and antiquark. Here, instead, $\vc\sigma$ acts on the initial quark or antiquark.

Although pair creation is not assumed explicitly, if one takes the quark-antiquark structure of the outgoing mesons seriously then the physical interpretation certainly requires it. The emission of a charged pion from a neutral D-flavoured meson
\be
(u\o c)\to \pi^+~(d\o c)
\ee
takes place in the pion-emission model by the interaction with a $\pi^+$-flavoured field which changes $u$ quark into an $d$ quark, but physically one visualises the creation of a $d\o d$ pair, 
\be
(u\o c)\to (u\o d)~(d\o c).
\ee
The operator $\vc\sigma$ suggests that if one interprets the pseudoscalar-meson emission model in terms of pair-creation, the created pair is in spin triplet. This will now be shown explicitly.

\begin{figure*}
 \vspace{1cm}
\rput(2.4,5.0){Pseudoscalar-meson emission}
 \rput(-0.3,2.2){$S$}
 \rput(5.0,0.0){$s$}
 \rput(12,5.0){Non-flip, triplet}
 \rput(9.4,2.2){$S$}
 \rput(14.2,4.4){$S$}
 \rput(14.7,3.9){$\o s$}
 \rput(14.7,0.0){$s$}
  \includegraphics[width=0.85\linewidth]{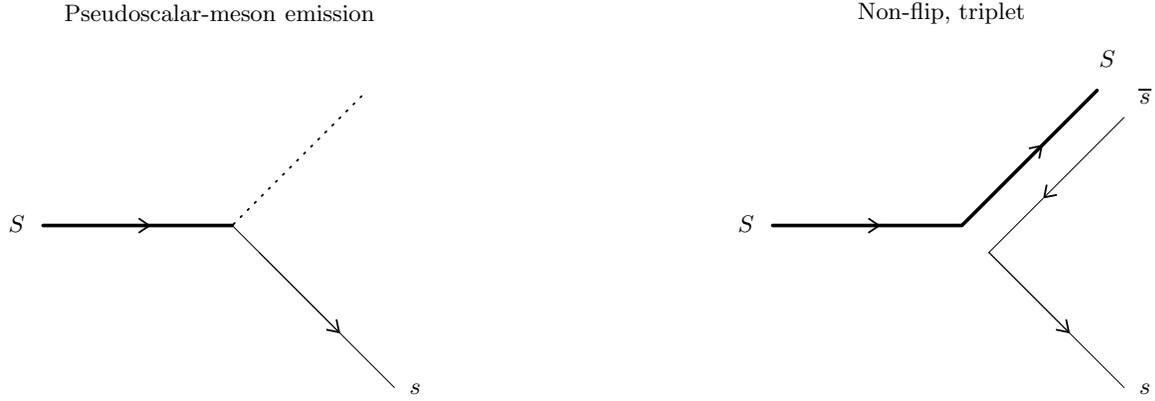}
\vspace{0.5cm}
\caption{The projections of quark spin in the pseudoscalar-meson emission model, and the corresponding process in non-flip, triplet models. }
 \label{pseudo:nonflip}
\end{figure*}

The correspondence between the spin degrees of freedom in the two approaches in shown in Fig. \ref{pseudo:nonflip}, for the case in which the pseudoscalar field interacts with a quark, as opposed to an antiquark. In both approaches the initial antiquark is treated as a spectator and so is not shown. The amplitude of interest involves incoming and outgoing quarks with spin projections $S$ and $s$ respectively. In the pseudoscalar-meson emission model the meson field acts on the spin of the initial quark, leading to a transition from $S$ and $s$. In non-flip, triplet models, the initial quark ends up in the pseudoscalar meson, and the projection $S$ of its spin is conserved; the outgoing quark with projection $s$ is that which emerges from the vacuum, flux tube, or microscopic interaction. 

In the pseudoscalar-meson emission model the matrix element of the $m$th component between quark spin projections $S$ and $s$ can be written
\be
\<s\|\sigma^{m} \|S\>=\sqrt{3}\<\tfrac{1}{2} S, 1m\|\tfrac{1}{2} s\>.
\ee
Reordering the angular momenta in the Clebsch-Gordan coefficient, this can be expressed in terms of the spin amplitude $\spinom_{s,-S}$ given by equation \rf{cgspin}, so that
\be
\<s\|\vc \sigma \|S\>=\sqrt{2}(-)^{\tfrac{1}{2}-S}\spino_{s,-S}.
\ee
The spin operator can then be written
\be
\vc \sigma =\sum_{Ss}\|s\> \sqrt{2}(-)^{\tfrac{1}{2}-S}\spino_{s,-S}\<S\|.
\label{spinop:pseudo}
\ee

To show the connection between this and the corresponding operator in non-flip, triplet models, insert a complete set in equation \rf{spinoeqn},
\be
\spino=\sum_{Ss\o s}\|Ss\o s\>\spino_{s\o s}\<S\|.
\ee
For the special case of pseudoscalar-meson emission, multiplying on the left by a state vector $\<00\|$, describing a spin singlet formed of the initial quark and the created antiquark, yields a Clebsch-Gordan coefficient
\be
\<00\|Ss\o s\>=\<00\|\tfrac{1}{2}S,\tfrac{1}{2}\o s\>\|s\>
\ee
which contains a delta function
\be
\<00\|\tfrac{1}{2}S,\tfrac{1}{2}\o s\>=\sqrt{\frac{1}{2}}(-)^{\tfrac{1}{2}-S}\delta_{S,-\o s},
\ee
so that the operator is
\be
\<00\|\spino=\sum_{Ss}\|s\>\sqrt{\frac{1}{2}}(-)^{\tfrac{1}{2}-S}\spino_{s, -S}\<S\|.
\ee
This is the same, apart from a factor of 2, as the operator \rf{spinop:pseudo}. Therefore if one interprets the pseudoscalar-meson emission model in terms of quark-pair creation, the created pair is in spin triplet.  This shows how the phenomenological pseudoscalar-emission model is related to more fundamental approaches based on quarks, flux tubes, and microscopic interactions. It also implies, since the overall operator structure can be written in the familiar $\spino\cdot\spato$ form, that the angular-momentum dependence of the pseudoscalar-meson emission model is identical to that of the models described previously. 

Le Yaouanc \etal. \cite{LeYaouanc:1972ae} identified a similarity between the matrix elements of the $\an 3P0$ and pion-emission models, for a particular baryon decay transition; the discussion above establishes the origin of this equivalence, and generalises it.

At this point it is worth making a short digression from the strong decay problem at hand. The pseudoscalar-meson emission operator features in models for the nucleon-nucleon potential arising due to the exchange of pions, or more generally, pseudoscalar mesons. The deuteron is the prototypical example: see ref. \cite{Ericson:1985hf} and references therein. Several authors have considered the possibility of mesonic analogues of the deuteron, namely molecules (deusons) formed of mesons and (anti-) mesons bound by pion exchange \cite{Voloshin:1976ap,Gutbrod:1978jb,Tornqvist:1991ks,Manohar:1992nd,Tornqvist:1993ng,Tornqvist:1993vu,Ericson:1993wy}. Within the last decade the idea has received renewed interest due to the $X(3872)$ and other states in the charmonia and bottomonia mass regions apparently correlated with meson-meson thresholds \cite{Swanson:2003tb,Close:2003sg,Pakvasa:2003ea,Swanson:2004cq,Swanson:2004pp,Tornqvist:2004qy,Swanson:2005tn,Suzuki:2005ha,AlFiky:2005jd,Liu:2008fh,Thomas:2008ja,Ding:2009vj,Dong:2009yp,Nielsen:2010ij,Oset:2011bq,Sun:2012sy,Li:2012cs,
Braaten:2013poa}.
It is interesting to consider the implications for these models of the correspondence between the pseudoscalar-meson emission and non-flip, triplet operators.

The binding or otherwise of meson-meson molecules is determined by a poorly-constrained cut-off parameter, and so it is difficult (or impossible) to predict {\it a priori} whether or not such states should exist. Nevertheless, one can make arguments as to the relative likelihood of binding in different isospin and $J^{PC}$ channels, based on their Relative Binding Number (RBN), a numerical factor determining the strength and sign of the interaction potential \cite{Tornqvist:1991ks}. 

Arguments based on RBNs are not definitive, partly because they involve only the central part of the potential and ignore the tensor part. For example, in the $np$ system the isoscalar $\an 3S1$ and isovector $\an 1S0$ states have the same RBN, but only the former (the deuteron) is bound: the extra attraction is due to the tensor potential \cite{Tornqvist:1991ks}. Nevertheless for meson-meson molecules a larger number of isospin and $J^{PC}$ combinations is possible, so the RBNs can at least discriminate the relative attraction or repulsion in a first order approximation.

The value of the RBN can be traced to the  $\vc\sigma\cdot \spato$ structure of the meson emission and absorption vertices. As has been shown, this operator can be interpreted as the creation of a $\qq$ pair in spin triplet, and is equivalent (apart from the spatial dependence) to all other non-flip, triplet models. Thus one can anticipate that the RBNs are not unique to pseudoscalar-meson exchange models, but are common to all non-flip, triplet models.

\section{Angular momentum coefficients}
\label{coefficients}

The $\an 3P0$, flux tube, $sKs$, $j^0Kj^0$, and pseudoscalar-meson emission models all have the same $\spino\cdot\spato$ operator. In this section a general expression is obtained for matrix elements in these models, generically termed non-flip, triplet models.

The relationships among matrix elements in different bases is discussed first, followed by some considerations concerning the two possible topologies for the transitions. The partial wave matrix element is then expressed in terms of a sum over coefficients $\xi$, containing the common angular-momentum dependence, multiplying spatial matrix elements which vary from model to model. The symmetries, orthogonality and zeroes of the coefficients are discussed, and tables of values, which are collected in the appendix, are introduced.

\subsection{The partial wave matrix element}
\label{matrixelement}

The aim is to obtain the matrix element for an arbitrary transition
\be
nSLJ\to n_1S_1L_1J_1+n_2S_2L_2J_2,
\ee
where the mesons are characterised by their radial quantum number $n$, and the spin, orbital and total angular momenta $S$, $L$ and $J$. Note that $S$ is used here to denote the spin (0 or 1) formed by coupling the intrinsic spins of the quark and antiquark in each meson; in the previous section $S$ and $S'$ referred to the projections of the spin of the initial $Q$ before and after scattering within microscopic models, but these labels are no longer needed since the rest of this paper applies only to models in which there is no spin flip. The colour degrees of freedom are ignored because they lead to a common factor for all matrix elements. In practical calculations one has also to include a flavour factor, which will not be discussed here, and a statistical factor to avoid double-counting identical final states.

The matrix element of $\spino\cdot\spato$ can be expressed as a sum over products of separate matrix elements of $\spino$ and $\spato$, corresponding to the spin and spatial degrees of freedom respectively.  The spatial part of the matrix element, discussed in Appendix \ref{spatial:sect}, produces a momentum-conserving delta function. The quantity relevant to  decay widths, mass shifts, and so on, is the overall matrix element modulo this delta function. It is convenient to work in the rest frame of the initial state, so that the outgoing mesons 1 and 2 have equal and opposite momenta, taken as $\mathbf p$ and $-\mathbf p$ respectively. 

Various matrix elements appear in the literature, differing according to the basis used for the two-particle state.  Suppressing on the right hand side the spin, orbital and radial quantum numbers, the matrix element in the plane-wave canonical basis
\be
M^{\vehat p}_{J_zJ_{1z}J_{2z}}\qnset
{n\p S\p L\p J\p\\
 n_1S_1L_1J_1\\
 n_2S_2L_2J_2}
=
\<J_1J_2,J_{1z}J_{2z}\vehat p|\spino\cdot\spato\|JJ_z\>
 \ee
has final states with a definite decay axis $\vehat p$, and the components $J_z$, $J_{1z}$ and $J_{2z}$ of angular momenta are measured with respect to a fixed $z$-axis.  The dependence on the magnitude $p$ of the outgoing momenta is suppressed. 

In the helicity basis the components $\lambda_1$ and $\lambda_2$ of angular momenta are measured with respect to the decay axis. Instead of the plane wave specified by $\vehat p$, one forms a spherical wave with total angular momentum $J'$ and $z$-component $J_z'$. The scalar nature of the decay operator implies that the matrix element contains delta-functions $\delta_{J'J}\delta_{J_z'J_z}$, and is independent of $J_z$; it is therefore convenient to compute the reduced matrix element,
\be
M_{\lambda_1\lambda_2}\qnset
{n\p S\p L\p J\p\\
 n_1S_1L_1J_1\\
 n_2S_2L_2J_2}=
\frac{1}{|J|}
\<J_1J_2,\lambda_1\lambda_2J\£\spino\cdot\spato\£J\>,
\ee
where $\|J\|=\sqrt{2J+1}$.

A third possibility is the partial wave basis, in which the outgoing mesons have relative orbital angular momentum $l$. A state of good total angular momentum is formed by coupling $J_1$ and $J_2$ to $j$, and then coupling $j$ and $l$ to $J$. The matrix element is again independent of $J_z$, so it is convenient to deal with the reduced matrix element,
\be
M_{jl}\qnset
{n\p S\p L\p J\p\\
 n_1S_1L_1J_1\\
 n_2S_2L_2J_2}=
 \frac{1}{|J|}
\<J_1J_2,jlJ\£\spino\cdot\spato\£J\>.
\label{pwbasis}
\ee

The partial wave basis is generally the most useful for making comparison with experiment, and it also the most convenient for the present purposes, because it lends itself to an expression in which the spatial degrees of freedom are isolated from everything else. Matrix elements in the various bases are related:
\begin{widetext}
\bea
M^{\vehat p}_{J_zJ_{1z}J_{2z}}\qnset
{n\p S\p L\p J\p\\
 n_1S_1L_1J_1\\
 n_2S_2L_2J_2}
&=&
\sum_{jj_z}
\<J_1J_{1z},J_2J_{2z}\|jj_z\>
\sum_{ll_z}
\<\mathbf{\hat p} \| ll_z\>
\<jj_z,ll_z\|JJ_z\>
M_{jl}
\qnset
{n\p S\p L\p J\p\\
n_1S_1L_1J_1\\
n_2S_2L_2J_2},
\\
M_{\lambda_1\lambda_2}\qnset
{n\p S\p L\p J\p\\
 n_1S_1L_1J_1\\
 n_2S_2L_2J_2}
&=&
\sum_{j}
\<J_1\lambda_1,J_2-\lambda_2\|j\lambda\>
\sum_{l}
\sf2l+1/{2J+1}
\<j\lambda,l0|J\lambda\>
M_{jl}
\qnset
{n\p S\p L\p J\p\\
n_1S_1L_1J_1\\
n_2S_2L_2J_2}.
\eea
\end{widetext}

Strong decay widths, as well as mass shifts in the unquenched quark model, involve sums and integrals over squared matrix elements. The relations among the relevant terms follow from the above:
\be
\int d^2\mathbf{\hat p}\sum_{J_{1z}J_{2z}}
\left|
M^{\vehat p}_{J_zJ_{1z}J_{2z}}
\right|^2
= 
\sum_{\lambda_1\lambda_2}
\left|
M_{\lambda_1\lambda_2}
\right|^2
= 
\sum_{jl}
\left|
M_{jl}
\right|^2.
\label{squaredrelations}
\ee

\subsection{Topologies}
\label{topologies}

\begin{figure*}
 \vspace{1cm}
\rput(2.4,5.0){Topology $(+)$}
 \rput(-0.5,2.2){$nSLJ$}
 \rput(5.3,4.4){$n_1S_1L_1J_1$}
 \rput(5.3,0.1){$n_2S_2L_2J_2$}
 \rput(12,5.0){Topology $(-)$}
 \rput(9.2,2.2){$nSLJ$}
 \rput(15,4.4){$n_2S_2L_2J_2$}
 \rput(15,0.1){$n_1S_1L_1J_1$}
  \includegraphics[width=0.85\linewidth]{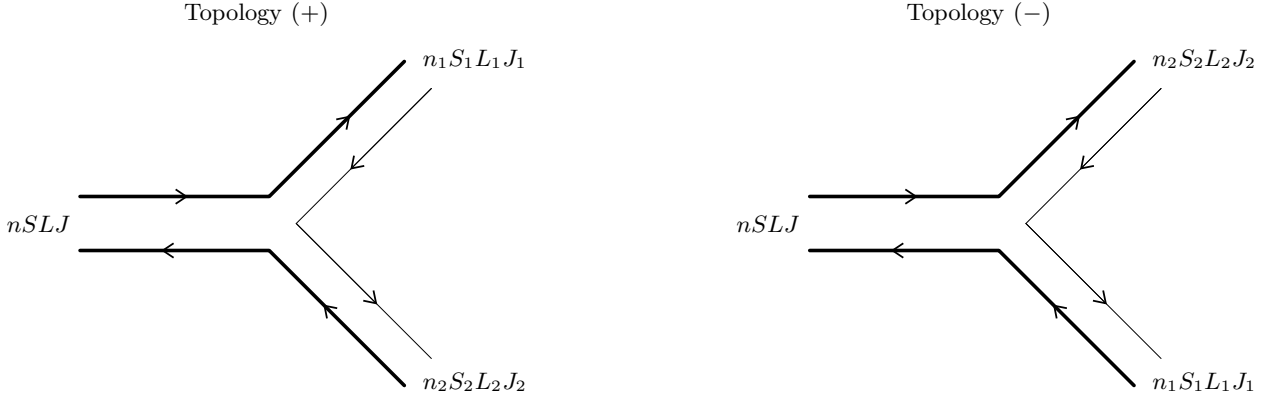}
\vspace{0.5cm}
 \caption{Quark line topologies. The two topologies differ in the arrangement of the quarks from the initial meson ($Q$ and $\Q$) and the created pair ($q$ and $\q$) in the final states. In the $(+)$ topology the mesons with quantum numbers $n_1S_1L_1J_1$ and $n_2S_2L_2J_2$ are respectively $Q\q$ and $q\Q$ states. The $(-)$ topology describes the opposite situation.}
\label{fig:topologies}
\end{figure*}

The transitions of interest are those which satisfy the OZI-rule,
\be
(Q\Q)\to(Q\q)~(q\Q).
\ee
Namely, one of the final states is formed of the initial $Q$ and created $\q$, while the other is formed of the initial $\Q$ and the created $q$. Note that the upper- and lower-case labels do not necessarily correspond to heavy and light quarks in this paper; they are used simply to distinguish the initial (anti)quarks from the created (anti)quarks. Certainly one is normally interested in the case that $q$ and $\q$ are light ($u$, $d$ or $s$ quarks), but $Q$ and $\Q$ can correspond to any flavour. Likewise, while $\q$ is certainly the antiparticle of $q$, the same need not be true of $\Q$ and $Q$.

For amplitudes with a heavy-heavy or heavy-light meson in the initial state, the final state can be defined in such a way that the coupling has a unique topology: it is always clear which of the two mesons contains the created $q$, and which the created $\q$. On the other hand, for amplitudes with solely light ($u$, $d$ or $s$) quarks in the initial state, in general it is not possible to define a final state which isolates a unique topology. In that case the total matrix element is given by the sum of two matrix elements, weighted by flavour factors.

The two topologies, denoted $(+)$ and $(-)$, are shown in Fig. \ref{fig:topologies}. In the $(+)$ topology the meson with quantum numbers $n_1S_1L_1J_1$ is formed of the initial quark and created antiquark ($Q\q$), while that with quantum numbers $n_2S_2L_2J_2$ is formed of the created quark and initial antiquark ($q\Q$). The $(-)$ topology describes the opposite situation.

To keep track of the different topologies, the spin and spatial parts of the decay operator will be written $\spino_\pm$ and $\spato_\pm$ as appropriate, although the labels will occasionally be omitted if they have no bearing on the results.

Note that the suppression of OZI-forbidden transitions
\be
(Q\Q)\to(Q\Q)~(q\q)
\ee
has a different explanation in the different models. In the $\an 3P0$ model it is assumed {\it a priori}. In the flux tube model it is because the created $q$ and $\q$ are at the ends of the broken sections of flux tube, each of which constitutes one of the outgoing mesons. In microscopic models, it is because the created $q\q$ are colour octet.

\subsection{General expression}
\label{expressionforthecoefficient}

The decay operator factorises into operators $\spino_\pm$ and $\spato_\pm$ acting separately on the spin and spatial degrees of freedom, which allows for a useful factorisation of the matrix element itself. The aim of this section is to arrive at an expression of the form

\be
M_{jl}
\qnset
{n\p S\p L\p J\p\\
n_1S_1L_1J_1\\
n_2S_2L_2J_2}_\pm
=
\sum_{L'l'}
\xi_{jl}^{L'l'}
\qnset
{S\p L\p J\p\\
S_1L_1J_1\\
S_2L_2J_2}_\pm
A_{l}^{L'l'}
\qnset
{n\p L\p\\
n_1L_1\\
n_2L_2}_\pm,
\label{eq:defining}
\ee
where the term $A$ is essentially the matrix element of the spatial operator $\spato_\pm$, and the coefficient $\xi$ contains the matrix element of the spin operator $\spino_\pm$, as well as some recoupling coefficients which depend on the angular momenta of the mesons involved. (The origin of the summation quantum numbers $L'$ and $l'$ will become apparent shortly.) This is a useful factorisation because the spatial part $A$ contains all of the model dependence, whereas the coefficients $\xi$ are common to all non-flip, triplet models. This is reflected in the arguments of the respective quantities; $A$ is a function of all of the quantum numbers (radial and orbital) associated with the spatial degrees of freedom, whereas $\xi$ is a function only of the angular momenta.

The spatial part $A$ is model-dependent in two senses. Firstly, it depends on the operator $\spato_\pm$, which distinguishes the various non-flip, triplet models. Secondly, because $A$ is essentially the overlap of the spatial wavefunctions of the initial and final states, there is model dependence in the choice of meson radial wavefunctions, and possible interactions between the final state pair.

By contrast, the angular momentum coefficients $\xi$ are common to all models, and only depend on the angular momenta of the mesons involved.

The coefficients can be derived most simply by working within a coupled formalism, to take advantage of vector recoupling coefficients and the Wigner-Eckart theorem. The spin and orbital degrees of freedom will be coupled in $SL$ order,
\be
\| (S\ot L)_J \>.
\label{slcoupling}
\ee
The first step towards the factorised form is to recouple the angular momenta of the outgoing meson pair so that the spin and spatial parts are collected together; this involves a sum over an addition spin quantum number $S'$, and two orbital quantum numbers $L'$ and $l'$. Writing out the ordering of the couplings explicitly, and suppressing the radial quantum numbers, the recoupling is
\begin{widetext}
\begin{multline}
\<(((S_1\ot L_1)_{J_1}\ot (S_2\ot L_2)_{J_2})_{j}\ot l)_J\|
\\=
\sum_{S'L'l'}
(-)^{S'+L'+l+J}
\|J_1J_2S'L'jl'\|
\ninej{S_1}{L_1}{J_1}{S_2}{L_2}{J_2}{S'}{L'}{j}
\sixj{S'}{L'}{j}{l}{J}{l'}
\<((S_1\ot S_2)_{S'}\ot ((L_1\ot L_2)_{L'}\ot l)_{l'})_J\|,
\label{recoupled[1]}
\end{multline}
where $\|J_1 J_2 \ldots l'|=\|J_1\£J_2\|\ldots\|l'\|$, and $|J_1|=\sqrt{2J_1+1}$ as before. This leaves a reduced matrix element which factors into its spin and spatial parts. Suppressing all quantum numbers irrelevant to this factorisation, the expression is 
\be
\<(S'\ot l')_J\£ \spino_\pm\cdot\spato_\pm\£ (S \ot L)_J \>
=
(-)^{J+S+l'}
\|J\|
\sixj{S }{L }{J }{l'}{S'}{1}
\<{S'}\£\spino_\pm\£S\>\<l'\£\spato_\pm\£L\>.
\label{slfactored}
\ee
The spatial part $A$ of the matrix element is to be identified with the final term in the above; restoring the previously suppressed quantum numbers, it is defined
\be
A_{l}^{L'l'}
\qnset
{n\p L\p\\
n_1L_1\\
n_2L_2}_\pm
=\frac{1}{|L|}\<((n_1L_1\ot n_2L_2)_{L'}\ot l)_{l'}\£\spato_\pm\£nL\>,
\label{spatialme}
\ee
where the factor $|L|$ is for later convenience. General formulae for $A$ in terms of integrals over meson spatial wavefunctions are given in Appendix \ref{spatial:sect}.

All of the angular momentum algebra, including the spin matrix element, in then collected into the $\xi$ coefficient,
\be
\xi_{jl}^{L'l'}
\qnset
{S\p L\p J\p\\
S_1L_1J_1\\
S_2L_2J_2}_\pm
=
\sum_{S'}(-)^{S+S'+l+l'+L'} 
|J_1J_2S'L'jl'L|
\ninej{S_1}{L_1}{J_1}{S_2}{L_2}{J_2}{S'}{L'}{j}
\sixj{S'}{L'}{j}{l}{J}{l'}
\sixj{S }{L }{J }{l'}{S'}{1}
\<(S_1\ot S_2)_{S'}\£\spino_\pm\£S\>,
\label{eq:xi}
\ee
\end{widetext}
and note that the factor of $|J|$ is absent because of equation \rf{pwbasis}.

The spin matrix element can be written in terms of a 9-$j$ coefficient expressing the spin coupling in the appropriate order. For the $(+)$ topology it is given by
\begin{multline}
\<(S_1\ot S_2)_{S'}\£\spino_+\£S\>=\\(-)^{S_2+1}|S_1S_2S'S1|\ninej{1/2}{1/2}{S_1}{1/2}{1/2}{S_2}{S}{1}{S'}.
\label{spinme}
\end{multline}
That of the $(-)$ topology is related to this by a phase,
\begin{multline}
  \<(S_1\ot S_2)_{S'}\£\spino_-\£S\>=\\(-)^{S+S_1+S_2+1} \<(S_1\ot S_2)_{S'}\£\spino_+\£S\>.
\end{multline}

An equivalent approach (using vector recoupling coefficients) is used by Bonnaz and Silvestre-Brac \cite{BonnazSilvestre-Brac99discussion}, in the specific context of the $\an 3P0$ model. One difference involves the sum over the quantum number $S'$. In their paper, the sum is contained in the equivalent of equation \rf{eq:defining}, so that their equivalent of $\xi$ carries the additional index $S'$. In this paper the sum over $S'$ is absorbed into the definition of $\xi$, so that ultimately there are fewer coefficients and a somewhat simpler formalism. 

Using symmetry properties of the 6-$j$ and 9-$j$ coefficients, one can show that the expressions above are identical to those of Bonnaz and Silvestre-Brac apart from a factor of $\sqrt 3$ due to the normalisation of the pair-creation operator, and a phase factor
\be
 (-)^{S+L+J+S_1+L_1+J_1+S_2+L_2+J_2+1}
\ee
which, apart from an overall factor of $-1$, is explained by their choice of $LS$ rather than $SL$ coupling for the meson wavefunctions.

The expression in Burns \etal. \cite{Burns:2007hk} is the same as the above, taking the spin matrix element for the $(-)$ topology. An earlier incarnation of the same expression \cite{Burns:2006rz} inadvertently omitted a factor of $(-)^{S+S_1}$  from the spin matrix element, although this has no bearing on any of the results presented there.

\subsection{Symmetries}
\label{section:symmetries}

This section concerns the symmetries relating the $\xi$ coefficients for the topologies $(+)$ and $(-)$, and under the interchange of the quantum numbers $S_1L_1J_1$ and $S_2L_2J_2$. Combining these with analogous relations for the spatial part of the matrix element leads to symmetries in the overall matrix element, and to the conservation of $C$- and $G$-parity.

Under the interchange of topologies, the only part of $\xi$ which changes is the spin part, so
\be
\xi_{jl}^{L'l'}
\qnset
{S\p L\p J\p\\
S_1L_1J_1\\
S_2L_2J_2}_\pm
=
(-)^{S+S_1+S_2+1}
\xi_{jl}^{L'l'}
\qnset
{S\p L\p J\p\\
S_1L_1J_1\\
S_2L_2J_2}_\mp.
\label{xisymm[1]}
\ee

Interchanging the quantum numbers of mesons 1 and 2 brings phase factors due to the ordering of the spin couplings, and the exchange of rows in the 9-$j$ coefficients, leading to
\begin{multline}
\xi_{jl}^{L'l'}
\qnset
{S\p L\p J\p\\
S_1L_1J_1\\
S_2L_2J_2}_\pm
=(-)^{S_1+L_1+J_1+S_2+L_2+J_2+S+L'+j+1}
\\
\times\xi_{jl}^{L'l'}
\qnset
{S\p L\p J\p\\
S_2L_2J_2\\
S_1L_1J_1\\}_\pm.
\label{xisymm[2]}
\end{multline}

This implies a selection rule for final states with the same spin, orbital and total angular momenta,
\be
\xi_{jl}^{L'l'}
\qnset
{S\p L\p J\p\\
S_1L_1J_1\\
S_1L_1J_1}_\pm
=0
\textrm{ if } (-)^{S+L'+j}=1.
\label{xisel}
\ee
This is discussed, along with other zeroes, in Section~\ref{zeroes}.

The symmetry relations for the spatial part of the matrix element $A$ are discussed in Appendix \ref{spatial:sect}, and these depend on whether the masses $M$ and $\o M$ of the initial quark and antiquark are degenerate. The discussion that follows does not, in general, apply to matrix elements with hybrid mesons; these are discussed in Section \ref{hybrids}.

In the general case ($M\ne \o M$) there are no symmetry relations in $A$ under separately interchanging either the topology, or the quantum numbers of mesons 1 and 2, but there is a relation, equation \rf{spatsymm:1}, under both interchanges combined. Together with the above symmetries in $\xi$, this gives the following symmetry for the full matrix element
\be
M_{jl}
\qnset
{n\p S\p L\p J\p\\
n_1S_1L_1J_1\\
n_2S_2L_2J_2}_\pm
=
(-)^{J_1+J_2+j+l}
M_{jl}
\qnset
{n\p S\p L\p J\p\\
n_2S_2L_2J_2\\
n_1S_1L_1J_1
}_\mp.
\label{fullsymm:1}
\ee
Thus, for example, the amplitude for ${D^*}'\to D_0\pi$ is not directly related that of ${D^*}'\to a_0 D$ (interchanging topology) or ${D^*}'\to D a_0$ (interchanging final state quantum numbers), but it is related to ${D^*}'\to \pi D_0$ (interchanging both).

For an initial meson with degenerate quarks ($M=\o M$) there are separate symmetry relations \rf{spatsymm:2} and \rf{spatsymm:3} under the interchange of the topology, or of the quantum numbers of mesons 1 and 2. This leads to the following symmetries in the full matrix element: for the topology,
\be
M_{jl}
\qnset
{n\p S\p L\p J\p\\
n_1S_1L_1J_1\\
n_2S_2L_2J_2}_\pm
=
(-)^{S+S_1+S_2+l+1}
M_{jl}
\qnset
{n\p S\p L\p J\p\\
n_1S_1L_1J_1\\
n_2S_2L_2J_2}_\mp,
\label{fullsymm:2}
\ee
and the meson quantum numbers,
\begin{multline}
M_{jl}
\qnset
{n\p S\p L\p J\p\\
n_1S_1L_1J_1\\
n_2S_2L_2J_2}_\pm
=
(-)^{S_1+J_1+S_2+J_2+S+j+1}\\
\times
M_{jl}
\qnset
{n\p S\p L\p J\p\\
n_2S_2L_2J_2\\
n_1S_1L_1J_1}_\pm.
\label{fullsymm:3}
\end{multline}
So, for example, the amplitude for $\psi'\to D_0\o D$ is related to both $\psi'\to \o D_0 D$ (interchanging topology) and $\psi'\to D \o D_0 $ (interchanging final state quantum numbers).

The second of these symmetries implies a selection rule which is closely related to \rf{xisel}, 
\be
M_{jl}
\qnset
{n\p S\p L\p J\p\\
n_1S_1L_1J_1\\
n_1S_1L_1J_1}_\pm
=0 \textrm{ if } (-)^{S+j}=1,
\ee
except that it only applies to the case $M=\o M$, and requires not only the same angular momenta quantum numbers in the final states, but also the same radial wavefunctions. This also follows from \rf{xisel} and a selection rule \rf{selspat} in the spatial matrix element.

The appendix describes how the spatial matrix element enforces the conservation of parity,
\be
(-)^{L+L_1+L_2}=(-1)^{l+1}.\label{normalparity}
\ee
Implementing this constraint, the symmetry relation between the two topologies can be written in terms of the products of the eigenvalues $C=(-)^{L+S}$ of the charge conjugation operator for neutral mesons,
\begin{multline}
M_{jl}
\qnset
{n\p S\p L\p J\p\\
n_1S_1L_1J_1\\
n_2S_2L_2J_2}_\pm
=
CC_1C_2
M_{jl}
\qnset
{n\p S\p L\p J\p\\
n_1S_1L_1J_1\\
n_2S_2L_2J_2}_\mp.
\label{cparity}
\end{multline}

As noted earlier, for transitions with heavy-heavy or heavy-light initial states, the final state can always be defined in such a way that only one of the topologies $(+)$ or $(-)$ contributes; it is always clear which of the final states contains the created $q$, and which the created $\q$. The same is not true, in general, for transitions with an initial state composed of solely light ($u$, $d$ and $s$) quarks. In such cases, the matrix elements $M_{jl}$ for the $(+)$ and $(-)$ topologies are summed, weighted by flavour factors.  

Tables of flavour factors are given elsewhere, for example in references \cite{Ackleh:1996yt,Barnes:2002mu}. If the initial meson and both of the final mesons are $C$-parity eigenstates (namely, neutral and non-strange), the flavour factors are the same for both topologies. The amplitude for such transitions is therefore proportional to 
\be
M_{jl}
\qnset
{n\p S\p L\p J\p\\
n_1S_1L_1J_1\\
n_2S_2L_2J_2}_+
+
M_{jl}
\qnset
{n\p S\p L\p J\p\\
n_1S_1L_1J_1\\
n_2S_2L_2J_2}_-,
\ee
so that equation \rf{cparity} enforces the conservation of $C$-parity.

More generally, for decays involving only non-strange states, the flavour factors for the two topologies are related by a phase $(-)^{I+I_1+I_2}$, where  $I$, $I_1$ and $I_2$ are the isospins of the mesons. The total amplitude is proportional to
\be
M_{jl}
\qnset
{n\p S\p L\p J\p\\
n_1S_1L_1J_1\\
n_2S_2L_2J_2}_+
+(-)^{I+I_1+I_2}
M_{jl}
\qnset
{n\p S\p L\p J\p\\
n_1S_1L_1J_1\\
n_2S_2L_2J_2}_-.
\ee
Each of the mesons in the transition is an eigenstate of $G$-parity, with eigenvalue $G=C(-)^I$. Equation \rf{cparity} then implies the conservation of $G$-parity. 

Note that the application of equation \rf{fullsymm:3} assumes that the initial meson has equal masses of quark and antiquark. For an isovector initial state, conservation of $G$-parity is only satisfied to the extent that the $u$ and $d$ quarks are degenerate, namely in the limit that $I$, and consequently $G$, are exactly conserved quantum numbers.

\subsection{Orthogonality}

Using the expression for the matrix element contained in ref. \cite{Burns:2007hk}, which is equivalent to the above formulation in terms of the $\xi$ coefficient, Barnes and Swanson \cite{Barnes:2007xu} derived theorems for mass shifts and mixing amplitudes in the unquenched quark model, and for meson total decay widths. Close and Thomas~\cite{Close:2009ii} applied similar ideas to mixing of hybrid and conventional mesons via coupling to meson-meson pairs.

These results arise from orthogonality properties of the 6-$j$ and 9-$j$ coefficients under summation of the spin and total angular momenta of the final states. In terms of the $\xi$ coefficients, the relevant orthogonality relation can easily be derived,
\be
\sum_{\substack{S_1S_2\\J_1J_2j}}
\xi_{jl}^{\widehat L' \widehat l'}
\qnset
{\widehat S\p\widehat L\p J\p\\
S_1L_1J_1\\
S_2L_2J_2}_\pm
\xi_{jl}^{L'l'}
\qnset
{S\p L\p J\p\\
S_1L_1J_1\\
S_2L_2J_2}_\pm
=\delta_{\widehat S S}\delta_{\widehat L L}\delta_{\widehat L' L'}\delta_{\widehat l' l'}.
\label{orthogrelation}
\ee
In Section \ref{applications} this will be used to formulate the above results in the context of $\xi$ coefficient. The orthogonality relation is also used to check the tabulated values of the $\xi$ coefficients contained in the appendix, discussed later.

\subsection{Angular momenta constraints}
\label{angularmomentaconstraints}

There are several constraints on the $\xi$s due to conservation of angular momenta, which is enforced by triangle relations in the 6-$j$ and 9-$j$ coefficients. The total angular momenta must satisfy
\be
\mathbf j+\mathbf l=\mathbf J\label{overallj}.
\ee

Further constraints arise in coupling the spatial degrees of freedom. Returning to the defining equation~\rf{eq:defining}, the matrix element is expressed as a sum over quantum numbers $L'$ and $l'$. Their allowed values are restricted, for a given set of orbital angular momenta $L$, $L_1$ and $L_2$ and a partial wave $l$,  by the triangular conditions,
\bea
\mathbf L_1+\mathbf L_2&=&\mathbf L'\label{orbitaltriangle[1]},\\
\mathbf L' +\mathbf l&=&\mathbf l'\label{orbitaltriangle[2]}, \\
\mathbf l'+\mathbf 1&=&\mathbf L\label{orbitaltriangle[3]}.
\eea
The first two are obvious from the angular momentum couplings in the state vector in equation \rf{recoupled[1]}, and the third is because $\spato$ is a vector quantity connecting $l'$ and $L$ in equation \rf{spatialme}.

In many cases these conditions imply that there is a single spatial matrix element for a given $l$ \cite{Burns:2007hk}. In particular, 
\begin{enumerate}
\item[(i)] if both of the final state mesons are S-wave ($L_1=L_2=0$) then $L'=0$ and $l'=l$ only,
\item[(ii)] if the initial state and one of the final states  are both S-wave ($L_2=L=0$) then $L'=L_1$ and $l'=1$, and
\item[(iii)] if one of the final state mesons is S-wave and the mesons are coupled in a relative S-wave ($L_2=l=0$) then $L'=L_1$ and $l'=L_1$. 
\end{enumerate}
In such cases the defining equation \rf{eq:defining} simplifies so that there is no sum over $l'$ or $L'$. This leads to direct relations among matrix elements for transitions involving mesons with the same spatial quantum numbers but different spin and total angular momenta; these are discussed in Section \ref{applications}. A simplified expression for the $\xi$ coefficient in the special case (i) is discussed in Section \ref{swavepairs}.

Additional triangle relations constrain the spin variables. In equation \rf{eq:xi} the $\xi$ coefficient is expressed as a sum over $S'$, whose allowed values are restricted
\bea
\mathbf S_1+\mathbf S_2 &=&\mathbf S',\label{spintriangle[1]}\\
\mathbf S'+\mathbf L'&=&\mathbf j,\label{dashtriangle[1]}\\
\mathbf S'+\mathbf l'&=&\mathbf J,\label{dashtriangle[2]}\\
\mathbf S'+\mathbf 1&=&\mathbf S. \label{spintriangle[2]}
\eea
The first three of these follow from the recoupling \rf{recoupled[1]}, and the last is due to the vector nature of the spin operator $\spino$ in equation \rf{spinme}. Some of these constraints lead to selection rules, discussed in Section \ref{zeroes}.

\subsection{Guide to the tables}
\label{guidetothetables}

Tables of the $\xi$ coefficients are presented in Appendix \ref{tables}, for all initial states in S-, P-, D- and F-wave. Tables \ref{tab:sss}--\ref{tab:fss} are for a final state consisting of a pair of S-wave mesons ($L_1=L_2=0$). Tables \ref{tab:sps}--\ref{tab:fps} involve a P-wave final state ($L_1=1$, $L_2=0$). Tables \ref{tab:sds}--\ref{tab:fds}  are for D-wave final states ($L_1=2$, $L_2=0$), restricted to the case that meson 2 is a pseudoscalar meson ($S_2=0$). The tables show the coefficients for the $(+)$ topology; those of the $(-)$ topology follow from equation \rf{xisymm[1]}. 

The tables are constructed to reflect the orbital angular momenta constraints \rf{orbitaltriangle[1]} and \rf{orbitaltriangle[2]}. In so doing, transitions which violate \rf{orbitaltriangle[3]} but which are otherwise allowed by angular momentum and parity emerge as selection rules, discussed in the next section.

For Tables \ref{tab:sss}--\ref{tab:fss} (which have $L_1=L_2=0$) the quantum numbers $L'$ and $l'$ are superfluous; these are an example of the special case (i) discussed in the previous section, for which there is a single spatial matrix element for a given partial wave $l$. In this case the defining equation \rf{eq:defining} no longer contains any summation variables; suppressing the quantum numbers $J_1=S_1$, $J_2=S_2$ and $L_1=L_2=0$, it reduces to
\be
M_{jl}
\qnset
{n\p S\p L\p J\p\\
n_1S_1\Pp\Pp\\
n_2S_2\Pp\Pp}_\pm
=
\xi_{jl}
\qnset
{S\p L\p J\p\\
S_1\Pp\Pp\\
S_2\Pp\Pp}_\pm
A_{l}
\qnset
{n\p L\p\\
n_1\Pp\\
n_2\Pp}_\pm.
\label{eq:definingss}
\ee
Transitions to pairs of S-wave mesons are the most important phenomenologically, and so the properties of these $\xi$ coefficients deserve special attention. In Section \ref{swavepairs} a simpler expression for these coefficients is given, along with a table of values for arbitrary $L$.

For the transitions in Tables \ref{tab:sps}--\ref{tab:fps} ($L_1=1$, $L_2=0$) and Tables \ref{tab:sds}--\ref{tab:fds} ($L_1=2$, $L_2=0$) the coefficient $L'$ is superfluous, but in general $l'$ can take on a range of values, so that the matrix element involves a summation,
\be
M_{jl}
\qnset
{n\p S\p L\p J\p\\
n_1S_1L_1J_1\\
n_2S_2\Pp\Pp}_\pm
=
\sum_{l'}
\xi^{l'}_{jl}
\qnset
{S\p L\p J\p\\
S_1L_1J_1\\
S_2\Pp\Pp}_\pm
A^{l'}_{l}
\qnset
{n\p L\p\\
n_1L_1\\
n_2\Pp}_\pm.
\label{eq:definingps}
\ee
In such cases the different values of $l'$ can be read across the table. Note, however, that some transitions with non-zero $L_1$ also fall into the categories of special cases (ii) and (iii), for which there is only one value of $l'$. Thus, for example, the transition $\an 3D1\to\, \an3P1\,\an 1S0$ involves a single S-wave coefficient
\be
\xi_{\uS}=\frac{1}{2}\sqrt{\frac{5}{6}},
\ee
but three D-wave coefficients
\bea
\xi_{\uD}^1&=&-\frac{1}{4}\sqrt{\frac{5}{6}},\\
\xi_{\uD}^2&=&\frac{3}{4}\sqrt{\frac{1}{2}},\\
\xi_{\uD}^3&=&0.
\eea

As discussed earlier, the conservation of parity is enforced by the spatial part of the matrix element, not the $\xi$s. Consequently there are $\xi$s which are non-zero but which correspond to transitions forbidden by parity. For obvious reasons it is convenient to tabulate only those $\xi$s which correspond to parity-allowed transitions. In the tables, the partial waves $l$ are chosen so that they satisfy the conservation of parity relation~\rf{normalparity}. Partial waves up to $l=6$ are shown.

(Hybrid mesons occur in parity doublets, and include states with ``abnormal'' parity, namely with $P=(-)^L$ rather than $P=(-)^{L+1}$. For transitions involving a single, or more generally an odd number, of ``abnormal'' parity hybrids, the allowed partial waves are opposite to those determined by equation \rf{normalparity}. The coefficients for these abnormal parity transitions are tabulated separately, as discussed in Section \ref{hybrids}.)

The tables in the appendix are constructed in such a way that the only entries are those which satisfy the conservation of angular momentum, equation \rf{overallj}. Explicit zeroes in the tables therefore indicate selection rules which are a consequence of the $\spino\cdot\spato$ structure; these  are discussed in the next section.

Two checks have been carried out on the numerical values in the tables. Firstly, the orthogonality relation~\rf{orthogrelation} has been verified for all tabulated values. 

Secondly, matrix elements $M_{jl}$ were obtained from the tabulated $\xi$s along with computed spatial matrix elements $A$, and then compared to the results of Barnes \etal.~\cite{Barnes:1996ff}. Examples of the calculation of $A$, in the approximation of harmonic oscillator wavefunctions of equal width, are given in Section \ref{sect:transitionamplitudes}. The computed matrix elements $M_{jl}$ were checked for a representative sample of decays from each of the following groups: $1\uS\to 1\uS\,1\uS$, $1\uP\to 1\uS\,1\uS$, $1\uD\to 1\uS\,1\uS$, $1\uF\to 1\uS\,1\uS$, $2\uS\to 1\uP\,1\uS$, $1\uD\to 1\uP\,1\uS$, $1\uF\to 1\uP\, 1\uS$, $1\uF\to 1\uD\, 1\uS$. 

Each of the matrix elements is consistent in magnitude with Barnes \etal. (Their tabulated results are larger by a factor of 2 because they sum the contributions of the two topologies.) The phases differ by a factor $(-)^{J_1+J_2+j}$ which, after accounting for their calculations corresponding to topology $(-)$ of this paper, and for the difference in $LS$ and $SL$ coupling, reduces to an irrelevant phase of $(-)^l$.

\subsection{Zeroes}
\label{zeroes}

Entries in the tables of Appendix \ref{tables} are transitions which are allowed by the conservation of angular momentum and parity. Any zeroes which appear are therefore selection rules: modes which are allowed by $J^P$ but forbidden by the decay model. The search for such decays in experiment is a test of decay model fundamentals. Assuming their validity, the selection rules also allow one to discriminate among possible interpretations of a state whose quark model classification is not fixed by $J^{PC}$ alone.  

Most of the zeroes can be understood as arising from simple angular momentum constraints. If all of mesons are spin singlets ($S=S_1=S_2=0$), the spin matrix element vanishes. This is the well-known spin-singlet selection rule, and is the obvious consequence of the action of a vector operator between scalar states, as in equations \rf{spintriangle[1]} and \rf{spintriangle[2]}. There are no examples of this rule for conventional meson decays to $\an 1S0$ pairs, since their unnatural parity ($0^-,1^+,2^-$, etc.) already forbids such transitions by $J^P$ conservation. The simplest examples therefore involve $\an1P1~ \an1S0$ and $\an 1D2 {}\an 1S0$ final states, and are marked ($^\star$) in the tables.

If both final states are spin triplets ($S_1=S_2=1$) then the sum over $S'$ runs over 0, 1 and 2. However due to zeroes in the spin matrix element \rf{spinme}, if $S=0$ only $S'=1$ contributes, whereas if $S=1$ only $S'=0$ and $S'=2$ contribute. In certain cases the values of $S'$ which survive these constraints do not satisfy the triangular condition \rf{dashtriangle[1]}, leading to a new spin-triplet selection rule. Zeroes of this nature, which are indicated ($^\dagger$) in the tables, are specific to a particular value of $j$, and therefore do not lead to absolute zeroes in decay widths if other values $j$ are possible.

For the particular cases of final states $\an 1S0{}\an 1S0$ or $\an 3S1~ \an 3S1$, both the spin-singlet and spin-triplet selection rules follow from a selection rule discussed earlier, equation \rf{xisel}.

Other zeroes  ($^\diamond$) arise because there is no combination of summation quantum numbers $S'$ and $l'$ which satisfy both triangular relations \rf{dashtriangle[1]} and \rf{dashtriangle[2]}. Such cases are again specific to a particular value of $j$.

There are further zeroes  ($^\bullet$) which arise due to cancellation between terms in the sum over $S'$. All such zeroes involve channels with $S=S_1=S_2=1$ because in no other case is there more than one term in the sum.

Additional zeroes ($^\triangle$) arise because a transition does not satisfy the triangular condition \rf{orbitaltriangle[3]} in the spatial degrees of freedom.  This will be referred to as the spatial-vector selection rule, because it is a consequence of the spatial part of the decay operator being a vector quantity. 

All remaining zeroes ($^\ddagger$) in the tables are due to an accidental zero in the 6-$j$ coefficient,
\be
\sixj{1}{2}{2}{3}{2}{2}=0.
\ee

\section{Spin-mixed states}
\label{spin-mixed}

Mesons with $J=L$ can have their intrinsic quark spins coupled to  singlet ($\an 1LL$) or triplet ($\an 3LL$). Those which are eigenstates of $G$-parity are diagonal in this basis, since the two components have opposite behaviour under $G$. Heavy-light mesons ($q\bar s$, $q\bar c$, $s\bar c$, etc.) are not $G$-parity eigenstates, so the two components mix. The aim of this section is to define the $\xi$ coefficient applicable to such states, and the discussion begins with some general properties of the  spin-mixed wavefunctions.

\subsection{Spin-mixed wavefunctions}
\label{sec:spinmix}

The physical states will be denoted by $\an ALL$ and $\an BLL$, defined in terms of a mixing matrix:
\bea
|\an ALL\> &=& |\an 3LL\>\<\an 3LL | \an ALL\> + |\an 1LL\>\<\an 1LL | \an ALL\>,\qquad \\
|\an BLL\> &=& |\an 3LL\>\<\an 3LL | \an BLL\> + |\an 1LL\>\<\an 1LL | \an BLL\>.\qquad
\eea
The mixing matrices for $\qQ$ mesons  ($K$, $\o D$, $B$) differ from those of the corresponding $\Qq$ mesons  ($\o K$, $D$, $\o B$). Denoting the former by $\an ALL$ and $\an BLL$, and the latter by $\an{A}{\overline L}{L}$ and $\an{B}{\overline L}{L}$, notice that because the spin triplet and singlet components have opposite behaviour under $C$, the requirement 
\be
C |\an XLL\>\to |\an{X}{\overline L}{L}\>,
\ee
where $X$ stands for either of the labels $A$ or $B$, implies that the mixing matrices for meson and anti-meson have a relative sign in either the spin triplet or spin singlet parts,
\be
\frac{\<\an 3LL | \an XLL \>}{\<\an 3LL |\an{X}{\overline L}{L} \>}=
-\frac{\<\an 1LL | \an XLL \>}{\<\an 1LL |\an{X}{\overline L}{L} \>}=
\pm 1.
\label{phasefactor[1]}
\ee
In what follows the positive sign will be adopted,
\be
\frac{\<\an 3LL | \an XLL \>}{\<\an 3LL |\an{X}{\overline L}{L} \>}=
-\frac{\<\an 1LL | \an XLL \>}{\<\an 1LL |\an{X}{\overline L}{L} \>}=
+ 1,
\label{phasefactor[2]}
\ee
which is the same convention as ref. \cite{Barnes:2002mu}.

In the heavy-quark limit the mixing matrices are fixed. Considering first the $q\Q$ states, in the limit that $m_{\Q}\to\infty$, the light quark angular momentum 
\be
\mathbf{J}_q=\mathbf{S}_q + \mathbf{L},
\ee
is a good quantum number, with eigenvalues  $J_q=L\pm 1/2$. The corresponding states are related to the usual states of total quark spin $S$ by recoupling
\begin{multline}
|((S_q\ot L)_{J_q} \ot S_{\Q})_{JM}\>
=
\\\sum_S
(-)^{1/2+L+S+J_q}|S,J_q|
\sixj{1/2}{1/2}{S}{L}{J}{J_q}
\\
\times
|((S_q\ot S_{\Q})_S \ot L)_{JM}\>,
\label{recoupler}
\end{multline}
The corresponding expression for $Q\q$ states differs by a phase $(-)^{S+1}$, so that sign of the $\an 1LL$ components is reversed; this is consistent with the choice \rf{phasefactor[2]}. 

Taking $\an ALL$ as the state with $J_q=L-1/2$ and $\an BLL$ as the state with $J_q=L+1/2$, the above gives the $q\Q$ wavefunctions 
\bea
|\an{A}{L}{L}\>&=&\sf{L+1}/{2L+1}|\an 3LL\> + \sf{L}/{2L+1}|\an 1LL\>,\qquad\label{hqlimit[1]}\\
|\an{B}{L}{L}\>&=&-\sf{L}/{2L+1}|\an 3LL\> + \sf{L+1}/{2L+1}|\an 1LL\>.\qquad\label{hqlimit[2]}
\eea


\subsection{The angular momentum coefficient}

The aim now is to define a new angular momentum coefficient so that the matrix elements for spin-mixed initial states can be written in the same form as before, namely
\be
M_{jl}
\qnset
{n\p X\p L\p L\p\\
n_1S_1L_1J_1\\
n_2S_2L_2J_2}_\pm
=
\sum_{L'l'}
\xi_{jl}^{L'l'}
\qnset
{X\p L\p L\p\\
S_1L_1J_1\\
S_2L_2J_2}_\pm
A_{l}^{L'l'}
\qnset
{n\p L\p\\
n_1L_1\\
n_2L_2}_\pm,
\ee
where $X$ stands for either of the labels $A$ or $B$. These matrix elements are the sums of those of the spin-singlet and spin triplet parts, weighted by the mixing angles,
\begin{widetext}
\be
M_{jl}
\qnset
{n\p X\p L\p L\p\\
n_1S_1L_1J_1\\
n_2S_2L_2J_2}_\pm
 = 
\<\an 3LL | \an XLL\> 
M_{jl}
\qnset
{n\p 3\p L\p L\p\\
n_1S_1L_1J_1\\
n_2S_2L_2J_2}_\pm
+ 
\<\an 1LL | \an XLL\> 
M_{jl}
\qnset
{n\p 1\p L\p L\p\\
n_1S_1L_1J_1\\
n_2S_2L_2J_2}_\pm,
\ee
which fixes the definition of the $\xi$ coefficient for the mixed states,
\be
\xi_{jl}^{L'l'}
\qnset
{X\p L\p L\p\\
S_1L_1J_1\\
S_2L_2J_2}_\pm
=
\<\an 3LL | \an XLL\>
\xi_{jl}^{L'l'}
\qnset
{3\p L\p L\p\\
S_1L_1J_1\\
S_2L_2J_2}_\pm
+\<\an 1LL | \an XLL\>
\xi_{jl}^{L'l'}
\qnset
{1\p L\p L\p\\
S_1L_1J_1\\
S_2L_2J_2}_\pm.
\label{mixedcoeff[1]}
\ee
Making use of this expression the coefficients for transitions of interest can be obtained from the tables in Appendix \ref{tables}.

Due to equations \rf{xisymm[1]} and \rf{xisymm[2]}, the spin triplet and singlet coefficients have opposite behaviour under the interchange of either the topology or the quantum numbers of the final states. Consequently the $\xi$ coefficient for the mixed state has no simple symmetry under either such transformation, although there is a symmetry under both transformations combined,
\be
\xi_{jl}^{L'l'}
\qnset
{X\p L\p L\p\\
S_1L_1J_1\\
S_2L_2J_2}_\pm
=(-)^{L_1+J_1+L_2+J_2+L'+j}
\xi_{jl}^{L'l'}
\qnset
{X\p L\p L\p\\
S_2L_2J_2\\
S_1L_1J_1}_\mp.
\ee
The symmetries of the spatial matrix element are discussed in the appendix, and for the case of heavy-light mesons ($M\ne \o M$) there is also no symmetry under either of the transformations separately, but there is a symmetry, equation \rf{spatsymm:1}, under the action of both transformations combined. This leads to the symmetry in the full matrix element, which is the same as equation \rf{fullsymm:1},
\be
M_{jl}
\qnset
{n\p X\p L\p L\p\\
n_1S_1L_1J_1\\
n_2S_2L_2J_2}_\pm
=
(-)^{J_1+J_2+j+l}
M_{jl}
\qnset
{n\p X\p L\p L\p\\
n_2S_2L_2J_2\\
n_1S_1L_1J_1}_\mp.
\ee

Compared to those of $\an XLL$ mesons, the  $\xi$ coefficients for $\an X{\o L}L$ mesons have a relative sign between the singlet and triplet parts. With the choice \rf{phasefactor[2]} these are given by
\be
\xi_{jl}^{L'l'}
\qnset
{ X\p \o L\p L\p\\
S_1L_1J_1\\
S_2L_2J_2}_\pm
=
\<\an 3LL | \an XLL\>
\xi_{jl}^{L'l'}
\qnset
{3\p L\p L\p\\
S_1L_1J_1\\
S_2L_2J_2}_\pm
-\<\an 1LL | \an XLL\>
\xi_{jl}^{L'l'}
\qnset
{1\p L\p L\p\\
S_1L_1J_1\\
S_2L_2J_2}_\pm.
\label{mixedcoeff[2]}
\ee
\end{widetext}

Clearly there is no direct symmetry relation between the $\xi$s for mesons and anti-mesons, but this could be anticipated because the amplitude for $D_1'\to D_0\pi$, for example, need not be related to the amplitude $\o D_1'\to a_0\o D$, obtained by swapping the initial meson for its antiparticle, but keeping the topology and final state quantum numbers the same.

However the opposite behaviour of the triplet and singlet parts leads to relations under the combined action of swapping the initial meson for its antiparticle, and interchanging either the topology,
\be
\xi_{jl}^{L'l'}
\qnset
{X\p L\p L\p\\
S_1L_1J_1\\
S_2L_2J_2}_\pm
=(-)^{S_1+S_2}
\xi_{jl}^{L'l'}
\qnset
{X\p \o L\p L\p\\
S_1L_1J_1\\
S_2L_2J_2}_\mp,
\ee
or the quantum numbers of mesons 1 and 2,
\begin{multline}
\xi_{jl}^{L'l'}
\qnset
{X\p L\p L\p\\
S_1L_1J_1\\
S_2L_2J_2}_\pm
=(-)^{S_1+L_1+J_1+S_2+L_2+J_2+L'+j}
\\
\times
\xi_{jl}^{L'l'}
\qnset
{X\p \o L\p L\p\\
S_2L_2J_2\\
S_1L_1J_1}_\pm.
\end{multline}
So, for example, the matrix element for $D_1'\to D_0\pi$ is related to that of $\o D_1'\to \o D_0\pi$ (interchanging topologies) and $\o D_1'\to \pi \o D_0$ (final state quantum numbers). The matrix element obtains an additional phase factor from the spatial matrix element due to the interchange of initial quark and antiquark masses.

The orthogonality relation for the spin-mixed $\xi$s,

\be
\sum_{\substack{S_1S_2\\J_1J_2j}}
\xi_{jl}^{\widehat L' \widehat l'}
\qnset
{\widehat X\p L\p L\p\\
S_1L_1J_1\\
S_2L_2J_2}_\pm
\xi_{jl}^{L'l'}
\qnset
{X\p L\p L\p\\
S_1L_1J_1\\
S_2L_2J_2}_\pm
=\delta_{\widehat X X}\delta_{\widehat L' L'}\delta_{\widehat l' l'}
\ee
follows from that of the ordinary $\xi$s, and the orthogonality of $\an ALL$ and $\an BLL$.

Coefficients can also be defined for transitions with final states of mixed spin, analogously to the above. Ref. \cite{Burns:2007hk} considered the angular-momentum dependence of decays of charmonia involving spin-mixed final states, such as $D_1\o D$, in the heavy-quark limit. Some selection rules arise which can be used to discriminate among $\an 3S1$, $\an 3D1$ and hybrid interpretations of vector charmonia, and these will be discussed in the language of the $\xi$s in Section \ref{applications}.

It is worth mentioning here an alternative approach to strong decays of heavy-light states, which assumes only the conservation of quark angular momentum and is valid in the heavy-quark limit. Consider the transition from one heavy-light $q\Q$ state to another, where the initial state has total and light quark angular momenta $J$ and $J_q$, and the final state $J'$ and $J_q'$. Their wavefunctions involve the vector couplings
\bea
\ve J &=&\ve J_q+\ve S_{\Q}, \\
\ve J'&=&\ve J_q'+\ve S_{\Q}.
\eea
A transition between the two via the emission of a light meson with total angular momentum $J_h$ requires the conservation of both the total and light quark angular momenta,
\bea
\ve J&=&\ve J_h+\ve J',\\
\ve J_q&=&\ve J_h+\ve J_q'.
\eea
The corresponding matrix element is therefore proportional to a vector recoupling coefficient \cite{Isgur:1991wq,Eichten:1993ub},
\begin{multline}
\< (J_h\ot(J_q'\ot S_{\Q})_{J'})_J \| ((J_h\ot J_q')_{J_q}\ot S_{\Q})_J\>
\\
=
(-)^{J_h+J_q'+1/2+J}
|J_q J'|
\sixj{1/2}{J_q'}{J'}{J_h}{J}{J_q}.
\label{hqrecouple}
\end{multline}
The approach yields relations among decay amplitudes. Since these are due to angular momentum conservation, the same relations must also be satisfied by the matrix elements of non-flip, triplet models. In the next section this is verified for the particular case of coupling to a pair of S-wave mesons. Nevertheless the two approaches are  not equivalent: the above approach involves no assumptions for the nature of the created $q\q$ pair which drives the transition, unlike in non-flip, triplet models. Whereas all relations valid in the above approach must also hold in non-flip, triplet models, the converse is not true in general. 

Several authors have supplemented the above formula with additional assumptions, in order to make absolute predictions. Chen \etal. \cite{Chen:2011rr,Chen:2012zk} assume the $\an 3P0$ model, so that their angular momentum algebra is equivalent to that described in this paper. Goity and Roberts \cite{Goity:1998jr} work within the chiral quark model.

\section{S-wave meson pairs}
\label{swavepairs}
\begingroup\squeezetable
\begin{table*}
\begin{tabular}{|>{$}l<{$}|>{$}l<{$}|>{$}c<{$}|>{$}c<{$}|>{$}c<{$}|>{$}c<{$}|>{$}c<{$}|>{$}c<{$}|}
\hline
&	&\an3L{L-1}	&\an1L{L  }		&\an3L{L  }	&\an3L{L+1}				&\an ALL		&\an BLL\\
\hline
\an 1S0{}\an 1S0	&\an1{L-1}~	&\nf1/2\sf{2L+1}/{2L-1}	&\phantom{space}\phantom{space}&	&			&\phantom{\sf1/1}\phantom{\sf1/1}	&\\	
\an 1S0{}\an 1S0	&\an1{L+1}~	&		&\phantom{space}\phantom{space}	&	&\nf1/2\sf{2L+1}/{2L+3}		&\phantom{\sf1/1}\phantom{\sf1/1}	&\\	
\hline
\an 3S1{}\an 1S0	&\an3{L-1}~	&\nf1/2\sf{(2L+1)(L-1)}/{L(2L-1)}&-\nf1/2	&\nf1/2\sf{L+1}/L&			&\nf 1/2\sf{1}/{L(2L+1)}	&-\sf{L+1}/{2L+1}	\\	
\an 3S1{}\an 1S0	&\an3{L+1}~	&		&-\nf1/2	&-\nf1/2\sf{L}/{L+1}&-\nf1/2\sf{(2L+1)(L+2)}/{(L+1)(2L+3)}	&-\sf L/{2L+1}		&-\nf1/2\sf{1}/{(L+1)(2L+1)}\\	
\hline
\an 1S0{}\an 3S1	&\an3{L-1}~	&-\nf1/2\sf{(2L+1)(L-1)}/{L(2L-1)}&-\nf1/2	&-\nf1/2\sf{L+1}/L&			&-\nf1/2\sf{2L+1}/{L}	&0	\\	
\an 1S0{}\an 3S1	&\an3{L+1}~	&		&-\nf1/2	&\nf1/2\sf{L}/{L+1}&\nf1/2\sf{(2L+1)(L+2)}/{(L+1)(2L+3)}	&0		&-\nf1/2\sf{2L+1}/{L+1}\\	
\hline
\an 3S1{}\an 3S1	&\an1{L-1}~	&\nf1/2\sf{2L+1}/{3(2L-1)}	&	&	&			&\phantom{\sf1/1}\phantom{\sf1/1}	&\\	
\an 3S1{}\an 3S1	&\an3{L-1}~	&0^\dag		&-\sf1/2	&0\dag	&			&-\sf{L}/{2(2L+1)}	&-\sf{L+1}/{2(2L+1)}\\	
\an 3S1{}\an 3S1	&\an5{L-1}~	&-\sf{(2L-3)(L-1)}/{6L(2L-1)}	&0^\dag	&-\sf{L-1}/{2L}	&-1			&-\sf{(L+1)(L-1)}/{2L(2L+1)}	&\sf{(L-1)}/{2(2L+1)}\\	
\an 3S1{}\an 3S1	&\an1{L+1}~	&		&	&	&\nf1/2\sf{2L+1}/{3(2L+3)}		&\phantom{\sf1/1}\phantom{\sf1/1}	&\\	
\an 3S1{}\an 3S1	&\an3{L+1}~	&		&-\sf1/2	&0\dag	&0^\dag			&-\sf{L}/{2(2L+1)}	&-\sf{L+1}/{2(2L+1)}\\	
\an 3S1{}\an 3S1	&\an5{L+1}~	&-1		&0^\dag	&-\sf{L+2}/{2(L+1)}&-\sf{(L+2)(2L+5)}/{6(L+1)(2L+3)}	&-\sf{L+2}/{2(2L+1)}	&\sf{L(L+2)}/{2(L+1)(2L+1)}\\	
\hline
\end{tabular}
\caption{The $\xi$ coefficients for coupling to a pair of S-wave mesons ($L_1=L_2=0$), for the $(+)$ topology. The final two columns are for the decay of spin-mixed states with mixing angle given by wavefunctions \rf{hqlimit[1]} and \rf{hqlimit[2]} corresponding to the heavy-quark limit; the state  $\an ALL$ has $J_q=L-1/2$, while $\an BLL$ has $J_q=L+1/2$. }
\label{xiss}
\end{table*}
\endgroup

Final states consisting of a pair of S-wave mesons ($L_1=L_2=0$) are particularly important phenomenologically. For strong decays, these channels constitute the vast majority of experimental data. In the unquenched quark model the strongest influence on meson masses and mixing is from channels whose threshold is nearby in mass, which, for most charmonia and bottomonia, are S-wave meson pairs.

These transitions are the first of three special cases (discussed in Section \ref{angularmomentaconstraints}) for which  there is a single spatial matrix element for a given partial wave $l$. Consequently there is no sum over $L'$ or $l'$ in the expression for $M_{jl}$, equation \rf{eq:definingss}. Zeroes in the 6-$j$ and 9-$j$ coefficients lead to a simpler form for the coefficient,
\begin{multline}
\xi_{jl}
\qnset
{S\p L\p J\p\\
S_1\Pp\Pp\\
S_2\Pp\Pp}_\pm
=(-)^{S+l+J}|L|\sixj{S}{L}{J}{l}{j}{1}\\
\times\<(S_1\ot S_2)_{j}\£\spino_\pm\£S\>.
\end{multline}
In Table \ref{xiss} these coefficients are tabulated as a function of $L$ for various initial states 
$\an3L{L-1}$,
$\an1L{L  }$,
$\an3L{L  }$ and
$\an3L{L+1}$. The allowed partial waves, assuming the transition involves only conventional mesons, are $l=L-1$ and $l=L+1$. 

 Table \ref{xiss} also shows the corresponding coefficients for spin-mixed states $\an ALL$ and $\an BLL$ with wavefunctions \rf{hqlimit[1]} and \rf{hqlimit[2]} from the heavy-quark limit.  Notice the zeros in the $\an1S0{}\an 3S1$ channel. The state with $J_q=L-1/2$ does not couple to the $L+1$ partial wave, while that with $J_q=L+1/2$ does not couple to the $L-1$ partial wave. These zeroes express the conservation of the light quark angular momentum, and are therefore not unique to non-flip, triplet models. The appearance of these zeroes has been discussed elsewhere in the specific contexts of the spin-mixed P-wave \cite{
Rosner:1985dx,Godfrey:1986wj,Blundell:1995au,Barnes:2002mu,Godfrey:2005ww,Falk:1992cx,Isgur:1991wq} and D-wave \cite{Close:2005se} mesons. The table shows that the selection rule generalises to arbitrary $L$.

In the previous section an alternative approach to heavy-light meson transitions was mentioned, which starts from the heavy-quark limit and assumes only the conservation of angular momentum, leading to matrix elements proportional to a recoupling coefficient, equation \rf{hqrecouple}. To make the connection between that approach and the non-flip, triplet coefficients of Table \ref{xiss}, consider a $q\Q$ meson decaying in topology $(+)$, in which case the outgoing light meson has intrinsic angular momentum $J_1$. The coefficients in Table \ref{xiss} have final states of definite $l$, whereas those of equation \rf{hqrecouple} have definite $J_h$, which is formed of the vector addition of $J_1$ and $l$. In the special case of pseudoscalar-meson emission ($J_1=0$) the two bases coincide, and the corresponding coefficients can be directly compared. In equation \rf{hqrecouple} one has either $J_q=L-1/2$ (for the decay of $\an 3L{L-1}$ and $\an ALL$) or $J_q=L+1/2$  ($\an 3L{L+1}$ and $\an BLL$),
 and for the S-wave final states of Table \ref{xiss}, $J_q'=1/2$. The relative strengths of the  amplitudes predicted by equation \rf{hqrecouple} are consistent with those of Table \ref{xiss}, up to phase factors.

\section{Hybrid mesons}
\label{hybrids}

This section concerns the angular-momentum dependence of matrix elements involving hybrid mesons. Within the flux tube model, transitions are described by the same $\xi$ coefficients as those discussed earlier, although for some transitions additional tables of values are required (and are presented in Appendix \ref{tables}). The angular-momentum dependence in constituent gluon models is in general different, although it turns out that for negative parity hybrids in most channels of interest the ordinary $\xi$s  are valid.

\subsection{Flux tube models}

States in the flux tube model are classified by excitations of both quark and flux tube degrees of freedom, and within the adiabatic approximation the radial energy dependence of the flux tube defines the potential energy of the quarks \cite{Isgur:1983wj,Isgur:1984bm,Merlin:1985mu,Isgur:1985vy,Barnes:1995hc}. Conventional mesons are those with the flux tube in its ground state, so that the orbital angular momentum $L$ of the quark-antiquark pair is a good quantum number. Hybrid mesons are those in which the flux tube carries angular momentum, and are no longer eigenstates of quark orbital angular momentum; they are classified by the projection $\Lambda$ of the flux tube angular momentum along the quark-antiquark axis, and of the total orbital angular momentum $L$, formed of the quark and flux tube angular momenta.

Hybrids occur in parity doublets, which arise from linear combinations of degenerate $\Lambda=|\Lambda|$ and $\Lambda=-|\Lambda|$ states. It is therefore convenient to label the states by $|\Lambda|$ and a parity quantum number $P$. Those with parity $P=(-)^{L+1}$ will be described as having normal parity, namely, the same as the corresponding conventional meson with the same $L$. Their partners with $P=(-)^L$ will be described as having abnormal parity. The quantum numbers $|\Lambda|=0,1,2\ldots$ are labelled $\Sigma, \Pi, \Delta\ldots$ in analogy to S, P, D\ldots, as in molecular physics.

The spectrum of hybrids is constrained by $|\Lambda|\leq L$, so the lightest hybrid states are those with  $|\Lambda|=1$ and $L=1$ ($\Pi {\uP}$ states). These are the hybrids which are of most phenomenological interest, although in principle there exist higher-lying states such as those with  $|\Lambda|=1$ and $L=2$ ($\Pi {\uD}$) or $|\Lambda|=2$ and $L=2$ ($\Delta {\uD}$).

In the lightest ($\Pi\uP$) hybrid family, labelling the states with an adapted atomic notation $^{2S+1}|\Lambda|L_J^P$, the $J^{PC}$ quantum numbers of the normal parity states are
\begin{align}
^1\Pi\uP_1^+ 
&&^3\Pi\uP_0^+ 
&&^3\Pi\uP_1^+ 
&&^3\Pi\uP_2^+ \nonumber\\
1^{++}
&&0^{+-}
&&1^{+-}
&&2^{+-}
\end{align}
and those of the abnormal parity states are
\begin{align}
^1\Pi\uP_1^- 
&&^3\Pi\uP_0^- 
&&^3\Pi\uP_1^- 
&&^3\Pi\uP_2^- \nonumber\\
1^{--}
&&0^{-+}
&&1^{-+}
&&2^{-+}
\label{abnormalspectrum}
\end{align}
Three states ($0^{+-}$, $1^{-+}$, $2^{+-}$) have exotic $J^{PC}$ not possible for conventional mesons.

While the spatial wavefunctions for hybrids differ from those of conventional mesons, involving Wigner $D$-functions in place of ordinary spherical harmonics, the critical point for this paper is that they are still tensors of rank $L$. Their overall wavefunctions can be written as a tensor product of $S$ and $L$ exactly as in the case of conventional mesons, equation \rf{slcoupling}, and the derivation of the $\xi$ coefficient assumes only this tensor structure. 

Consequently hybrid mesons with orbital angular momentum $L$ can be treated on the same footing as conventional mesons with the same $L$. In particular, one can write an expression analogous to equation \rf{eq:defining}
\begin{multline}
M_{jl}
\qnset
{n\p |\Lambda|\p P\p S\p L\p J\p\\
n_1|\Lambda_1|P_1S_1L_1J_1\\
n_2|\Lambda_2|P_2S_2L_2J_2}_\pm
\\=
\sum_{L'l'}
\xi_{jl}^{L'l'}
\qnset
{S\p L\p J\p\\
S_1L_1J_1\\
S_2L_2J_2}_\pm
A_{l}^{L'l'}
\qnset
{n\p|\Lambda|\p P\p L\p\\
n_1|\Lambda_1|P_1L_1\\
n_2|\Lambda_2|P_2L_2}_\pm,
\label{eq:defining:flux}
\end{multline}
in which the  $\xi$ coefficient is given by the usual expression \rf{eq:xi}, and the quantum numbers $|\Lambda|$ and $P$ describing the flux tubes enter only into the spatial part $A$.

For present purposes the only important feature of the spatial part $A$ is that it must enforce the conservation of parity --  namely, it is zero unless
\be
PP_1P_2=(-)^l.
\ee
If all of the states in the transition (conventional or hybrid) have normal parity, or indeed if there is an even number of abnormal parity mesons, then the constraint reduces to equation \rf{normalparity},
\be
(-)^{L+L_1+L_2}=(-)^{l+1}.\label{nparity}
\ee
In Appendix \ref{spatial:sect} this parity relation is shown to arise from the spatial matrix element for the case of transitions involving only conventional mesons. On the other hand if there is an odd number of abnormal parity mesons in a transition, the spatial matrix element leads to the opposite constraint,
\be
(-)^{L+L_1+L_2}=(-)^l.\label{aparity}
\ee

Because all transitions involving conventional mesons are normal parity transitions, the tables in Appendix \ref{tables}, as well as Table \ref{xiss}, are constructed in such a way that the partial waves $l$ are those which satisfy equation \rf{nparity}. From the above discussion, it follows that these coefficients can also be applied to some transitions involving hybrid mesons, namely those in which there is an even number of abnormal parity states. Thus for example the $\xi$ coefficients for transitions 
\begin{align}
& \Pi\uP^+\to \uS + \uS,\\
& \Pi\uP^+\to \uP + \uS,\\
& \Pi\uP^+\to \uD + \uS,
\end{align}
can be read off Tables \ref{tab:pss}, \ref{tab:pps} and \ref{tab:pds}, and likewise those for hybrid production
\begin{align}
& \uS\to \Pi\uP^+ + \uS,\\
& \uP\to \Pi\uP^+ + \uS,\\
& \uD\to \Pi\uP^+ + \uS,\\
& \uF\to \Pi\uP^+ + \uS,
\end{align}
from Tables \ref{tab:sps}, \ref{tab:pps}, \ref{tab:dps} and \ref{tab:fps}. Similarly, for the ``cascade'' transitions
\begin{align}
& \Pi\uP^+\to \Pi\uP^+ + \uS,\\
& \Pi\uP^-\to \Pi\uP^- + \uS,
\end{align}
the coefficients are those of Table \ref{tab:pps}.

On the other hand, for abnormal parity transitions satisfying equation \rf{aparity}, additional tables are required. Those for transitions
\begin{align}
& \Pi\uP^-\to \uS + \uS,\\
& \Pi\uP^-\to \uP + \uS,\\
& \Pi\uP^-\to \uD + \uS,
\end{align}
are in Tables \ref{tab:pss:abnormal}, \ref{tab:pps:abnormal} and \ref{tab:pds:abnormal}, and note that Table  \ref{tab:pps:abnormal} can also be used for the hybrid cascades
\begin{align}
& \Pi\uP^+\to \Pi\uP^- + \uS,\\
& \Pi\uP^-\to \Pi\uP^+ + \uS.
\end{align}
The coefficients for negative parity hybrid production
\begin{align}
& \uS\to \Pi\uP^- + \uS,\\
& \uP\to \Pi\uP^- + \uS,\\
& \uD\to \Pi\uP^- + \uS,
\end{align}
are not tabulated.

It is worth emphasising that the $\xi$ coefficients apply not only to the ordinary flux tube model for hybrid transitions \cite{Dowrick:1986ub,Isgur:1985vy, Close:1994hc}, but also to modified models with a different spatial operator \cite{Page:1998gz,Swanson:1997wy}, and indeed to any non-flip, triplet models which treat hybrids as states with orbital angular momentum $L$. Close and Dudek \cite{Close:2003af} computed the decays of hybrids to $\an 3S1{}\an1S0$ in the pion emission model, and noted that the angular-momentum dependence of their amplitudes correlated with those of the ordinary flux tube model. The correspondence between the two is a consequence of their both being non-flip, triplet models. 

The discussion of the spatial matrix element $A$ in Appendix \ref{spatial:sect} applies to transitions involving conventional mesons only. Generalising this to hybrid mesons is beyond the scope of this paper, however some remarks are in order. 

The symmetry relations \rf{spatsymm:1}, \rf{spatsymm:2} and \rf{spatsymm:3} for the spatial matrix elements were used, in Section~\ref{section:symmetries}, to derive symmetry relations in the full matrix element \rf{fullsymm:1}--\rf{fullsymm:3}, and these are responsible for the conservation of $C$- and $G$-parity. The corresponding symmetry relations can be derived for transitions involving hybrid mesons, and lead to the same conservation laws.

The selection rule \rf{selspat} for conventional meson transitions can easily be generalised to the case of hybrids in the initial state \cite{Burns07production}, and is responsible for the well-known result that $\Pi\uP$ hybrids are forbidden to decay to identical S-wave meson pairs.

\subsection{Constituent gluon models}
\label{constglue}

In the constituent gluon model, the gluonic degrees of freedom of a hybrid meson are manifest as a massive vector particle. This leads to a simple model for decays in which the gluon annihilates into a $\qq$ pair~\cite{LeYaouanc:1984gh,Donnachie:1993hm,Kalashnikova:1993xb,Iddir:1998yc,Kalashnikova:2008qr}. In general the angular-momentum dependence of the matrix elements differs from that of non-flip, triplet models, but it turns out that for the cases of most phenomenological interest, the dependence is the same. This is particularly interesting because the spectrum of light hybrids in a dynamical lattice calculation appears to be more consistent with the constituent gluon model than with the  flux tube model~\cite{Dudek:2011bn}. The connection between the two is discussed in references \cite{Buisseret:2006wc,Buisseret:2007ed}.

Symbolically the state vector for a general hybrid state has the form
\be
\| (S\ot (L_{\QQ} \ot (L_g\ot 1)_{J_g})_L )_J\>,
\ee
where $L_{\QQ}$ is the orbital angular momentum of the $\QQ$ pair, and $J_q$ is the angular momentum of the gluon, formed by coupling its orbital angular momentum $L_g$ and its intrinsic angular momentum (represented by ``1'' in the above). The gluon annihilates into a $\qq$ pair which, due to the vector quantum numbers of the gluon, is in a $\an3S1$ state. Consequently in computing the transition amplitude, the ``1'' in the above state vector can be replaced by $\spino_\pm$, the spin triplet wavefunction of the $\qq$ pair.

In practice one is usually interested in the case in which the quark-antiquark pair is in S-wave ($L_{\QQ}=0$), so that the state vector has the form
\be
\| (S\ot (L_g\ot \spino_\pm)_{J_g})_J\>\label{cgstate}.
\ee

One might expect the lightest hybrids in this picture to have the gluon in S-wave ($L_g=0$), namely with $J_g^{P_gC_g}=1^{--}$. This leads to a spectrum of states with the same $J^{PC}$ quantum numbers as conventional P-wave mesons. 

In practice, in most approaches the lightest hybrid has the gluon in P-wave ($L_g=1$), namely with $J_g^{P_gC_g}=1^{+-}$. This is a feature of the bag model~\cite{Barnes:1982tx,Chanowitz:1982qj}, Coulomb-gauge QCD \cite{Guo:2008yz}, and calculations based on lattice adiabatic potentials \cite{Bali:2003jq}, and is generally assumed in constituent gluon models~\cite{Donnachie:1993hm,Kalashnikova:1993xb}. The spectrum of $J^{PC}$ quantum numbers of such states is identical to that of the abnormal parity $\Pi\uP$ multiplet, equation \rf{abnormalspectrum}. It turns out that the angular-momentum dependence of their decay amplitudes is also the same.

To see how this comes about, consider the reduced matrix element of the general state vector \rf{cgstate} with an arbitrary final state. The first step is to recouple the outgoing meson wavefunctions into states of good total spin and orbital angular momenta $S'$ and $L'$, as in equation \rf{recoupled[1]}. The resulting reduced matrix element factorises into spin and spatial parts, 
\begin{multline}
\<(S'\ot l')_J\£ (S\ot (L_g\ot \spino_\pm)_{J_g})_J\>
=\\
(-)^{S+J}\frac{|1,J|}{|l'|}
\sixj{S}{1}{J}{L_g}{S'}{1}
\<S'\£\spino_\pm\£S\>\<l'\£L_g\>,
\end{multline}
and the expression is very similar to equation \rf{slfactored}. Apart from the 6-$j$ coefficient, the expressions are identical up to numerical factors which can be absorbed into a definition of the spatial matrix element, in analogy to equation \rf{spatialme}. Due to the conservation of orbital angular momentum, the spatial matrix element will contain $\delta_{l'L_g}$, so that after summation over $l'$ the only remaining difference between equation \rf{slfactored} and the above expression is a single entry in the 6-$j$ coefficients, which reads $L$ in equation \rf{slfactored} and 1 in the above. The $\Pi\uP$ multiplet in the flux tube model has $L=1$, so the expressions are equivalent.

Thus if the spatial matrix element is defined appropriately, the matrix element for transitions from  the $J_g^{P_gC_g}=1^{+-}$ hybrids in the constituent gluon model can be expressed in terms of the same $\xi$ coefficient as that of the corresponding $\Pi\uP$ multiplet in the flux tube model. The validity of the result is confirmed by comparing the $\xi$ coefficients in Tables \ref{tab:pss:abnormal} and \ref{tab:pps:abnormal} with the spin-recoupling coefficients tabulated in references~\cite{Donnachie:1993hm,Kalashnikova:1993xb,Kalashnikova:2008qr}. Some implications of the correspondence between the different approaches are discussed in the next section.

Hybrid decays have also been discussed in a relativistic model \cite{Poplawski:2004qj} and with QCD sum rules \cite{Chen:2010ic,Huang:2010dc}. The angular-momentum dependence of these approaches is more complicated than that of non-flip, triplet models.

\section{Some applications}
\label{applications}

The $\xi$ coefficients have many possible applications, and this final section gives an introduction to the terrain. The discussion focusses on several results which have already appeared in the literature in other contexts. The purpose of doing this is to introduce the range of possible physics questions which can be addressed with the $\xi$s, and to demonstrate that  the existing results are more general than the particular models within which they have been derived: they are common to all non-flip, triplet models. Further applications of the $\xi$s will be discussed in future work.
\\
\subsection{Transition matrix elements}
\label{sect:transitionamplitudes}

The coefficients can be used as a practical tool for calculating transition matrix elements, the basic ingredients of strong decay widths and, in the unquenched quark model, mass shifts, meson-meson wavefunctions, and loop-induced spectroscopic mixing. By means of the $\xi$s the computation of a partial wave matrix element $M_{jl}$ reduces to that of calculating the spatial part $A$. The various non-flip, triplet models involve different $A$s, but the $\xi$s are common to all models.

In Appendix \ref{spatial:sect} a general expression for $A$ is given, as an integral (in both position and momentum space) over spatial wavefunctions. This can be used as a starting point for future calculations. It applies to $\an 3P0$ and flux tube models, regardless of the functional form of the spatial pair-creation amplitude. The corresponding expressions for microscopic and pseudoscalar-meson emission models are similar.

As an example of the approach, consider a transition in the $\an 3P0$ model, using the spatial operator \rf{acklehspatial}. From equation \rf{spatialamplitude}, the spatial matrix element is given by 
\begin{widetext}
\be
A_{l}^{L'l'}
 \qnset
 {n\p L\p\\
 n_1L_1\\
 n_2L_2}_+
=
-\frac{2^{3/2}\gamma}{|L|}\int d^3\ve q \int d^3\ve k
 \<((n_1 L_1 \ot n_2 L_2)_{L'}\ot p l)_{l'}\£\half(\ve q+\ve k),\half(\ve q+\ve k),\ve q-\ve k\>
\ve k
\<\ve q\£ nL\>
\ee
\end{widetext}
if all quark masses are taken to be the same. Using harmonic oscillator wavefunctions as given in, for example, ref. \cite{Barnes:1996ff}, the integrals can be evaluated analytically. Specialising to the case of equal wavefunction size, all matrix elements are proportional to a common factor
\be
\frac{\gamma}{\pi^{1/4}\beta^{1/2}}e^{-x^2/12},
\ee
where $\beta$ is the parameter controlling the size of the harmonic oscillator wavefunction, and $x=p/\beta$.

For illustration consider transitions of the type $1\uD \to 1\uP\,1\uS$. Parity allows transitions in S-, D-, G- and I-wave, but the last of these is forbidden by the spatial-vector selection rule. The non-zero spatial matrix elements are, modulo the common factor above,
\bea
A_{\uS}&=&\frac{2^6}{3^{7/2}}\left(1-\frac{5x^2}{18}+\frac{x^4}{135}\right),	\\
A_{\uD}^1&=&\frac{2^{15/2}}{3^{11/2} \cdot 5}x^2\left(1-\frac{x^2}{6}\right),\\
A_{\uD}^2&=&\frac{2^{11/2}}{3^4\cdot  5^{1/2}}x^2,\\
A_{\uD}^3&=&\frac{2^{11/2}\cdot 7^{1/2}}{3^4\cdot  5}x^2\left(1-\frac{x^2}{21}\right),\\
A_{\uG}&=&\frac{2^{13/2}}{3^{11/2}\cdot 5 \cdot 7^{1/2}}x^4,	
\eea
where the numerical labels on the $A_{\uD}$ refer to the different values of $l'$. 

Combining these with the $\xi$s of Table \ref{tab:dps} gives the full matrix elements $M_{jl}$ for any transition from the family $1\uD \to 1\uP\,1\uS$. Thus, for example, for $1\an 3D3 \to 1\an 3P2~1\an 3S1$,
\begin{align}
M_{^7\uS}&=-\frac{2^6}{3^{7/2}}\left(1-\frac{5x^2}{18}+\frac{x^4}{135}\right),	\\
M_{^3\uD}&=\frac{2^{9/2}}{3^{11/2}}x^2\left(1-\frac{p^2}{105}\right),\\
M_{^5\uD}&=\frac{2^5}{3^{11/2}\cdot 5^{1/2}}x^2\left(1+\frac{x^2}{21}\right),\\
M_{^7\uD}&=-\frac{2^6}{3^5}x^2\left(1-\frac{4x^2}{105}\right),\\
M_{^3\uG}&=\frac{2^{9/2}}{3^5\cdot 5\cdot 7}x^4,\\
M_{^5\uG}&=-\frac{2^{9/2}}{3^{13/2}\cdot 7}x^4,\\
M_{^7\uG}&=-\frac{2^{13/2}\cdot 11^{1/2}}{3^{13/2}\cdot 5 \cdot 7}x^4.
\end{align}
Up to a phase factor, which was addressed in Section \ref{guidetothetables}, these matrix elements are consistent with Barnes \etal. (Their results are larger by a factor of two because they tabulate matrix elements for the sum of two topologies.) As discussed previously, a large number of matrix elements has been calculated in this way, and compared to the results of Barnes \etal., as a numerical check on the tabulated  $\xi$s. 

For the particular case of harmonic oscillator wavefunctions and the $\an 3P0$ model, many matrix elements of interest are already available in the literature. The $\xi$s will be useful in computing matrix elements which cannot be found elsewhere, either because they have not be written down, or because they require numerical computation. Spatial matrix elements using more realistic wavefunctions, for example, or for certain spatial operators, cannot be obtained analytically. In such cases the $\xi$ coefficients can be used for efficient numerical computation of the full matrix elements, either starting from the general expression, or by drawing on values from the tables in Appendix \ref{tables}.

References \cite{Burns:2006rz,Burns07production} presented formulae for the spatial matrix elements involving arbitrary radial and orbital excitations, in the $\an 3P0$ model with harmonic oscillator wavefunctions. In future work these will be generalised to go beyond the approximation of equal meson widths and degenerate quark masses, and the resulting closed form expressions tabulated. These can be used in conjunction with the $\xi$ coefficients to calculate any matrix element of interest.

\subsection{Spin-singlet selection rule}

Ref. \cite{Page:2003za} showed that the spin-singlet selection rule is valid not only for the $\an 3P0$ and flux tube models, but also the pseudoscalar-meson emission model. This can now be generalised to any non-flip, triplet model. As discussed in Section \ref{zeroes}, the rule is a simple property of the $\xi$s arising from angular momentum constraints.

The simplest non-trivial example in conventional meson decays is the transition  $\pi_2(1670)\to b_1\pi$ \cite{Page:2003za}, for which there is a strong upper limit on the experimental branching fraction~\cite{Beringer:1900zz},
\be
B(\pi_2(1670)\to b_1(1235)\pi)<1.9\times 10^{-3}~(97.7\% \textrm{ C.L.}).
\ee
The limit is particularly impressive given how strongly $\pi_2(1670)$ decays to another P-wave meson with similar mass but different spin,
\be
B(\pi_2(1670)\to f_2(1270)\pi)=(56.3\pm 3.2)\%.
\ee

For hybrid meson decays the rule has been noted in both the flux tube and constituent gluon models~\cite{Close:1994pr,Close:1994hc,Barnes:1996ff,Close:1997dj,Page:1998gz,Burns:2007hk,Burns:2006rz,Burns:2007dc,Kalashnikova:2008qr}. The rule could help to discriminate between hybrid and conventional meson interpretations of states with non-exotic $J^{PC}$. For example, the hybrid states with quantum numbers $0^{-+}$, $1^{--}$, $2^{-+}$, $1^{++}$ and $1^{+-}$ in the flux tube model each have opposite quark spin to the corresponding conventional meson with the same $J^{PC}$.

\subsection{Spin-triplet selection rule} 

The spin-triplet selection rule of Section \ref{section:symmetries} appears not to have been discussed in the literature before, although some of the corresponding zeroes in decay amplitudes can be seen in expressions given elsewhere, for example in references \cite{Barnes:2005pb,Barnes:1996ff,Barnes:2002mu} . These zeroes can be tested experimentally and are predictions common to non-flip, triplet models. The transition $\an 3P1\to {}\an3S1{}\an 3S1$ is a good example. As shown in Table \ref{tab:pss}, angular momentum allows this transition in both S- and D-wave, but in non-flip, triplet models the S-wave amplitude vanishes,
\be
\xi_{\uS}
\qnset{\an 3P1\\ \an 3S1 \\ \an 3S1}=0.
\ee

Applications of this, and other spin-triplet zeroes, will be discussed in future work.

\subsection{Spatial-vector selection rule}

Some of the zeroes in decay amplitudes noted by other authors can be explained in terms of the spatial-vector selection rule discussed in Section \ref{zeroes}. In most cases these zeroes have been discussed in the context of the $\an 3P0$ model, but they can now be generalised to any non-flip, triplet model. For example, Barnes \etal.~\cite{Barnes:2002mu} noted that the $\an 3P2$ kaon couples weakly to $\rho K^*$, $\omega K^*$ and $\phi K^*$; this is due to the vanishing S-wave amplitude, which can be read off Table~\ref{tab:fss},
\be
\xi_{\uS}
\qnset{\an 3F2\\ \an 3S1 \\ \an 3S1}=0.
\ee

Similarly  the coupling of $\an 3S1$ charmonia to $D^*\o D^*$ in F-wave vanishes \cite{Barnes:2005pb}; this is likewise due to the spatial-vector selection rule, as shown Table \ref{tab:sss},
\be
\xi_{\uF}
\qnset{\an 3S1\\ \an 3S1 \\ \an 3S1}=0.
\ee

Other zeroes of this nature, and their implications, will be discussed in future work.

\subsection{Selection rules for spin-mixed states}

References \cite{Eichten:1978tg,Eichten:1979ms,Barnes:2005pb,Burns:2007hk} identified selection rules for the decays of $\an 3S1$ charmonia  to $D_1\o D$ and $D_1'\o D$, where $D_1$ and $D_1'$ are admixtures of $\an 3P1$ and $\an 1P1$ determined by the heavy-quark limit. By analogy with the discussion in Section~\ref{spin-mixed}, the $\xi$ coefficients for decays involving the states $\an AP1$ (with $J_q=1/2$) and $\an BP1$ (with $J_q=3/2$) are
\bea
\xi_{l}
\qnset
{\an 3S1\\
\an AP1\\
\an 1S0}_+
&=&
\sf2/3
\xi_{l}
\qnset
{\an 3S1\\
\an 3P1\\
\an 1S0}_+
-
\sf1/3
\xi_{l}
\qnset
{\an 3S1\\
\an 1P1\\
\an 1S0}_+,
\\
\xi_{l}
\qnset
{\an 3S1\\
\an BP1\\
\an 1S0}_+
&=&
-\sf1/3
\xi_{l}
\qnset
{\an 3S1\\
\an 3P1\\
\an 1S0}_+
-
\sf2/3
\xi_{l}
\qnset
{\an 3S1\\
\an 1P1\\
\an 1S0}_+.
\eea
The origin of the signs in the above is because for the $(+)$ topology the initial mesons are $Q\q$, rather than $q\Q$ states; the distinction is discussed in Section \ref{sec:spinmix}. The selection rules follow using the $\xi$s in Table \ref{tab:sps},
\bea
\xi_{\uD}
\qnset
{\an 3S1\\
\an AP1\\
\an 1S0}_+
&=&0,
\\
\xi_{\uS}
\qnset
{\an 3S1\\
\an BP1\\
\an 1S0}_+
&=&0.
\eea
Although both S- and D-wave decays are allowed in general, each of the states is forbidden in one wave and allowed in the other. This can also be understood in terms of conservation of the quark angular momenta. A similar selection rule \cite{Burns:2007hk} for the decays of $\an 3D1$ charmonia also follows from the $\xi$s:
\be
\xi_{\uS}
\qnset
{\an 3D1\\
\an AP1\\
\an 1S0}_+
=0.
\ee

\subsection{Ratios of amplitudes}
Absolute width predictions are complicated by several factors such as overall normalisation uncertainties, flavour wavefunctions, $SU$(3)-breaking in pair creation, and ambiguity in the treatment of phase space. These complications cancel out in ratios of amplitudes, which allow models to be tested directly. 

In Section \ref{angularmomentaconstraints}, three special cases were identified in which there is a single spatial matrix element for a given $l$. For such cases the defining equation \rf{eq:defining} simplifies as there is no sum over quantum numbers $l'$ and $L'$. If two decay modes are expressed in terms of the same spatial matrix element, then in their ratio the spatial matrix element cancels and the result depends only upon the $\xi$s. 

The simplest example is the ratio of two matrix elements involving the same mesons, but with different values of $j$. Barnes \etal.~ \cite{Barnes:2005pb,Barnes:2002mu}  identified several such ratios in the $\an 3P0$ model and applied these to the decays of charmonia and light mesons; these predictions can now be seen to be more general, valid for any non-flip, triplet model. One of their examples is the ratio of $j=2$ to $j=0$ amplitudes in the $D^*\o D^*$ decays of vector charmonia, which can be used to distinguish $\an 3S1$ and $\an 3D1$ interpretations. Reading off Tables \ref{tab:sss} and \ref{tab:dss} the relevant ratios are, for $\an 3S1\to{}\an 3S1{}\an3S1$,
\bea
\frac{M_{^5\uP}}{M_{^1\uP}}&=&-2\sqrt{5},
\eea
and for $\an 3D1\to{}\an 3S1{}\an3S1$,
\bea
\frac{M_{^5\uP}}{M_{^1\uP}}&=&-\frac{1}{\sqrt{5}}.
\eea
Other ratios of this type will be discussed in future work.

\subsection{Ratios-of-ratios of amplitudes}

The ratio of amplitudes in two different partial waves is another in which most of the complications cancel out.  An initial state with non-zero $L$  decaying into pairs of S-wave mesons is an example of case~(i); in general there are two partial waves $l=L\pm 1$ available, and the ratio of these can be measured in experiment. With the aid of the $\xi$s, the ratio of matrix elements $M_{jl}$ for a given channel
\be
\frac{M_{j,L+1}}{M_{j,L-1}}=\frac{\xi_{j,L+1}}{\xi_{j,L-1}}\frac{A_{L+1}(p)}{A_{L-1}(p)}
\ee
probes directly the ratio of spatial matrix elements for the $L+1$ and $L-1$ partial waves at the given decay momentum $p$. Since the spatial matrix elements contain all of the model dependence, this ratio allows the different models to be compared directly. Historically, such ratios have been useful in discriminating among different decay models \cite{Geiger:1994kr,Ackleh:1996yt,DeQuadros:2010ai,Li:2011qb}. 

For example, for the transitions $a_1\to\rho\pi$ and $b_1\to\omega\pi$, with matrix elements $M$ and $M'$ respectively, the ratios can obtained from Table \ref{tab:pss},
\bea
\frac{M_D}{M_S}&=&-\frac{1}{2}\frac{A_D}{A_S},\\
\frac{M_D'}{M_S'}&=&\frac{A_D}{A_S}.
\eea

A more direct test of the angular-momentum dependence of decay models is the ratio of two such ratios for different decay channels, chosen so that the spatial matrix elements cancel out. This was first discussed~\cite{Ackleh:1996yt} in the context of $a_1\to\rho\pi$ and $b_1\to\omega\pi$. To the extent that the mesons under comparison ($a_1$ cf. $b_1$, $\rho$ cf. $\omega$) have the same radial wavefunction, and that the two decays have the same momenta (a good approximation in this case), the spatial matrix elements are common to both processes and cancel. This leaves a ratio of  D-to-S ratios which depends only on the $\xi$s, 
\be
\frac{M_{\uD}}{M_{\uS}}\frac{M'_{\uS}}{M'_{\uD}}=-\frac{1}{2}.\label{ratiob}
\ee
This result was derived in ref. \cite{Ackleh:1996yt} in the $\an 3P0$ model, and was subsequently generalised to the flux tube model \cite{Burns:2007hk}. Ref. \cite{Ackleh:1996yt} noted that the ratio is associated with the absence of a spin-flip component in the amplitude, and the formalism of this paper confirms this observation: it can now be seen to be a generic feature common to all non-flip, triplet models.   Thus one can see deviations from the above ratio in the expressions of references \cite{daSilva:2008rp,DeQuadros:2010ai,Li:2011qb}, which are not non-flip, triplet models.

The current PDG averages \cite{Beringer:1900zz} for the ratios, 
\bea
\frac{M_{\uD}}{M_{\uS}} &=& -0.062 \pm 0.020, \\
\frac{M'_{\uD}}{M'_{\uS}} &=& +0.277 \pm 0.027,
\eea
imply considerable disagreement with the above prediction,
\be
\frac{M_{\uD}}{M_{\uS}}\frac{M'_{\uS}}{M'_{\uD}}=-0.22\pm 0.09.
\ee

However the PDG average for the $a_1\to\rho\pi$ mode is very strongly influenced by an incorrect measurement from FOCUS \cite{Link:2007fi}. Their Table III quotes the amplitudes
\bea
M_{\uS}&=& 1\textrm{ (fixed)},\\
M_{\uD}&=&0.241 \pm 0.033 \pm 0.024,
\eea
which implies
\be
\frac{M_{\uD}}{M_{\uS}}=0.241 \pm 0.033 \pm 0.024.
\ee
Their quoted value for the D-to-S ratio, which is used in the calculation of the PDG average, is
\be
\frac{M_{\uD}}{M_{\uS}}=-0.043 \pm 0.009 \pm 0.005.
\ee
This value is obtained by incorrectly scaling the data by a factor of $-1/\sqrt{32}$, due to a misinterpretation of a theoretical result. 

Both interpretations of the FOCUS data differ markedly from the average of the remaining three experimental measurements in the PDG,
\be
\frac{M_{\uD}}{M_{\uS}}=-0.108 \pm 0.016.
\ee
If one uses this average instead, the experimental ratio-of-ratios is much closer to the model prediction:
\be
\frac{M_{\uD}}{M_{\uS}}\frac{M'_{\uS}}{M'_{\uD}}= -0.39 \pm 0.10.
\ee

 Other ratios-of-ratios will be discussed in future work.

\subsection{Relations among widths}

The ratio-of-ratios in the previous section relies on the assumption that the decays $a_1\to\rho\pi$ and $b_1\to\omega\pi$ are controlled by the same spatial matrix elements. This can be justified by the approximately equal decay momenta (in turn due to the approximate degeneracy of $a_1$ and $b_1$, and of $\rho$ and $\omega$) and the reasonable assumption that the corresponding mesons in each decay ($a_1$ cf. $b_1$, and $\rho$ cf. $\omega$) have the same radial wavefunctions.

It is useful to generalise these arguments with the concept of a spatial multiplet, meaning a family of mesons which share the same spatial quantum numbers $n$ and $L$, but differ in either or both of $S$ and $J$. Thus $a_1$ and $b_1$ belong to the 1P multiplet, while $\rho$ and $\omega$ belong to the 1S multiplet. It can often be useful, as above, to make the approximation that mesons within the same multiplet are degenerate (so that decay momenta are the same in different channels) and have the same spatial (radial and orbital) wavefunctions. Both approximations are consistent with the usual treatment of spin splittings as perturbations. 

To the extent that these approximations are valid, different transitions are described in terms of the same spatial matrix elements, which can be eliminated to obtain relations among decay amplitudes, as in the previous section, or widths, discussed now. 

The width is a sum over $j$ and $l$ of the square of the matrix element, multiplied by a phase space factor, 
\be
\Gamma
\qnset
{n\p S\p L\p J\p\\
n_1S_1L_1J_1\\
n_2S_2L_2J_2}
=2\pi\frac{pE_1E_2}{m}\sum_{jl}
\left|
M_{jl}
\qnset
{n\p S\p L\p J\p\\
n_1S_1L_1J_1\\
n_2S_2L_2J_2}
\right|^2,
\label{width}
\ee
where $m$ is the mass of the initial meson and $E_1$ and $E_2$ are the energies of the outgoing mesons with momentum $p$. Different parametrisations of phase space are discussed in the literature, but these have no bearing on the discussion that follows. Note also that in practice the matrix element should be weighted by a flavour factor, which is suppressed in this paper, and for light meson decays there is (in general) a weighted sum over two topologies; the arguments below can easily be generalised to include these additional features.

In general, the expansion of $M_{jl}$ in terms of $\xi$ coefficients and spatial matrix elements $A$ involves a sum over two quantum numbers $L'$ and $l'$ which, on substituting in the above, yields a fairly complicated expression: see equation \rf{widthexpanded} later. The situation is much simpler for the special cases discussed in Section \ref{angularmomentaconstraints}, for which there is no sum over $L'$ or $l'$. In these cases the width is linear in the squares of the spatial matrix elements $A$. For illustration, consider the decay to pairs of S-wave mesons,
 \begin{multline}
\Gamma
 \qnset
 {n\p S\p L\p J\p\\
 n_1S_1\Pp\Pp\\
 n_2S_2\Pp\Pp}
=2\pi\frac{pE_1E_2}{m}
\\\times \sum_{l}
\bigg(\sum_j
\xi_{jl}
 \qnset
 {S\p L\p J\p\\
 S_1\Pp\Pp\\
 S_2\Pp\Pp}^2
\bigg)
\left|
A_{l}
 \qnset
 {n\p L\p\\
 n_1\Pp\\
 n_2\Pp}
\right|^2.
\end{multline}

In the approximations described above, different decay channels are described in terms of common spatial matrix elements, which can be eliminated in favour of linear relations among decay widths. As an example, consider the decays of F-wave mesons to $\an 3S1{}\an 1S0$, which are allowed in D- and G-wave. From Table \ref{tab:fss} the widths are, modulo the phase space factor, 
\bea
\Gamma[\an 3F2]&=&\frac{7}{30}|A_D|^2,\\
\Gamma[\an 1F3]&=&\frac{1}{ 4}|A_D|^2+\frac{1}{ 4}|A_G|^2,\\
\Gamma[\an 3F3]&=&\frac{1}{3 }|A_D|^2+\frac{3}{16}|A_G|^2,\\
\Gamma[\an 3F4]&=&\frac{35}{144}|A_G|^2.
\eea
If the mesons in the F-wave multiplet are degenerate and have the same radial wavefunctions, the spatial matrix elements are common to all channels and can be eliminated.  As a system of four linear equations with two unknowns, this leads to two independent linear relations among the widths.  Thus, for example,
\bea
35\Gamma[\an 3F3]&=&50\Gamma[\an 3F2]+27\Gamma[\an 3F4],\\
70\Gamma[\an 1F3]&=&75\Gamma[\an 3F2]+72\Gamma[\an 3F4].
\eea
One can readily check that these relations (and others like them) are satisfied by calculations already in the literature. For example, the widths of ref. \cite{Eichten:2004uh} in the Cornell model with degenerate 1F charmonia are consistent with the above. The 1F and 2F charmonia in ref. \cite{Barnes:2005pb} have different masses with spin splittings determined by potential models, and so they are not operating within the limit of degeneracy across the spatial multiplet; nevertheless their $\an 3P0$ model decay widths are in good agreement with the above. 

Further examples of such relations for charmonia and light meson decays will be discussed in subsequent work. In general such relations are most useful for highly excited initial states for which the spin splittings are small. 

The same principle can be applied to decays involving the same final states but different initial states. Ref. \cite{Burns:2007dc} used relations of this type in the flux tube model to propose a model for the decay of charmonia to light meson pairs, and the approach can now immediately be generalised to the context of any non-flip, triplet model. References \cite{Burns:2007hk,Close:2008xn} derived relations among the $\epem$ cross sections for the production of $\chi_{cJ} J/\psi$ and $h_c\eta_c$, modelled as the decay of a virtual $\an 3S1$ state.

Another example of the same approach is the relation among the widths of a $\an 3S1$ state to combinations of pseudoscalar and vector mesons. In this case only one partial wave is allowed, so instead of a linear relation among the channels, there is a direct relation between each of the channels. This follows immediately from the $\xi$s,
\begin{multline}
 \Gamma[\an 1S0{}\an 1S0]~:~(\Gamma[\an 1S0{}\an 3S1]+\Gamma[\an 3S1{}\an 1S0])~:~\Gamma[\an 3S1{}\an 3S1]=\\1~:~4~:~7.
\end{multline}

These ratios have been quoted by many authors using different models \cite{Eichten:1979ms,LeYaouanc:1977ux,Lane:1976yh,Chaichian:1978th,Jaronski:1986gd,Ono:1980js,Alcock:1983gb}, and their common origin can now be understood as a consequence of their non-flip, triplet operators. It is not a good application of the approach since the symmetry in the final state multiplets is badly broken, so that the decays under comparison have quite different momenta. For this reason ref. \cite{Eichten:1979ms} refer to this as the ``infamous 1:4:7 ratio''.

\subsection{Lattice decays}
\label{latticesection}

Decays of the spin triplet $1^{-+}$ hybrid meson $\pi_1$ to $b_1\pi$ and $f_1\pi$ have been computed in lattice QCD~\cite{McNeile:2006bz}, with specially chosen masses so that the outgoing mesons have zero momenta. Ref. \cite{Burns:2006wz} compared these calculations to the predictions of the flux tube model, and found consistency in both the ratio of decay widths, and their magnitudes.

In general, the decays under comparison are allowed in both S- and D-wave, but in the limit that the decay momentum goes to zero, only the S-wave amplitude survives. This is an example of  the special case (iii) discussed  in Section \ref{angularmomentaconstraints}, for which there is a single spatial matrix element. Because the $b_1$ and $f_1$ belong to the same spatial multiplet but differ in their intrinsic quark spins (singlet and triplet respectively), if the radial wavefunctions of $b_1$ and $f_1$ are the same, the spatial matrix element is common to both transitions, and so cancels in their ratio. Consequently the relative strength of the decays is determined by the $\xi$ coefficients of Table~\ref{tab:pps:abnormal},
\bea
\xi_{\uS}
\qnset{\an 3P1\\ \an 1P1 \\ \an 1S0}
&=&-\frac{1}{2},
\\
\xi_{\uS}
\qnset{\an 3P1\\ \an 3P1 \\ \an 1S0}
&=&\frac{1}{2\sqrt 2},
\eea
along with a flavour factor which, assuming $f_1$ is $(u\o u + d\o d)/\sqrt 2$, enhances the $b_1\pi$ channel by a factor of $\sqrt 2$ in amplitude. The end result is the ratio
\be
\frac{M_{\uS}[b_1]}{M_{\uS}[f_1]}=2,
\ee
which is consistent with the lattice calculation. (The lattice amplitudes are extracted from decay widths, and so the sign of the ratio has no significance.)

This result was discussed in the context of the flux tube model \cite{Burns:2006wz}, and was subsequently noted to be a feature of spin-triplet pair creation models more generally \cite{Burns:2007hk}. The discussion in Section \ref{constglue} implies that the result is even more general than that, applying also to the constituent gluon model. Indeed the factor of two is evident in the constituent gluon model calculations of ref. \cite{Kalashnikova:1993xb}.

That the lattice calculation is restricted to zero momentum turns out to be an advantage in this type of comparison. In this limit any decays (provided one of the outgoing mesons is S-wave) are expressed in terms of a single spatial matrix element, which cancels in the limit that the mesons under comparison have the same radial wavefunction. The ratio of two such decays is uniquely determined by the $\xi$ coefficient, along with flavour factors.

Recently Lang \etal. \cite{Lang:2014tia} computed the S-wave decay $a_1\to\rho\pi$ in lattice QCD, and extracted the coupling constant
\be
g_{a_1\rho\pi}=1.71\pm 0.39 \GeV.
\ee
The authors note that future lattice calculations with improved statistical accuracy will be used to compute the corresponding coupling constant for $b_1\to\omega\pi$. From experiment they determine
\be
g_{b_1\omega\pi}=0.787\pm 0.25 \GeV.
\ee

The two decays are directly related in the model approach, assuming that  $b_1$ and $a_1$, and separately $\rho$ and $\omega$, have the same radial wavefunctions. From the $\xi$ coefficients, the $b_1$ decay matrix element is smaller by a factor of $\sqrt{2}$,
\bea
\xi_{\uS}
\qnset
{\an 3P1\\
\an 3S1\\
\an 1S0}
&=& \frac{1}{\sqrt 2},\label{a1}\\
\xi_{\uS}
\qnset
{\an 1P1\\
\an 3S1\\
\an 1S0}
&=&- \frac{1}{2}\label{b1}.
\eea
The $b_1$ decay is further suppressed by a factor of $\sqrt 2$ due to flavour. (The coupling constant above is extracted from the decay to both $\rho^-\pi^0$ and $\rho^0\pi^-$.) Thus from the results of Lang \etal., the prediction of non-flip, triplet models for the $b_1$ coupling constant is,
\be
g_{b_1\omega\pi}=0.86\pm 0.20 \GeV,
\ee
consistent with the experimental value.

\subsection{Width sum rule}

Barnes and Swanson \cite{Barnes:2007xu} obtained three theorems which form the basis of the discussion in this section and the two that follow. The theorems are easily derived in the formalism of the $\xi$ coefficients, and are due to the orthogonality relation \rf{orthogrelation}. The derivations in ref. \cite{Barnes:2007xu} involve charmonia, for which flavour considerations play no role. The discussion here will also ignore flavour, but the results can easily be generalised and shown to be valid in the appropriate limits of $SU(2)$ or $SU(3)$ symmetry.

Starting from equation \rf{width}, the decay width can be expressed in terms of $\xi$ coefficients and spatial matrix elements (ignoring, as usual, flavour and colour factors):
 \begin{widetext}
\be
\Gamma
\qnset
{n\p S\p L\p J\p\\
n_1S_1L_1J_1\\
n_2S_2L_2J_2}
=2\pi\frac{pE_1E_2}{m}
\sum_{{lL'l'\widehat L' \widehat l'}}
\left(
\sum_j
\xi_{jl}^{L'l'}
\qnset
{S\p L\p J\p\\
S_1L_1J_1\\
S_2L_2J_2}
\xi_{jl}^{\widehat L' \widehat l'}
\qnset
{S\p L\p J\p\\
S_1L_1J_1\\
S_2L_2J_2}
\right)
A_{l}^{L'l'}
\qnset
{n\p L\p\\
n_1L_1\\
n_2L_2}^*
A_{l}^{\widehat L' \widehat l'}
\qnset
{n\p L\p\\
n_1L_1\\
n_2L_2}
.
\label{widthexpanded}
\ee
\end{widetext}

We are interested in the total width of the meson $nSLJ$ to all of the mesons in the spatial multiplets specified by $n_1L_1$ and $n_2L_2$, which is obtained by summing the above over $S_1$, $J_1$, $S_2$ and $J_2$. In general the decay momentum $p$, the energies $E_1$ and $E_2$, and the spatial matrix elements $A$ depend upon $S_1$, $J_1$, $S_2$ and $J_2$, due to spin-splittings in the masses and different radial wavefunctions in the final state multiplets. If, however, one assumes that the final state multiplets are completely symmetric (meaning their mesons are degenerate and have the same radial wavefunctions), then only the $\xi$ coefficients depend on the summation quantum numbers, and the summation can be done with the help of equation \rf{orthogrelation},
\be
\sum_{S_1J_1S_2J_2}
\Gamma
\qnset
{n\p S\p L\p J\p\\
n_1S_1L_1J_1\\
n_2S_2L_2J_2}
=2\pi\frac{pE_1E_2}{m}
\sum_{{lL'l'}}
\left|
A_{l}^{L'l'}
\qnset
{n\p L\p\\
n_1L_1\\
n_2L_2}
\right|^2.
\ee
If the initial state multiplet is also symmetric in the sense described above, then this expression is independent of $S$ and $J$. Thus all of the mesons within a given $nL$ multiplet have equal widths  \cite{Barnes:2007xu}.

Jaronski and Robson \cite{Jaronski:1986gd} noted, within the flux tube model, the completeness relations which lead to the above result.
\\
\subsection{Mass renormalisation in the unquenched quark model}
\label{massrenorm}

Within the unquenched quark model, the coupling to meson-meson channels modifies physical hadron properties, and notably induces a mass shift with respect to the mass of the bare hadron state. Such effects are expected to be particularly pronounced near threshold, and may be responsible for the unusual properties of some $X$, $Y$ and $Z$ mesons in the charmonium and bottomonium mass regions. However, it is important to establish that these modified hadron properties remain consistent with the successful predictions of the naive (quenched) quark model. For this reason the second loop theorem of Barnes and Swanson \cite{Barnes:2007xu} is interesting.

For a bare $Q\Q$ state $nSLJ$ below threshold, coupling to  the meson pair $n_1S_1L_1J_1$ and $n_2S_2L_2J_2$ induces a downward mass shift whose magnitude is given in second-order perturbation theory by 
\begin{widetext}
\be
\Delta E
\qnset
{n\p S\p L\p J\p\\
n_1S_1L_1J_1\\
n_2S_2L_2J_2}
=
\int_0^\infty
\frac{dp~p^2}
{E_1(p)+E_2(p)-m}
\sum_{jl}
\left|
M_{jl}
\qnset
{n\p S\p L\p J\p\\
n_1S_1L_1J_1\\
n_2S_2L_2J_2}
\right|^2,
\ee
where $m$ is the rest mass of the bare state and $E_1$ and $E_2$ are the energies of the mesons in the loop. This can be obtained from the more familiar expression in terms of the plane wave matrix element using equation \rf{squaredrelations}. Expanding in terms of $\xi$ and $A$ gives
\be
\Delta E
\qnset
{n\p S\p L\p J\p\\
n_1S_1L_1J_1\\
n_2S_2L_2J_2}
=
\sum_{l L'l'\widehat L' \widehat l'}
\left(
\sum_j
\xi_{jl}^{L'l'}
\qnset
{S\p L\p J\p\\
S_1L_1J_1\\
S_2L_2J_2}
\xi_{jl}^{\widehat L' \widehat l'}
\qnset
{S\p L\p J\p\\
S_1L_1J_1\\
S_2L_2J_2}
\right)
\int_0^\infty
\frac{dp~ p^2}{E_1(p)+E_2(p)-m}
A_{l}^{L'l'}
\qnset
{n\p L\p\\
n_1L_1\\
n_2L_2}^*
A_{l}^{\widehat L' \widehat l'}
\qnset
{n\p L\p\\
n_1L_1\\
n_2L_2}
.
\ee

The total contribution to the mass shift from coupling to all members of the multiplets $n_1L_1$ and $n_2L_2$ is obtained by summing over $S_1$, $J_1$, $S_2$ and $J_2$. Analogously to the previous section, if one assumes that the spatial multiplets $n_1L_1$ and $n_2L_2$ are symmetric, namely that within each multiplet the mesons are degenerate and have the same radial wavefunctions, then the integrand is independent of the summation variables, and the sum can be done using the orthogonality relation, leaving
\be
\sum_{S_1J_1S_2J_2}
\Delta E
\qnset
{n\p S\p L\p J\p\\
n_1S_1L_1J_1\\
n_2S_2L_2J_2}
=
\int_0^\infty
\frac{dp ~ p^2}{E_1(p)+E_2(p)-m}
\sum_{l L'l'}
\left|
A_{l}^{L'l'}
\qnset
{n\p L\p\\
n_1L_1\\
n_2L_2}
\right|^2.
\ee
\end{widetext}

Assuming that the mesons within the $nL$ multiplet are also symmetric in the above sense, the expression is independent of $S$ and $J$. In this case all mesons within the $nL$ multiplet are subject to the same downward mass shift due to their coupling to the multiplets $n_1L_1$ and $n_2L_2$: initially degenerate mesons remain degenerate after including loop effects, and there are no induced spin-dependent splittings across an $nL$ multiplet. Since the overall mass scale in potential models is controlled by the quark masses and an arbitrary additive term, the result implies that the effect of coupling to meson-meson channels can, to a large extent, be absorbed into a redefinition of model parameters. 

The theorem applies to all non-flip, triplet models. For the particular case of coupling to pairs of S-wave mesons, the above result was observed in the $\an 3P0$ model by T\"ornqvist \cite{Tornqvist85quarkonium} and Kalashnikova \cite{Kalashnikova:2005ui}. It has also been discussed in a more general context~\cite{Kalashnikova:1993tv}. Within the constituent gluon model, Kalashnikova and Nefediev \cite{Kalashnikova:2008qr} have observed the same mechanism for the S-wave couplings of hybrid mesons to channels with one P- and one S-wave meson. 

Using the results of Section \ref{spin-mixed} the theorem can be generalised to the case of spin-mixed states.

\subsection{Mixing via loops}

The third theorem of Barnes and Swanson \cite{Barnes:2007xu} involves the mixing of different quarkonia states via the coupling to meson-meson channels, which is a second order effect,
\be
nSLJ\to
(n_1S_1L_1J_1)~
(n_2S_2L_2J_2)
\to
\widehat n \widehat S \widehat L J.
\ee

The amplitude to find the basis state $\widehat n \widehat S \widehat L J$ in what was, prior to mixing, a pure $nSLJ$ state, is given by second-order perturbation theory,
\begin{widetext}
\be
a(nSLJ;\widehat n \widehat S \widehat L J)
=
\frac{1}{\widehat m - m}
\sum_{\substack{n_1S_1L_1J_1\\n_2S_2L_2J_2}}
\int_0^\infty\frac{dp~ p^2 }{E_1(p)+E_2(p)-m}
\sum_{jl}
M_{jl}
\qnset
{n\p S\p L\p J\p\\
n_1S_1L_1J_1\\
n_2S_2L_2J_2}^*
M_{jl}
\qnset
{\widehat n\p \widehat S\p \widehat L\p J\p\\
n_1S_1L_1J_1\\
n_2S_2L_2J_2},
\ee
where $m$ and $\widehat m$ are the masses of the bare quarkonia states. (The mixing conserves $J$, as can easily be seen by starting from the expression in terms of the plane wave matrix element and expressing this in terms of the partial wave matrix element.) In terms of $\xi$ coefficients and spatial matrix elements $A$,
\begin{multline}
a(nSLJ;\widehat n \widehat S \widehat L J)
=\\
\frac{1}{\widehat m - m}
\sum_{\substack{n_1S_1L_1J_1\\n_2S_2L_2J_2}}
\sum_{lL'l'\widehat L'\widehat l'}
\left(
\sum_j
\xi_{jl}^{L'l'}
\qnset
{S\p L\p J\p\\
S_1L_1J_1\\
S_2L_2J_2}
\xi_{jl}^{\widehat L' \widehat l'}
\qnset
{\widehat S\p \widehat L\p J\p\\
S_1L_1J_1\\
S_2L_2J_2}
\right)
\int_0^\infty\frac{dp~ p^2 }{E_1(p)+E_2(p)-m}
A_{l}^{L'l'}
\qnset
{n\p L\p\\
n_1L_1\\
n_2L_2}^*
A_{l}^{\widehat L' \widehat l'}
\qnset
{\widehat n\p \widehat L\p\\
n_1L_1\\
n_2L_2}
.
\end{multline}

Incorporating the now familiar assumptions that the mesons within each of the multiplets $n_1L_1$ and $n_2L_2$ are degenerate and have the same radial wavefunctions, the integrand is independent of $S_1$, $J_1$, $S_2$ and $J_2$, so that one can again use the orthogonality relation \rf{orthogrelation}, to find
\be
a(nSLJ;\widehat n \widehat S \widehat L J)
=
\delta_{S\widehat S}
\delta_{L\widehat L}
\frac{1}{\widehat m - m}
\sum_{\substack{n_1L_1\\n_2L_2}}
\int_0^\infty\frac{dp~p^2 }{E_1(p)+E_2(p)-m}
\sum_{l L'l'}
A_{l}^{L'l'}
\qnset
{n\p L\p\\
n_1L_1\\
n_2L_2}^*
A_{l}^{L' l'}
\qnset
{\widehat n\p \widehat L\p\\
n_1L_1\\
n_2L_2}
.
\ee
\end{widetext}
So under these assumptions, spectroscopic mixing is forbidden between states which differ in spin or orbital angular momentum.  The discussion in ref. \cite{Barnes:2007xu} concentrated on the latter: there is no loop-induced mixing between $\an 3S1$ and $\an 3D1$ states, for example, or between $\an 3P2$ and $\an 3F2$. In the Cornell model Eichten \etal.~\cite{Eichten:1978tg} observed the result for the particular case of mixing via loops of S-wave meson pairs.

Close and Thomas \cite{Close:2009ii} highlighted the implications for mixing between hybrid and conventional mesons. States with quantum numbers $1^{++}$, $1^{+-}$, $1^{--}$, $0^{-+}$ and $2^{-+}$ exist within the spectra of both conventional and hybrid mesons, but in each case the corresponding states have opposite quark spin. (This is true at least for the lightest $\Pi\uP$ multiplet within the flux tube model.) According to the above theorem, there is no loop mixing between such states.

If the external mesons (as opposed to those in the loops) are charge conjugation eigenstates, then $C$-parity forbids mixing between $\an 1LL$ and $\an 3LL$. The above theorem is consistent with this, but is only valid in the approximation that the loop mesons are degenerate across their spatial multiplets. A more general theorem is required which ensures that $C$-parity is conserved by loop mixing even beyond this approximation, and this will be discussed in future work.

\subsection{Hyperfine splitting of P-wave mesons}

A challenge for the unquenched quark model is that it must retain the empirically successful features of the ordinary (quenched) quark model. For example the prediction for zero hyperfine splitting of P-wave mesons,
\be
\frac{1}{9}\left(M_{\an 3P0}+3M_{\an 3P1}+5M_{\an 3P2}\right)-M_{\an 1P1}=0,\label{zerohyperfine}
\ee
which follows from assumption of a point-like spin-spin interaction, is strikingly confirmed in experiment for 1P charmonia, and 1P and 2P bottomonia. In charmonia, for example \cite{Dobbs:2008ec},
\be
\o M_{\chi_c(1\uP)}-M_{h_c(1\uP)} =+0.02\pm 0.19\pm 0.13 \MeV.
\ee

The second loop theorem (discussed in Section~\ref{massrenorm}) states that if the external and loop mesons belong to symmetric spatial multiplets, the coupling to meson-meson pairs does not induce any spin-dependent splittings; thus an initially degenerate family of $\an 1P1$, $\an 3P0$, $\an 3P1$ and $\an 3P2$ remains degenerate in the unquenched quark model, and equation \rf{zerohyperfine} holds trivially.

In practice this symmetry limit is not realised, due to spin splittings among both the external mesons and those in the loop. Consequently mesons from the same spatial multiplet experience different mass shifts, leading to induced spin-dependent splittings which threaten to spoil the nice, model-independent prediction above. Typically such induced splittings are of the order of tens of MeV, so that {\it a priori} one would expect equation \rf{zerohyperfine} to be violated at approximately the same order. This raises the unpleasant spectre of relying on fine tuning to reproduce the experimental results.

Remarkably, it turns out that across a large range of models, despite large and different mass shifts for each of the $\an 1P1$, $\an 3P0$, $\an 3P1$ and $\an 3P2$ states, their net contribution to the hyperfine splitting conspires to be very small. The shifts for 1P charmonia from Li \etal.~\cite{Li:2009ad} are a typical example of the effect; their contribution (in MeV) to the hyperfine splitting is
\be
\frac{1}{9}(131+3\times152+5\times175)-162=0.4.
\ee
More recently, Ferretti and Santopinto \cite{Ferretti:2013vua} have computed mass shifts for bottomonia, and their results also show the mechanism in action; their splittings for 1P and 2P bottomonia lead to the net contributions
\bea
\frac{1}{9}(108 + 3\times 114 + 5\times 117) - 115 &=& 0,\\
\frac{1}{9}(137 + 3\times 144 + 5\times 149) - 146 &=& 0.
\eea
The same effect appears in many other papers in the $\an 3P0$ model \cite{Barnes:2007xu,Kalashnikova:2005ui,Ono:1983rd,Liu:2011yp}, and also the Cornell model~\cite{Yang:2010am}.

This mechanism was observed and explained in references \cite{Burns:2011fu,Burns:2011jv}, exploiting the angular-momentum dependence common to non-flip, triplet models. The proof considers mass shifts due to coupling to combinations of $\an 3S1$ and $\an 1S0$ mesons. The derivation involves a power-series expansion in a parameter which is small provided that the mass differences among the bare states, and among the mesons in the loops, are small compared to the binding energy. The result is a hierarchy of scales in the mass shifts: while the overall shifts can be large, the induced spin splitting between any pair of states is suppressed to first order in the expansion parameter, while the induced hyperfine splitting of the entire multiplet is suppressed by a further power of the parameter.

The proof was derived using the same general expression which defined the $\xi$s in this paper, an as such it applies to any non-flip, triplet model. To translate the derivations into the formalism used in this paper, the coefficients in Table II of ref. \cite{Burns:2011fu} are sums over squares of $\xi$ coefficients,
\be
C^l=\sum_j
\left(\xi_{jl}\right)
^2.
\ee

The relation equivalent to (\ref{zerohyperfine}) for D-wave mesons is also protected by the same mechanism. Using that relation one can predict the mass of the missing $\an 1D2$ bottomonium, which turns out to be consistent with the prediction of a string model \cite{Burns:2010qq}.

Note that models \cite{Pennington:2007xr,Shmatikov:1998dy} which do not incorporate coupling to each of the members of the spatial multiplet in the loop are not protected by this mechanism.

\subsection{Hyperfine splitting of S-wave mesons}

In the ordinary (quenched) quark model the $\epem$ widths of $\an 3S1$ states and the $\an 3S1 - {\an 1S0}$ hyperfine splitting are both proportional to the square of the wavefunction at the origin. This leads to the model-independent prediction 
\be
\frac{\hyperfine{2}}{\hyperfine{1}}=\frac{\epemwidth{2}}{\epemwidth{1}}\label{hypo}
\ee
connecting the widths and hyperfine splittings $\Delta M$ of different radial levels. The relation is satisfied for charmonia, and it has has recently been shown to be satisfied for bottomonia \cite{Burns:2012pc}, following the discovery of the $\eta_b(2\uS)$ at Belle \cite{Mizuk:2012pb}. 

It is therefore interesting to establish whether the relation survives the effect of coupling to meson-meson channels, which alters the quantities on both sides of the equation. The $\epem$ widths of the $n\an 3S1$ levels are suppressed by the probability $P(n)$ that the physical states are $(Q\Q)$, rather than $(Q\q)(q\Q)$. To establish the effect on the hyperfine splitting, one can perform a power series expansion making use of the $\xi$ coefficients, analogous to that described in the previous section and reported in references \cite{Burns:2011fu,Burns:2011jv}. The derivation, which is somewhat more straightforward than in the P-wave case, will be discussed in future work. The end result is that the different mass shifts of the  $\an 1S0$ and $\an 3S1$ states lead to a suppression of the hyperfine splittings $\hyperfine{n}$ by the same factor $P(n)$. Consequently the relation \rf{hypo} remains valid (to first order in the expansion parameter) within the unquenched quark model, with any non-flip, triplet 
coupling.

\subsection{The $X(3872)$ as a coupled-channel effect}
Several authors \cite{Kalashnikova:2005ui,Suzuki:2005ha,Zhang:2009bv,Li:2009ad,Danilkin:2009hr,Kalashnikova:2009gt,Coito:2010if,Ortega:2010qq,Coito:2012ka,Coito:2012vf,Coito:2012ka,Ortega:2012rs,Zhou:2013ada,Ferretti:2013faa,Ferretti:2014xqa} have investigated the possibility that the $1^{++}$ state $X(3872)$ arises from the S-wave coupling between the $2\an 3P1$ charmonium and the $D^*\o D$ channel. In this scenario, one might also expect a corresponding $1^{+-}$ state, since the $2\an 1P1$ charmonium also couples $D^*\o D$ in S-wave. Kalashnikova \cite{Kalashnikova:2005ui} has proposed an explanation for the uniqueness of the $1^{++}$ channel, using heavy quark spin conservation: the coupling strength in the $1^{++}$ is enhanced by a factor of $\sqrt 2$ compared to $1^{+-}$. The result follows from the $\xi$s of equations \rf{a1} and \rf{b1}.

\section{Conclusions}
\label{conclusions}

This article has brought together several apparently disparate approaches to strong decay and the unquenched quark model, parametrising their common angular-momentum dependence in terms of coefficients. Everything follows from the observation that transitions in such models involve the creation of a $q\q$ pair in spin-triplet, with the spins of the initial quarks acting as spectators. This is well-known to be true of the $\an 3P0$ and flux tube models, and as we have seen it also applies to the Cornell model, to a subset of more general microscopic models, and to the pseudoscalar-meson emission model. 

A solution has been obtained for arbitrary matrix elements in these non-flip, triplet models. The idea is to separate the (model-dependent) spatial matrix element from the (model-independent) angular momentum algebra, parametrising the latter in terms of coefficients. The properties of these $\xi$  coefficients have been studied, their values tabulated, and an introduction to some applications given.

Symmetries under the interchange of topologies or meson quantum numbers are immediate, and enforce the conservation of $C$- and $G$-parity. An orthogonality relation, which underlies theorems presented elsewhere in the literature, emerges simply. Angular momentum constraints allow one very simply  to classify the independent spatial matrix elements. Several new selection rules are derived.

The coefficients are easily generalised to mesons with mixed spin, and their symmetry properties are particularly useful in dealing with subtleties of phase conventions. The connection with an approach based on heavy-quark spin conservation is immediate, and well-known selection rules are confirmed and generalised. This article has concentrated on initial states of mixed spin, but a similar approach can be applied to final states.

Within the flux tube model, hybrid meson wavefunctions have the same overall spin and orbital structure as their conventional counterparts, so their transitions are determined by the same $\xi$ coefficients. For normal parity transitions these can be read off the tables used for conventional mesons; for abnormal parity transitions the coefficients are tabulated separately. In the constituent gluon model, hybrid transitions have a different angular-momentum dependence in general, but for the important special case of the lightest hybrids with negative parity the dependence is the same, and the ordinary $\xi$s apply.

The coefficients can be used as a practical tool for future calculations. Typically, matrix elements are computed by summing over Clebsch-Gordan coefficients which couple the various angular momenta, and taking spin and spatial matrix elements  for every combination of magnetic quantum numbers. The expression in terms of $\xi$s and spatial matrix elements $A$ is straightforward by comparison, involving fewer summation variables. This simplicity is due to the Wigner-Eckart theorem and vector recoupling coefficients which, given their obvious advantages, are used surprisingly rarely.

The tables in Appendix \ref{tables} cover all transitions of obvious physical interest, for initial S-, P-, D- and F-wave mesons coupling to combinations of S-wave meson pairs, and to a P- or a D-wave meson along with and an S-wave meson. Additional tables are supplied for hybrid meson transitions not already covered by the above. In the event that other $\xi$s are required, these can be computed straightforwardly using equation \rf{eq:xi}. More generally, for numerical calculations of a large number of transitions, it would be more practical to use the general expression than to extract coefficients individually from the tables.

The remaining ingredient  is the spatial matrix element $A$, for which a general expression, applicable to the $\an 3P0$ and flux tube models, is presented in Appendix \ref{spatial:sect}, in terms of integrals over spatial wavefunctions. These integrals can be computed analytically in the case of harmonic oscillator wavefunctions with the $\an 3P0$ model operator. In future work, a compendium of these spatial matrix elements will be presented which can be combined with the $\xi$s of this paper to give full matrix elements $M_{jl}$. In the limit of equal wavefunction size and degenerate quark masses, closed form expressions for many $M_{jl}$ are already available in the literature \cite{Barnes:1996ff}. (Comparison to these expressions is a useful check on the tabulated $\xi$s, as discussed earlier.) Applications to heavy quarkonia and heavy-flavoured mesons require the aforementioned approximations to be relaxed. Whilst general formulae for spatial integrals appear elsewhere 
in the literature \cite{Roberts:1992js,BonnazSilvestre-Brac99discussion,Burns:2006rz}, it would be useful to have tables of $A$s presented in closed-form, which can readily be combined with the $\xi$s in practical calculations.

Selection rules are manifest as zeroes in the $\xi$ coefficient for transitions which are allowed by angular momentum and parity. The spin-singlet selection rule is an example which has been much-discussed in the literature. In this article a further five classes of zeroes have been identified, including the spin-triplet and spatial-vector selection rules, which explain zeroes observed elsewhere in the literature. Indeed, many zeroes appearing within specific models can ultimately be traced to the $\xi$s, which implies that they have more general origin, common to all non-flip, triplet models. Experimental violations of these selection rules would indicate that decays involve operators beyond the $\spino\cdot\spato$ structure.

Experimental tests of amplitude ratios (for a single channel with different values of $j$) and ratios-of-ratios (for different channels and in two partial waves $l$) are useful tests of model fundamentals, because phase space, flavour, and other complications cancel out. Predictions for such ratios can be read off the tables of $\xi$ coefficients. Such ratios have been discussed elsewhere in the literature within the context of specific models, but these can now be seen to be common to all non-flip, triplet models.

In the approximation that mesons from the same spatial multiplet are degenerate and have the same radial wavefunctions, different transitions are expressed in terms of the same spatial matrix element, which can be treated as a free parameter and eliminated in favour of linear relations among the different channels. The relations follow immediately from the $\xi$s. The approach is reminiscent of the original quark model paper on the strong decay \cite{Micu:1968mk}.

The first lattice QCD calculations of strong decay are restricted to zero momentum, and in the model approach such transitions are expressed in terms of a single S-wave spatial matrix element. This leads immediately to relations among different channels, which are consistent with all available lattice data. Model predictions can be tested against future lattice calculations using the tables of coefficients. 

As well as decays, the $\xi$s can be applied to the calculation of other meson properties in the unquenched quark model. The importance of the coupling to meson-meson channels has been brought into focus in recent years by the proliferation of $X$, $Y$ and $Z$ mesons, whose masses (and decays) disagree with model predictions, and which are often correlated to thresholds. The meson-meson coupling causes mass shifts which can potentially reconcile models with experiment, but it is important to establish that the same shifts do not spoil other results.

According to the loop theorems of Barnes and Swanson \cite{Barnes:2007xu}, mass shifts due to the meson-meson coupling can be absorbed (to within a first order approximation) into a redefinition of potential model parameters. In the formalism of this paper, the theorem (and other related results) are due to the orthogonality of the $\xi$ coefficients. 

The theorem requires full symmetry across spatial multiplets: relaxing that approximation leads to induced spin-splittings which are large in comparison to the observed hyperfine splittings of P-wave quarkonia, which are consistent with zero. This potentially threatens the quark model prediction of zero splitting, but there is a mechanism, which arises from  the $\xi$ coefficients, which protect protects the result \cite{Burns:2011fu,Burns:2011jv}. The relation between the $e^+e^-$ widths of $\an 3S1$ mesons and the corresponding $\an 3S1-{\an 1S0}$ hyperfine splitting, recently confirmed in bottomonia, is protected by a similar mechanism \cite{Burns:2012pc}.

\begin{acknowledgments}
Discussions with Eric Swanson, Alberto Correa dos Reis and Yulia Kalashnikova are gratefully acknowledged.
\end{acknowledgments}

\appendix

\section{The spatial matrix element}
\label{spatial:sect}

In this section expressions are derived for the spatial part $A$ of the matrix element in momentum and position space. No assumptions are made for the functional forms of the pair-creation amplitudes $\spato(\ve k,\ve{\o k})$ and $\spato(\ve x)$, so that the derivations apply to essentially any implementation of the $\an 3P0$ model and, for conventional mesons, the flux tube model. For hybrid mesons in the flux tube model, the matrix elements involve integrations over Wigner $D$-functions and so differ from those discussed here. Likewise the matrix elements in microscopic models involve additional integrations due to momentum transfer in the scattering matrix element. In pseudoscalar-meson emission models the matrix element differs because the quark-antiquark degrees of freedom of the emitted meson are ignored.

The state vector characterising a quark at position $\ve X$ satisfies orthogonality and completeness relations,
\bea
\< \ve X\| \ve X'\>&=&\delta^3(\ve X'-\ve X),\label{delta:pos}\\
1&=&\int d^3\ve X \|\ve X\>\<\ve X\|,\label{completeness:pos}
\eea
and likewise for a quark of momentum $\ve K$,
\bea
\< \ve K\| \ve K'\>&=&\delta^3(\ve K'-\ve K),\label{delta:mom}\\
1&=&\int d^3\ve K \|\ve K\>\<\ve K\|.\label{completeness:mom}
\eea
The position and momentum space state vectors are related
\be
\< \ve X \| \ve K\> =\frac{1}{(2\pi)^{3/2}}e^{i\ve K\cdot \ve X}.\label{pos:mom}
\ee

The wavefunction for a quark-antiquark system in a central potential factorises into a centre-of-mass part (depending on the meson momentum $\ve P$) and a relative part (depending on the spatial quantum numbers $n$, $L$, and $L_z$),
\bea
\<\ve X,\vb X \|\ve P nLL_z\>&=&\<\ve R \|\ve P\>\<\ve r\|nLL_z\>\label{wavefunction:pos},\\
\<\ve K,\vb K \|\ve P nLL_z\>&=&\<\ve Q \|\ve P\>\<\ve q\|nLL_z\>\label{wavefunction:mom},
\eea
where  the centre-of-mass and relative coordinates are, for  quark and antiquark masses $M$ and $\o M$,
\bea
\pmat{\ve R \\ \ve r}
&=&
\pmat{\nf M/{M+\o M} & \nf {\o M}/{M+\o M}\\1 & -1}
\pmat{\ve X \\ \vb X},
\label{coordinates:pos}\\
\pmat{\ve Q \\ \ve q}
&=&
\pmat{1&1\\ \nf {\o M}/{M+\o M} & \nf {-M}/{M+\o M}}
\pmat{\ve K \\ \vb K}.
\label{coordinates:mom}
\eea
Equation \rf{wavefunction:mom} can be obtained from equation \rf{wavefunction:pos} by integrating a pair of factors of the form \rf{completeness:pos} using \rf{pos:mom}.

The final state is a direct product of wavefunctions of the above form, namely
\begin{widetext}
\bea
\<\ve X_1,\vb X_1,\ve X_2,\vb X_2 \|\ve P_1 n_1L_1L_{1z},\ve P_2 n_2L_2L_{2z}\>
&=&\<\ve R_1 \|\ve P_1\>\<\ve R_2 \|\ve P_2\>\<\ve r_1\|n_1L_1L_{1z}\>\<\ve r_2\|n_2L_2L_{2z}\>,\\
\<\ve K_1,\vb K_1,\ve K_2,\vb K_2 \|\ve P_1 n_1L_1L_{1z},\ve P_2 n_2L_2L_{2z}\>
&=&\<\ve Q_1 \|\ve P_1\>\<\ve Q_2 \|\ve P_2\>\<\ve q_1\|n_1L_1L_{1z}\>\<\ve q_2\|n_2L_2L_{2z}\>,
\eea
\end{widetext}
where $\ve R_i$ and $\ve Q_i$ are defined in terms of $\ve X_i$, $\vb X_i$, $\ve K_i$ and $\vb K_i$ by equations analogous to \rf{coordinates:pos} and \rf{coordinates:mom}. It is more convenient to work with states of total and relative momenta $\ve P'$ and $\ve p$, rather than individual meson momenta $\ve P_1$ and $\ve P_2$,
\be
\pmat{\ve P'\\ \ve p}=
\pmat{1&1\\ \nf{M_2}/{M_1+M_2}& \nf{-M_1}/{M_1+M_2}}
\pmat{\ve P_1 \\ \ve P_2},
\label{finalmomenta}
\ee
where $M_1$ and $M_2$ are the masses of mesons 1 and 2. Defining
\be
\pmat{\ve R'\\ \ve r'}
=
\pmat{\nf{M_1}/{M_1+M_2}&\nf{M_2}/{M_1+M_2} \\ 1&-1}
\pmat{\ve R_1 \\ \ve R_2},
\label{coordinates:pos2}
\ee

\be
\pmat{\ve Q'\\ \ve q'}
=
\pmat{1&1\\ \nf{M_2}/{M_1+M_2}&\nf{-M_1}/{M_1+M_2}}
\pmat{\ve Q_1 \\ \ve Q_2},
\label{coordinates:mom2}
\ee
the product of the two centre-of-mass factors can be replaced by an overall centre-of-mass factor and a relative wavefunction,
\bea
\<\ve R_1 \| \ve P_1\>\<\ve R_2 \| \ve P_2 \>
&=&
\<\ve R' \| \ve P'\>\<\ve r' \| \ve p \>,
\\
\<\ve Q_1 \| \ve P_1\>\<\ve Q_2 \| \ve P_2 \>
&=&
\<\ve Q' \| \ve P'\>\<\ve q' \| \ve p \>.
\eea
States of good orbital angular momentum $l$,
\be
\| \ve p \>=\sum_{ll_z} \|pll_z\>\<ll_z\|\vh p\>,
\ee
and determined by the overlaps
\bea
\<\ve r'\|pll_z\>&=&4\pi i^l j_l(r'p)\<\vh r'\| ll_z\>,\\
\<\ve q'\|pll_z\>&=&\frac{\delta(p-q')}{q'^2}\<\vh q'\| ll_z\>,
\eea
where $j_l$ is the spherical Bessel function. The corresponding final state is
\begin{widetext}
\bea
\<\ve X_1,\ve X_1, \ve X_2, \ve X_2 \|\ve P' , n_1 L_1 L_{1z} ,n_2 L_2 L_{2z}, p ll_z \>
&=&
\<\ve R' \| \ve P'\>
\<\ve r_1\| n_1L_1L_{1z}\>
\<\ve r_2\| n_2L_2L_{2z}\>
\<\ve r'\|pll_z\>,
\\
\<\ve K_1,\ve K_1, \ve K_2, \ve K_2 \|\ve P' ,n_1 L_1 L_{1z} ,n_2 L_2 L_{2z},  pll_z \>
&=&
\<\ve Q' \| \ve P'\>
\<\ve q_1\| n_1L_1L_{1z}\>
\<\ve q_2\| n_2L_2L_{2z}\>
\<\ve q'\|pll_z\>.
\eea
Finally, as we are interested in the reduced matrix element, the final state should be coupled to total angular momentum $l'$ as in equation \rf{spatialme},
\bea
\<\ve X_1,\ve X_1, \ve X_2, \ve X_2 \|\ve P' ((n_1 L_1 \ot n_2 L_2)_{L'}\ot p l)_{l'l_z'} \>
&=&
\<\ve R' \| \ve P'\>
\<\ve r_1,\ve r_2 ,\ve r'\| ((n_1 L_1 \ot n_2 L_2)_{L'}\ot p l)_{l'l_z'} \>,
\label{finalstate[1]}
\\
\<\ve K_1,\ve K_1, \ve K_2, \ve K_2 \|\ve P' ((n_1 L_1 \ot n_2 L_2)_{L'}\ot p l)_{l'l_z'} \>
&=&
\<\ve Q' \| \ve P'\>
\<\ve q_1,\ve q_2 ,\ve q'\| ((n_1 L_1 \ot n_2 L_2)_{L'}\ot p l)_{l'l_z'} \>.
\label{finalstate[2]}
\eea

In position space the transition operator creates a quark-antiquark pair at a point $\ve x$, and it is convenient to characterise the amplitude in terms of the vector $\vc \rho=\ve x-\ve R$ measured with respect to the centre-of-mass of the initial meson. The two topologies have operators of the form
\bea
\spato_+&=&\int d^3\ve X \int d^3\vb X \int d^3\ve x \| \ve X ,\ve x ,\ve x, \vb X \> \spato (\vc \rho,\ve r ) \<\ve X, \vb X \|,\\
\spato_-&=&\int d^3\ve X \int d^3\vb X \int d^3\ve x \| \ve x ,\vb X ,\ve X, \ve x \> \spato (\vc \rho,\ve r ) \<\ve X, \vb X \|.
\eea
These are generalisations of equation \rf{posop}, obtained by inserting complete sets of states describing the initial $Q$ and $\Q$ at $\ve X$ and $\ve {\o X}$. The two operators have the same amplitude $\spato (\vc \rho,\ve r )$ for pair creation, but differ in the arrangement of the quarks and antiquarks in the final state. For $\spato_+$ the quark and antiquark of meson 1 are at $\ve X$ and $\ve x$ respectively, and those of meson 2 are at $\ve x$ and $\vb X$. The operator $\spato_-$ corresponds to the opposite situation.

In momentum space the transition operator creates a quark-antiquark pair with equal and opposite momenta $\ve k$ and $-\ve k$ respectively, leading to the generalisations of equation \rf{singlespat},
\bea
\spato_+&=&\int d^3\ve K \int d^3\vb K \int d^3\ve k \| \ve K ,-\ve k ,\ve k, \vb K \> \spato (\ve k,\ve q) \<\ve K, \vb K \|,\\
\spato_-&=&\int d^3\ve K \int d^3\vb K \int d^3\ve k \| \ve k , \vb K ,\ve K, -\ve k \> \spato (\ve k, \ve q ) \<\ve K, \vb K \|.
\eea
\end{widetext}
The operators satisfy the symmetry relations:
\bea
\spato (\vc \rho,\ve r )&=&\spato (-\vc \rho,\ve r ),\label{spatosymmetry:1}\\
\spato (\vc \rho,\ve r )&=&-\spato (\vc \rho,-\ve r ),\label{spatosymmetry:2}
\eea
and
\bea
\spato (\ve k,\ve q) &=& -\spato (-\ve k,\ve q ),\label{spatosymmetry:3}\\
\spato (\ve k,\ve q) &=& \spato (\ve k,-\ve q).\label{spatosymmetry:4}
\eea

In taking the matrix element of the spatial operators, it is convenient to replace the integration variables
\bea
\int d^3\ve X \int d^3\vb X \int d^3\ve x &\to& 
\int d^3\ve R \int d^3\ve r \int d^3\vc \rho, \qquad\\
\int d^3\ve K \int d^3\vb K \int d^3\ve k &\to &
\int d^3\ve Q \int d^3\ve q \int d^3\ve k,\qquad
\eea
and the next step is to express the final state wavefunctions \rf{finalstate[1]} and \rf{finalstate[2]} in terms of these new variables. In position space, $\ve X_1$, $\vb X_1$,  $\ve X_2$, $\vb X_2$ are replaced by $\ve X$, $\vb X$ and $\ve x$ as appropriate to the particular topology, and these are expressed in terms of the integration variables by inverting equation \rf{coordinates:pos}, and using $\ve x=\vc \rho +\ve R$. Finally the arguments $\ve R'$, $\ve r'$, $\ve r_1$ and $\ve r_2$ of the final state wavefunction are obtained using equation \rf{coordinates:pos2}, and the analogues of equation \rf{coordinates:pos} for $\ve R_1$ and $\ve R_2$, with quark and antiquark masses $M$, $\o M$ and $m$ appropriate to the given topology. The analogous procedure  is applied to the momentum-space matrix element.

The resulting coordinate transformations are, for the operator $\spato_+$,
\bea
\pmat{\ve R'\\ \ve r' \\ \ve r_1 \\ \ve r_2}
&=&
\pmat{1& \cdots & \cdots \\
0 & \gamma & \Delta	\\
0 & \o \lambda & -1 \\
0 & \lambda & 1 }
\pmat{\ve R\\ \ve r \\ \vc \rho},
\\
\pmat{\ve Q'\\ \ve q' \\ \ve q_1 \\ \ve q_2}
&=&
\pmat{1 & 0 & 0 \\
\cdots & 1 & -1 \\
\cdots & \lambda_l & \lambda_h \\
\cdots & \o\lambda_l & \o\lambda_h}
\pmat{\ve Q\\ \ve q \\ \ve k},
\eea
where the ellipsis ($\cdots$) indicate terms which will, for reasons that will be outlined shortly, play no role, the $\lambda$s are dimensionless mass ratios,
\begin{align}
 \lambda&=\nf{M}/{M+\o M}\qquad  &\o\lambda&=\nf{\o M}/{M+\o M},\\
\lambda_h&=\nf{M}/{M+m}\qquad  &\o\lambda_h&=\nf{\o M}/{\o M+ m },\\
 \lambda_l&=\nf{m}/{M+m}\qquad  &\o\lambda_l&=\nf{m}/{\o M+m },
 \end{align}
and
\bea
\gamma&=&\lambda_h \o \lambda +\o \lambda_h \lambda,\\
\Delta&=&\lambda_l-\o \lambda_l=\o\lambda_h-\lambda_h.
\eea
In the above the labels $h$ and $l$ indicate heavy and light, as short hands for the initial, and created, quarks and antiquarks. (The initial quarks need not be heavy, but the created quarks are certainly light.)

Integration over $\ve R'$ in position space, or $\ve Q'$ in momentum space, produces a momentum conserving delta function $\<\ve P'\| \ve P\>$. There remains, however, a dependence on the total momenta through the $\cdots$ terms, which spoils Galilean invariance \cite{Roberts:1992js}. The usual prescription is to choose $\ve P=0$, and then from equation \rf{finalmomenta} the outgoing mesons have equal and opposite momenta $p=|\ve P_1|=|\ve P_2|$. The quantity of interest is the full matrix element modulo the delta function
\begin{widetext}
\be
\<\ve P' ((n_1 L_1 \ot n_2 L_2)_{L'}\ot p l)_{l'}\£ \spato_\pm \£ \ve P=0, nL\>
=
\delta^3(\ve P')
\<((n_1 L_1 \ot n_2 L_2)_{L'}\ot p l)_{l'}\£ \spato_\pm \£  nL\>
\ee
and in both position and momentum space it involves integration over two variables. For the operator $\spato_+$, 
\begin{multline}
\<((n_1 L_1 \ot n_2 L_2)_{L'}\ot p l)_{l'}\£ \spato_\pm \£  nL\>
 \\
 \begin{aligned}
 &= \int d^3\ve r \int d^3\vc \rho
\<((n_1 L_1 \ot n_2 L_2)_{L'}\ot p l)_{l'}\£\o\lambda\ve r -\vc\rho,\lambda\ve r+\vc\rho,\gamma\ve r+\Delta\vc\rho\>
 \spato (\vc \rho,\ve r )
\<\ve r\£nL\>
 \\
 &=\int d^3\ve q \int d^3\ve k
 \<((n_1 L_1 \ot n_2 L_2)_{L'}\ot p l)_{l'}\£\lambda_l\ve q+\lambda_h\ve k,\o\lambda_l\ve q+\o\lambda_h\ve k,\ve q-\ve k\>
 \spato (\ve k,\ve q )
\<\ve q\£ nL\>.
 \end{aligned}
 \label{spatialamplitude}
 \end{multline}
 \end{widetext}

The spatial matrix element $A$ is defined in terms of the above by equation \rf{spatialme}. It enforces the conservation of parity,
\be
(-)^{L+L_1+L_2+l+1}=1,
\ee
as can be seen by reversing the orientations of the integration variables, and re-writing the integrand using the symmetries of the spherical harmonics and of the pair creation amplitude,
\bea
 \spato (-\vc \rho,-\ve r )&= &-\spato (\vc \rho,\ve r ),\\
 \spato (-\ve k,-\ve q )&=& -\spato (\ve k,\ve q ).
 \eea
Comparing the coordinate transformations for $\spato_-$, 
 \bea
\pmat{\ve R'\\ \ve r' \\ \ve r_1 \\ \ve r_2}
&=&
\pmat{1& \cdots & \cdots \\
0 & -\gamma & -\Delta	\\
0 & \lambda & 1 \\
0 & \o \lambda & -1 }
\pmat{\ve R\\ \ve r \\ \vc \rho},
\\
\pmat{\ve Q'\\ \ve q' \\ \ve q_1 \\ \ve q_2}
&=&
\pmat{1 & 0 & 0 \\
\cdots & -1 & 1 \\
\cdots & \o\lambda_l & \o \lambda_h \\
\cdots & \lambda_l & \lambda_h}
\pmat{\ve Q\\ \ve q \\ \ve k},
\eea
with those of  $\spato_+$ above, leads to a symmetry relation under the interchange of both the topology, and of the quantum numbers of mesons 1 and 2,
\be
A_{l}^{L'l'}
\qnset
{n\p L\p\\
n_1L_1\\
n_2L_2}_\pm
=
(-)^{L_1+L_2+L'+l}
A_{l}^{L'l'}
\qnset
{n\p L\p\\
n_2L_2\\
n_1L_1}_\mp.
\label{spatsymm:1}
\ee
For the special case of the initial quark and antiquark having the same mass ($M=\o M$), the integrals simplify with 

\bea
\lambda&=&\o\lambda=1/2,\\
\lambda_h&=&\o\lambda_h=\gamma,\\
\lambda_l&=&\o\lambda_l,\\
\Delta&=&0 ,
\eea
leading to separate symmetry relations under the interchange either of the topology, 
\be
A_{l}^{L'l'}
\qnset
{n\p L\p\\
n_1L_1\\
n_2L_2}_\pm
=
(-)^l
A_{l}^{L'l'}
\qnset
{n\p L\p\\
n_1L_1\\
n_2L_2}_\mp,
\label{spatsymm:2}
\ee
or the quantum numbers of mesons 1 and 2,
\be
A_{l}^{L'l'}
\qnset
{n\p L\p\\
n_1L_1\\
n_2L_2}_\pm
=
(-)^{L_1+L_2+L'}
A_{l}^{L'l'}
\qnset
{n\p L\p\\
n_2L_2\\
n_1L_1}_\pm.
\label{spatsymm:3}
\ee
In position space these symmetries use equation \rf{spatosymmetry:1}.

The second of these implies a selection rule for final states with the same spatial wavefunctions,
\be
A_{l}^{L'l'}
\qnset
{n\p L\p\\
n_1L_1\\
n_1L_1}_\pm
=0 \textrm{ if } (-)^{L'+1}=1.
\label{selspat}
\ee

\section{Tables of coefficients}
\label{tables}

Tables of $\xi$ coefficients are presented below, for transitions with a single S-wave meson ($L_2=0$) and for various combinations of $L$ and $L_1$. The tables show coefficients for the $(+)$ topology; those of the $(-)$ topology are given by the symmetry relation \rf{xisymm[1]}. 

The first tables are for normal parity transitions, namely those involving only conventional mesons, or more generally, an even number of abnormal parity states. Tables \ref{tab:sss}--\ref{tab:fss} are for transitions of S-, P-, D-, and F-wave states, respectively, to all pairs of S-wave final states. Likewise Tables \ref{tab:sps}--\ref{tab:fps} are for the same initial states decaying to any combination of a P- and an S-wave state. Tables \ref{tab:sds}--\ref{tab:fds}, for decay into a D-wave state, are restricted to $S_2=0$. 

The remaining Tables \ref{tab:pss:abnormal}--\ref{tab:pds:abnormal} are for abnormal parity transitions, with an initial P-wave coupling respectively to a pair of S-wave states, a P- and an S-wave state, and a D- and an S-wave state.

The columns in each table are labelled by the quantum numbers $SLJ$, and the rows by $S_1L_1J_1$, $S_2L_2J_2$ and $jl$. The standard spectroscopic notation is used for the mesons, and also for the quantum numbers $j$ and $l$, which are denoted  $^{2j+1}l$. Partial waves up to $l=6$ are included.

\begingroup\squeezetable
\begin{table}
\vspace{1cm}
\begin{tabularx}{\columnwidth}{|>{$}r<{$}|>{$}X<{$}|>{$}X<{$}|}

\hline

-&\slj1S0~&\slj3S1~\\

\hline

\slj1S0~\slj1S0~\sl1P&&\dfrac{1}{2 \sqrt{3}}\\

\hline

\slj3S1~\slj1S0~\sl3P&\dfrac{-1}{2}&\dfrac{-1}{\sqrt{6}}\\

\hline

\slj1S0~\slj3S1~\sl3P&\dfrac{-1}{2}&\dfrac{1}{\sqrt{6}}\\

\hline

\slj3S1~\slj3S1~\sl1P&&\dfrac{1}{6}\\

\sl3P&\dfrac{-1}{\sqrt{2}}&0^\dag\\

\sl5P&&\dfrac{-\sqrt{5}}{3}\\

\sl5F&&0^\triangle\phrac\\

\hline

\end{tabularx}
\caption{The $\xi$ coefficients for $\uS\to\uS\,\uS$.}
\label{tab:sss}
\end{table}

\begin{table}
\begin{tabularx}{\columnwidth}{|>{$}r<{$}|>{$}X<{$}|>{$}X<{$}|>{$}X<{$}|>{$}X<{$}|}

\hline

-&\slj3P0~&\slj1P1~&\slj3P1~&\slj3P2~\\

\hline

\slj1S0~\slj1S0~\sl1S&\dfrac{\sqrt{3}}{2}&&&\\

\sl1D&&&&\dfrac{\sqrt{3}}{2 \sqrt{5}}\\

\hline
\slj3S1~\slj1S0~\sl3S&&\dfrac{-1}{2}&\dfrac{1}{\sqrt{2}}&\\

\sl3D&&\dfrac{-1}{2}&\dfrac{-1}{2 \sqrt{2}}&\dfrac{-3}{2 \sqrt{10}}\\

\hline
\slj1S0~\slj3S1~\sl3S&&\dfrac{-1}{2}&\dfrac{-1}{\sqrt{2}}&\\

\sl3D&&\dfrac{-1}{2}&\dfrac{1}{2 \sqrt{2}}&\dfrac{3}{2 \sqrt{10}}\\

\hline
\slj3S1~\slj3S1~\sl1S&\dfrac{1}{2}&&&\\

\sl3S&&\dfrac{-1}{\sqrt{2}}&0^\dag&\\

\sl5S&&&&-1\phrac\\

\sl1D&&&&\dfrac{1}{2 \sqrt{5}}\\

\sl3D&&\dfrac{-1}{\sqrt{2}}&0^\dag&0^\dag\\

\sl5D&-1&0^\dag&\dfrac{-\sqrt{3}}{2}&\dfrac{-\sqrt{7}}{2 \sqrt{5}}\\

\sl5G&&&&0^\triangle\phrac\\

\hline

\end{tabularx}
\caption{The $\xi$ coefficients for $\uP\to\uS\,\uS$.}
\label{tab:pss}
 \end{table}

 \begin{table}
\begin{tabularx}{\columnwidth}{|>{$}r<{$}|>{$}X<{$}|>{$}X<{$}|>{$}X<{$}|>{$}X<{$}|}

\hline

-&\slj3D1~&\slj1D2~&\slj3D2~&\slj3D3~\\

\hline

\slj1S0~\slj1S0~\sl1P&\dfrac{\sqrt{5}}{2 \sqrt{3}}&&&\\

\sl1F&&&&\dfrac{\sqrt{5}}{2 \sqrt{7}}\\

\hline
\slj3S1~\slj1S0~\sl3P&\dfrac{\sqrt{5}}{2 \sqrt{6}}&\dfrac{-1}{2}&\dfrac{\sqrt{3}}{2 \sqrt{2}}&\\

\sl3F&&\dfrac{-1}{2}&\dfrac{-1}{\sqrt{6}}&\dfrac{-\sqrt{5}}{\sqrt{21}}\\

\hline
\slj1S0~\slj3S1~\sl3P&\dfrac{-\sqrt{5}}{2 \sqrt{6}}&\dfrac{-1}{2}&\dfrac{-\sqrt{3}}{2 \sqrt{2}}&\\

\sl3F&&\dfrac{-1}{2}&\dfrac{1}{\sqrt{6}}&\dfrac{\sqrt{5}}{\sqrt{21}}\\

\hline
\slj3S1~\slj3S1~\sl1P&\dfrac{\sqrt{5}}{6}&&&\\

\sl3P&0^\dag&\dfrac{-1}{\sqrt{2}}&0^\dag&\\

\sl5P&\dfrac{-1}{6}&0^\dag&\dfrac{-1}{2}&-1\\

\sl1F&&&&\dfrac{\sqrt{5}}{2 \sqrt{21}}\\

\sl3F&&\dfrac{-1}{\sqrt{2}}&0^\dag&0^\dag\\

\sl5F&-1&0^\dag&\dfrac{-\sqrt{2}}{\sqrt{3}}&\dfrac{-\sqrt{2}}{\sqrt{7}}\\

\sl5H&&&&0^\triangle\phrac\\

\hline

\end{tabularx}
\caption{The $\xi$ coefficients for $\uD\to\uS\,\uS$.}
 \label{tab:dss}
 \end{table}

 \begin{table}
\begin{tabularx}{\columnwidth}{|>{$}r<{$}|>{$}X<{$}|>{$}X<{$}|>{$}X<{$}|>{$}X<{$}|}

\hline

-&\slj3F2~&\slj1F3~&\slj3F3~&\slj3F4~\\

\hline

\slj1S0~\slj1S0~\sl1D&\dfrac{\sqrt{7}}{2 \sqrt{5}}&&&\\

\sl1G&&&&\dfrac{\sqrt{7}}{6}\\

\hline
\slj3S1~\slj1S0~\sl3D&\dfrac{\sqrt{7}}{\sqrt{30}}&\dfrac{-1}{2}&\dfrac{1}{\sqrt{3}}&\\

\sl3G&&\dfrac{-1}{2}&\dfrac{-\sqrt{3}}{4}&\dfrac{-\sqrt{35}}{12}\\

\hline
\slj1S0~\slj3S1~\sl3D&\dfrac{-\sqrt{7}}{\sqrt{30}}&\dfrac{-1}{2}&\dfrac{-1}{\sqrt{3}}&\\

\sl3G&&\dfrac{-1}{2}&\dfrac{\sqrt{3}}{4}&\dfrac{\sqrt{35}}{12}\\

\hline
\slj3S1~\slj3S1~\sl5S&0^\triangle&&&\\

\sl1D&\dfrac{\sqrt{7}}{2 \sqrt{15}}&&&\\

\sl3D&0^\dag&\dfrac{-1}{\sqrt{2}}&0^\dag&\\

\sl5D&\dfrac{-1}{\sqrt{15}}&0^\dag&\dfrac{-1}{\sqrt{3}}&-1\\

\sl1G&&&&\dfrac{\sqrt{7}}{6 \sqrt{3}}\\

\sl3G&&\dfrac{-1}{\sqrt{2}}&0^\dag&0^\dag\\

\sl5G&-1&0^\dag&\dfrac{-\sqrt{5}}{2 \sqrt{2}}&\dfrac{-\sqrt{55}}{6 \sqrt{6}}\\

\sl5I&&&&0^\triangle\phrac\\

\hline

\end{tabularx}
\caption{The $\xi$ coefficients for $\uF\to\uS\,\uS$.}
 \label{tab:fss}\end{table}
\begin{table}
\begin{tabularx}{\columnwidth}{|>{$}r<{$}|>{$}X<{$}|>{$}X<{$}|}

\hline

-&\slj1S0~&\slj3S1~\\
\hline
\slj3P0~\slj1S0~\sl1S&\dfrac{-1}{2}&\\
\hline

\slj1P1~\slj1S0~\sl3S&&\dfrac{1}{2 \sqrt{3}}\\

\sl3D&&\dfrac{1}{2 \sqrt{3}}\\

\hline
\slj3P1~\slj1S0~\sl3S&&\dfrac{-1}{\sqrt{6}}\\

\sl3D&&\dfrac{1}{2 \sqrt{6}}\\

\hline
\slj3P2~\slj1S0~\sl5D&\dfrac{-1}{2}&\dfrac{-1}{2 \sqrt{2}}\\

\hline
\slj3P0~\slj3S1~\sl3S&&\dfrac{-1}{2}\\

\sl3D&&0^\bullet\phrac\\

\hline
\slj1P1~\slj3S1~\sl1S&\dfrac{-1}{2}&\\

\sl3S&&\dfrac{-1}{\sqrt{6}}\\

\sl3D&&\dfrac{1}{2 \sqrt{6}}\\

\sl5D&\dfrac{-1}{2}&\dfrac{1}{2 \sqrt{2}}\\

\hline
\slj3P1~\slj3S1~\sl1S&\dfrac{1}{\sqrt{2}}&\\

\sl3S&&\dfrac{1}{\sqrt{3}}\\

\sl3D&&\dfrac{1}{4 \sqrt{3}}\\

\sl5D&\dfrac{-1}{2 \sqrt{2}}&\dfrac{1}{4}\\

\hline
\slj3P2~\slj3S1~\sl3S&&0^\bullet\phrac\\

\sl3D&&\dfrac{1}{4 \sqrt{5}}\\

\sl5D&\dfrac{-\sqrt{3}}{2 \sqrt{2}}&\dfrac{-1}{4 \sqrt{3}}\\

\sl7D&&\dfrac{-\sqrt{7}}{\sqrt{15}}\\

\sl7G&&0^\triangle\phrac\\

\hline

\end{tabularx}
\caption{The $\xi$ coefficients for $\uS\to\uP\,\uS$.}
\label{tab:sps}
 \end{table}
 
\begin{table*}
\begin{tabularx}{\textwidth}{|>{$}r<{$}|>{$}X<{$}>{$}X<{$}>{$}X<{$}|>{$}X<{$}>{$}X<{$}>{$}X<{$}|>{$}X<{$}>{$}X<{$}>{$}X<{$}|>{$}X<{$}>{$}X<{$}>{$}X<{$}|}

\hline

-&\slj3P0~&&&\slj1P1~&&&\slj3P1~&&&\slj3P2&&~\\

\hline
\slj3P0~\slj1S0~\sl1P&&&&\dfrac{-1}{6}&\dfrac{1}{2 \sqrt{3}}&\dfrac{-\sqrt{5}}{6}&\dfrac{1}{3 \sqrt{2}}&\dfrac{-1}{2 \sqrt{6}}&\dfrac{-\sqrt{5}}{6 \sqrt{2}}&&&\\

\hline
\slj1P1~\slj1S0~\sl3P&\dfrac{\sqrt{3}}{2}&0^\diamond&0^\diamond&0^\star&0^\star&0^\star&0^\diamond&\dfrac{-1}{2}&0^\diamond&0^\diamond&0^\diamond&\dfrac{\sqrt{3}}{2 \sqrt{5}}\\

\sl3F&&&&&&&&&&\dfrac{\sqrt{3}}{2 \sqrt{5}}&&\\

\hline
\slj3P1~\slj1S0~\sl3P&0^\diamond&\dfrac{1}{\sqrt{2}}&0^\diamond&\dfrac{1}{2 \sqrt{3}}&\dfrac{-1}{4}&\dfrac{-\sqrt{5}}{4 \sqrt{3}}&\dfrac{-1}{\sqrt{6}}&\dfrac{1}{4 \sqrt{2}}&\dfrac{-\sqrt{5}}{4 \sqrt{6}}&0^\diamond&\dfrac{1}{4 \sqrt{2}}&\dfrac{-3 \sqrt{3}}{4 \sqrt{10}}\\

\sl3F&&&&&&&&&&\dfrac{\sqrt{3}}{2 \sqrt{10}}&&\\

\hline
\slj3P2~\slj1S0~\sl5P&&&&\dfrac{-\sqrt{5}}{6}&\dfrac{-\sqrt{5}}{4 \sqrt{3}}&\dfrac{-1}{12}&\dfrac{\sqrt{5}}{3 \sqrt{2}}&\dfrac{\sqrt{5}}{4 \sqrt{6}}&\dfrac{-1}{12 \sqrt{2}}&0^\diamond&\dfrac{-\sqrt{3}}{4 \sqrt{2}}&\dfrac{-3}{4 \sqrt{10}}\\

\sl5F&&&&\dfrac{-1}{2}&&&\dfrac{-1}{2 \sqrt{2}}&&&\dfrac{-\sqrt{3}}{2 \sqrt{5}}&&\\

\hline
\slj3P0~\slj3S1~\sl3P&\dfrac{1}{6}&0^\diamond&\dfrac{-\sqrt{5}}{3}&\dfrac{1}{3 \sqrt{2}}&\dfrac{-1}{2 \sqrt{6}}&\dfrac{-\sqrt{5}}{6 \sqrt{2}}&0^\diamond&\dfrac{1}{4 \sqrt{3}}&\dfrac{-\sqrt{5}}{4}&\dfrac{-1}{3}&\dfrac{\sqrt{3}}{4}&\dfrac{-\sqrt{5}}{12}\\

\sl3F&&&&&&&&&&0^\bullet\phrac&&\\

\hline
\slj1P1~\slj3S1~\sl1P&&&&\dfrac{-1}{6}&\dfrac{1}{2 \sqrt{3}}&\dfrac{-\sqrt{5}}{6}&\dfrac{-1}{3 \sqrt{2}}&\dfrac{1}{2 \sqrt{6}}&\dfrac{\sqrt{5}}{6 \sqrt{2}}&&&\\

\sl3P&0^\diamond&\dfrac{1}{\sqrt{2}}&0^\diamond&\dfrac{-1}{2 \sqrt{3}}&\dfrac{1}{4}&\dfrac{\sqrt{5}}{4 \sqrt{3}}&\dfrac{-1}{\sqrt{6}}&\dfrac{1}{4 \sqrt{2}}&\dfrac{-\sqrt{5}}{4 \sqrt{6}}&0^\diamond&\dfrac{1}{4 \sqrt{2}}&\dfrac{-3 \sqrt{3}}{4 \sqrt{10}}\\

\sl5P&&&&\dfrac{-\sqrt{5}}{6}&\dfrac{-\sqrt{5}}{4 \sqrt{3}}&\dfrac{-1}{12}&\dfrac{-\sqrt{5}}{3 \sqrt{2}}&\dfrac{-\sqrt{5}}{4 \sqrt{6}}&\dfrac{1}{12 \sqrt{2}}&0^\diamond&\dfrac{\sqrt{3}}{4 \sqrt{2}}&\dfrac{3}{4 \sqrt{10}}\\

\sl3F&&&&&&&&&&\dfrac{\sqrt{3}}{2 \sqrt{10}}&&\\

\sl5F&&&&\dfrac{-1}{2}&&&\dfrac{1}{2 \sqrt{2}}&&&\dfrac{\sqrt{3}}{2 \sqrt{5}}&&\\

\hline
\slj3P1~\slj3S1~\sl1P&&&&\dfrac{1}{3 \sqrt{2}}&\dfrac{-1}{\sqrt{6}}&\dfrac{\sqrt{5}}{3 \sqrt{2}}&0^\dag&0^\dag&0^\dag&&&\\

\sl3P&\dfrac{1}{2 \sqrt{3}}&0^\diamond&\dfrac{\sqrt{5}}{2 \sqrt{3}}&\dfrac{1}{2 \sqrt{6}}&\dfrac{-1}{4 \sqrt{2}}&\dfrac{-\sqrt{5}}{4 \sqrt{6}}&0^\diamond&\dfrac{-3}{8}&\dfrac{\sqrt{15}}{8}&\dfrac{1}{2 \sqrt{3}}&\dfrac{-3}{8}&\dfrac{11}{8 \sqrt{15}}\\

\sl5P&&&&\dfrac{-\sqrt{5}}{6 \sqrt{2}}&\dfrac{-\sqrt{5}}{4 \sqrt{6}}&\dfrac{-1}{12 \sqrt{2}}&0^\diamond&\dfrac{-\sqrt{15}}{8}&\dfrac{-3}{8}&\dfrac{1}{2}&\dfrac{-\sqrt{3}}{8}&\dfrac{-7}{8 \sqrt{5}}\\

\sl3F&&&&&&&&&&\dfrac{\sqrt{3}}{4 \sqrt{5}}&&\\

\sl5F&&&&\dfrac{-1}{2 \sqrt{2}}&&&\dfrac{1}{4}&&&\dfrac{\sqrt{3}}{2 \sqrt{10}}&&\\

\hline
\slj3P2~\slj3S1~\sl3P&\dfrac{\sqrt{5}}{6}&0^\diamond&\dfrac{-1}{6}&\dfrac{-\sqrt{5}}{6 \sqrt{2}}&\dfrac{\sqrt{5}}{4 \sqrt{6}}&\dfrac{5}{12 \sqrt{2}}&0^\diamond&\dfrac{-\sqrt{5}}{8 \sqrt{3}}&\dfrac{-1}{8}&\dfrac{-1}{6 \sqrt{5}}&\dfrac{\sqrt{3}}{8 \sqrt{5}}&\dfrac{13}{120}\\

\sl5P&&&&\dfrac{-\sqrt{5}}{2 \sqrt{6}}&\dfrac{-\sqrt{5}}{4 \sqrt{2}}&\dfrac{-1}{4 \sqrt{6}}&0^\diamond&\dfrac{\sqrt{5}}{8}&\dfrac{\sqrt{3}}{8}&\dfrac{-1}{2 \sqrt{3}}&\dfrac{1}{8}&\dfrac{7}{8 \sqrt{15}}\\

\sl7P&&&&&&&&&&\dfrac{-\sqrt{7}}{\sqrt{15}}&\dfrac{-\sqrt{7}}{2 \sqrt{5}}&\dfrac{-\sqrt{7}}{10 \sqrt{3}}\\

\sl3F&&&&&&&&&&\dfrac{3}{20}&&\\

\sl5F&&&&\dfrac{-\sqrt{3}}{2 \sqrt{2}}&&&\dfrac{-1}{4 \sqrt{3}}&&&\dfrac{-1}{2 \sqrt{10}}&&\\

\sl7F&-1&&&0^\diamond&&&\dfrac{-\sqrt{2}}{\sqrt{3}}&&&\dfrac{-\sqrt{6}}{5}&&\\

\sl7H&&&&&&&&&&0^\triangle\phrac&&\\

\hline

\end{tabularx}
\caption{The $\xi$ coefficients for $\uP\to\uP\,\uS$.}
\label{tab:pps}
 \end{table*}

 \begin{table*}
\begin{tabularx}{\textwidth}{|>{$}r<{$}|>{$}X<{$}>{$}X<{$}>{$}X<{$}|>{$}X<{$}>{$}X<{$}>{$}X<{$}|>{$}X<{$}>{$}X<{$}>{$}X<{$}|>{$}X<{$}>{$}X<{$}>{$}X<{$}|}

\hline

-&\slj3D1~&&&\slj1D2~&&&\slj3D2~&&&\slj3D3~&&\\

\hline
\slj3P0~\slj1S0~\sl1D&&&&\dfrac{-1}{2 \sqrt{5}}&\dfrac{1}{2 \sqrt{3}}&\dfrac{-\sqrt{7}}{2 \sqrt{15}}&\dfrac{\sqrt{3}}{2 \sqrt{10}}&\dfrac{-1}{6 \sqrt{2}}&\dfrac{-\sqrt{7}}{3 \sqrt{10}}&&&\\

\hline
\slj1P1~\slj1S0~\sl3S&\dfrac{\sqrt{5}}{2 \sqrt{3}}&&&&&&&&&&&\\

\sl3D&\dfrac{\sqrt{5}}{2 \sqrt{3}}&0^\diamond&0^\diamond&0^\star&0^\star&0^\star&0^\diamond&\dfrac{-1}{2}&0^\diamond&0^\diamond&0^\diamond&\dfrac{\sqrt{5}}{2 \sqrt{7}}\\

\sl3G&&&&&&&&&&\dfrac{\sqrt{5}}{2 \sqrt{7}}&&\\

\hline
\slj3P1~\slj1S0~\sl3S&\dfrac{\sqrt{5}}{2 \sqrt{6}}&&&&&&&&&&&\\

\sl3D&\dfrac{-\sqrt{5}}{4 \sqrt{6}}&\dfrac{3}{4 \sqrt{2}}&0^\diamond&\dfrac{3}{4 \sqrt{5}}&\dfrac{-1}{4 \sqrt{3}}&\dfrac{-\sqrt{7}}{2 \sqrt{15}}&\dfrac{-3 \sqrt{3}}{4 \sqrt{10}}&\dfrac{1}{12 \sqrt{2}}&\dfrac{-\sqrt{7}}{3 \sqrt{10}}&0^\diamond&\dfrac{1}{3 \sqrt{2}}&\dfrac{-\sqrt{10}}{3 \sqrt{7}}\\

\sl3G&&&&&&&&&&\dfrac{\sqrt{5}}{2 \sqrt{14}}&&\\

\hline
\slj3P2~\slj1S0~\sl5S&&&&\dfrac{-1}{2}&&&\dfrac{\sqrt{3}}{2 \sqrt{2}}&&&&&\\

\sl5D&\dfrac{\sqrt{5}}{4 \sqrt{2}}&\dfrac{\sqrt{3}}{4 \sqrt{2}}&0^\diamond&\dfrac{-\sqrt{7}}{4 \sqrt{5}}&\dfrac{-\sqrt{7}}{4 \sqrt{3}}&\dfrac{-1}{2 \sqrt{15}}&\dfrac{\sqrt{21}}{4 \sqrt{10}}&\dfrac{\sqrt{7}}{12 \sqrt{2}}&\dfrac{-1}{3 \sqrt{10}}&0^\diamond&\dfrac{-1}{3}&\dfrac{-\sqrt{5}}{3 \sqrt{7}}\\

\sl5G&&&&\dfrac{-1}{2}&&&\dfrac{-1}{\sqrt{6}}&&&\dfrac{-5}{2 \sqrt{42}}&&\\

\hline
\slj3P0~\slj3S1~\sl3S&0^\bullet\phrac&&&&&&&&&&&\\

\sl3D&\dfrac{1}{4 \sqrt{5}}&\dfrac{1}{4 \sqrt{3}}&\dfrac{-\sqrt{7}}{\sqrt{15}}&\dfrac{\sqrt{3}}{2 \sqrt{10}}&\dfrac{-1}{6 \sqrt{2}}&\dfrac{-\sqrt{7}}{3 \sqrt{10}}&\dfrac{-1}{4 \sqrt{5}}&\dfrac{5}{12 \sqrt{3}}&\dfrac{-2 \sqrt{7}}{3 \sqrt{15}}&\dfrac{-1}{\sqrt{5}}&\dfrac{2}{3 \sqrt{3}}&\dfrac{-\sqrt{7}}{6 \sqrt{15}}\\

\sl3G&&&&&&&&&&0^\bullet\phrac&&\\

\hline
\slj1P1~\slj3S1~\sl3S&\dfrac{\sqrt{5}}{2 \sqrt{6}}&&&&&&&&&&&\\

\sl5S&&&&\dfrac{-1}{2}&&&\dfrac{-\sqrt{3}}{2 \sqrt{2}}&&&&&\\

\sl1D&&&&\dfrac{-1}{2 \sqrt{5}}&\dfrac{1}{2 \sqrt{3}}&\dfrac{-\sqrt{7}}{2 \sqrt{15}}&\dfrac{-\sqrt{3}}{2 \sqrt{10}}&\dfrac{1}{6 \sqrt{2}}&\dfrac{\sqrt{7}}{3 \sqrt{10}}&&&\\

\sl3D&\dfrac{-\sqrt{5}}{4 \sqrt{6}}&\dfrac{3}{4 \sqrt{2}}&0^\diamond&\dfrac{-3}{4 \sqrt{5}}&\dfrac{1}{4 \sqrt{3}}&\dfrac{\sqrt{7}}{2 \sqrt{15}}&\dfrac{-3 \sqrt{3}}{4 \sqrt{10}}&\dfrac{1}{12 \sqrt{2}}&\dfrac{-\sqrt{7}}{3 \sqrt{10}}&0^\diamond&\dfrac{1}{3 \sqrt{2}}&\dfrac{-\sqrt{10}}{3 \sqrt{7}}\\

\sl5D&\dfrac{-\sqrt{5}}{4 \sqrt{2}}&\dfrac{-\sqrt{3}}{4 \sqrt{2}}&0^\diamond&\dfrac{-\sqrt{7}}{4 \sqrt{5}}&\dfrac{-\sqrt{7}}{4 \sqrt{3}}&\dfrac{-1}{2 \sqrt{15}}&\dfrac{-\sqrt{21}}{4 \sqrt{10}}&\dfrac{-\sqrt{7}}{12 \sqrt{2}}&\dfrac{1}{3 \sqrt{10}}&0^\diamond&\dfrac{1}{3}&\dfrac{\sqrt{5}}{3 \sqrt{7}}\\

\sl3G&&&&&&&&&&\dfrac{\sqrt{5}}{2 \sqrt{14}}&&\\

\sl5G&&&&\dfrac{-1}{2}&&&\dfrac{1}{\sqrt{6}}&&&\dfrac{5}{2 \sqrt{42}}&&\\

\hline
\slj3P1~\slj3S1~\sl3S&\dfrac{\sqrt{5}}{4 \sqrt{3}}&&&&&&&&&&&\\

\sl5S&&&&\dfrac{-1}{2 \sqrt{2}}&&&\dfrac{-\sqrt{3}}{4}&&&&&\\

\sl1D&&&&\dfrac{1}{\sqrt{10}}&\dfrac{-1}{\sqrt{6}}&\dfrac{\sqrt{7}}{\sqrt{30}}&0^\dag&0^\dag&0^\dag&&&\\

\sl3D&\dfrac{7}{8 \sqrt{15}}&\dfrac{-1}{8}&\dfrac{\sqrt{7}}{2 \sqrt{5}}&\dfrac{3}{4 \sqrt{10}}&\dfrac{-1}{4 \sqrt{6}}&\dfrac{-\sqrt{7}}{2 \sqrt{30}}&\dfrac{\sqrt{3}}{8 \sqrt{5}}&\dfrac{-11}{24}&\dfrac{\sqrt{7}}{3 \sqrt{5}}&\dfrac{\sqrt{3}}{2 \sqrt{5}}&\dfrac{-1}{3}&\dfrac{11}{6 \sqrt{35}}\\

\sl5D&\dfrac{1}{8 \sqrt{5}}&\dfrac{-5}{8 \sqrt{3}}&\dfrac{-\sqrt{7}}{2 \sqrt{15}}&\dfrac{-\sqrt{7}}{4 \sqrt{10}}&\dfrac{-\sqrt{7}}{4 \sqrt{6}}&\dfrac{-1}{2 \sqrt{30}}&\dfrac{\sqrt{21}}{8 \sqrt{5}}&\dfrac{-\sqrt{7}}{8}&\dfrac{-1}{\sqrt{5}}&\dfrac{\sqrt{3}}{\sqrt{10}}&0^\ddag&\dfrac{-3}{\sqrt{70}}\\

\sl3G&&&&&&&&&&\dfrac{\sqrt{5}}{4 \sqrt{7}}&&\\

\sl5G&&&&\dfrac{-1}{2 \sqrt{2}}&&&\dfrac{1}{2 \sqrt{3}}&&&\dfrac{5}{4 \sqrt{21}}&&\\

\hline
\slj3P2~\slj3S1~\sl3S&\dfrac{1}{4}&&&&&&&&&&&\\

\sl5S&&&&\dfrac{-\sqrt{3}}{2 \sqrt{2}}&&&\dfrac{1}{4}&&&&&\\

\sl7S&&&&&&&&&&-1&&\\

\sl3D&\dfrac{11}{40}&\dfrac{1}{8 \sqrt{15}}&\dfrac{-\sqrt{7}}{10 \sqrt{3}}&\dfrac{-\sqrt{3}}{4 \sqrt{2}}&\dfrac{\sqrt{5}}{12 \sqrt{2}}&\dfrac{\sqrt{7}}{6 \sqrt{2}}&\dfrac{-1}{40}&\dfrac{-13}{24 \sqrt{15}}&\dfrac{-\sqrt{7}}{15 \sqrt{3}}&\dfrac{-1}{10}&\dfrac{1}{3 \sqrt{15}}&\dfrac{19}{30 \sqrt{21}}\\

\sl5D&\dfrac{-1}{8 \sqrt{15}}&\dfrac{5}{24}&\dfrac{\sqrt{7}}{6 \sqrt{5}}&\dfrac{-\sqrt{21}}{4 \sqrt{10}}&\dfrac{-\sqrt{7}}{4 \sqrt{2}}&\dfrac{-1}{2 \sqrt{10}}&\dfrac{-\sqrt{7}}{8 \sqrt{5}}&\dfrac{\sqrt{7}}{8 \sqrt{3}}&\dfrac{1}{\sqrt{15}}&\dfrac{-1}{\sqrt{10}}&0^\ddag&\dfrac{\sqrt{3}}{\sqrt{70}}\\

\sl7D&\dfrac{-\sqrt{7}}{10 \sqrt{3}}&\dfrac{-\sqrt{7}}{6 \sqrt{5}}&\dfrac{-1}{15}&0^\dag&0^\dag&0^\dag&\dfrac{-\sqrt{7}}{5 \sqrt{2}}&\dfrac{-\sqrt{7}}{\sqrt{30}}&\dfrac{-\sqrt{2}}{5 \sqrt{3}}&\dfrac{-\sqrt{6}}{5}&\dfrac{-\sqrt{2}}{\sqrt{5}}&\dfrac{-2 \sqrt{2}}{5 \sqrt{7}}\\

\sl3G&&&&&&&&&&\dfrac{\sqrt{3}}{4 \sqrt{7}}&&\\

\sl5G&&&&\dfrac{-\sqrt{3}}{2 \sqrt{2}}&&&\dfrac{-1}{6}&&&\dfrac{-5}{12 \sqrt{7}}&&\\

\sl7G&-1&&&0^\dag&&&\dfrac{-\sqrt{5}}{3}&&&\dfrac{-\sqrt{11}}{3 \sqrt{7}}&&\\

\sl7I&&&&&&&&&&0^\triangle\phrac&&\\

\hline

\end{tabularx}
\caption{The $\xi$ coefficients for $\uD\to\uP\,\uS$.}
\label{tab:dps}
 \end{table*}
 
 \begin{table*}
\begin{tabularx}{\textwidth}{|>{$}r<{$}|>{$}X<{$}>{$}X<{$}>{$}X<{$}|>{$}X<{$}>{$}X<{$}>{$}X<{$}|>{$}X<{$}>{$}X<{$}>{$}X<{$}|>{$}X<{$}>{$}X<{$}>{$}X<{$}|}

\hline

-&\slj3F2~&&&\slj1F3~&&&\slj3F3~&&&\slj3F4~&&\\

\hline
\slj3P0~\slj1S0~\sl1F&&&&\dfrac{-\sqrt{5}}{2 \sqrt{21}}&\dfrac{1}{2 \sqrt{3}}&\dfrac{-\sqrt{3}}{2 \sqrt{7}}&\dfrac{\sqrt{5}}{3 \sqrt{7}}&\dfrac{-1}{12}&\dfrac{-3}{4 \sqrt{7}}&&&\\

\hline
\slj1P1~\slj1S0~\sl3P&\dfrac{\sqrt{7}}{2 \sqrt{5}}&&&&&&&&&&&\\

\sl3F&\dfrac{\sqrt{7}}{2 \sqrt{5}}&0^\diamond&0^\diamond&0^\star&0^\star&0^\star&0^\diamond&\dfrac{-1}{2}&0^\diamond&0^\diamond&0^\diamond&\dfrac{\sqrt{7}}{6}\\

\sl3H&&&&&&&&&&\dfrac{\sqrt{7}}{6}&&\\

\hline
\slj3P1~\slj1S0~\sl3P&\dfrac{\sqrt{7}}{2 \sqrt{10}}&&&&&&&&&&&\\

\sl3F&\dfrac{-\sqrt{7}}{3 \sqrt{10}}&\dfrac{\sqrt{2}}{3}&0^\diamond&\dfrac{\sqrt{5}}{\sqrt{42}}&\dfrac{-1}{4 \sqrt{6}}&\dfrac{-3 \sqrt{3}}{4 \sqrt{14}}&\dfrac{-\sqrt{10}}{3 \sqrt{7}}&\dfrac{1}{24 \sqrt{2}}&\dfrac{-9}{8 \sqrt{14}}&0^\diamond&\dfrac{3}{8 \sqrt{2}}&\dfrac{-5 \sqrt{7}}{24 \sqrt{2}}\\

\sl3H&&&&&&&&&&\dfrac{\sqrt{7}}{6 \sqrt{2}}&&\\

\hline
\slj3P2~\slj1S0~\sl5P&\dfrac{\sqrt{7}}{2 \sqrt{30}}&&&\dfrac{-1}{2}&&&\dfrac{1}{\sqrt{3}}&&&&&\\

\sl5F&\dfrac{\sqrt{7}}{3 \sqrt{5}}&\dfrac{1}{3}&0^\diamond&\dfrac{-1}{\sqrt{14}}&\dfrac{-\sqrt{5}}{4 \sqrt{2}}&\dfrac{-\sqrt{5}}{4 \sqrt{14}}&\dfrac{\sqrt{2}}{\sqrt{21}}&\dfrac{\sqrt{5}}{8 \sqrt{6}}&\dfrac{-\sqrt{15}}{8 \sqrt{14}}&0^\diamond&\dfrac{-\sqrt{15}}{8 \sqrt{2}}&\dfrac{-\sqrt{35}}{8 \sqrt{6}}\\

\sl5H&&&&\dfrac{-1}{2}&&&\dfrac{-\sqrt{3}}{4}&&&\dfrac{-\sqrt{7}}{4 \sqrt{3}}&&\\

\hline
\slj3P0~\slj3S1~\sl3P&0^\bullet\phrac&&&&&&&&&&&\\

\sl3F&\dfrac{\sqrt{5}}{6 \sqrt{21}}&\dfrac{1}{3 \sqrt{3}}&\dfrac{-\sqrt{3}}{\sqrt{7}}&\dfrac{\sqrt{5}}{3 \sqrt{7}}&\dfrac{-1}{12}&\dfrac{-3}{4 \sqrt{7}}&\dfrac{-\sqrt{5}}{3 \sqrt{21}}&\dfrac{11}{24 \sqrt{3}}&\dfrac{-5 \sqrt{3}}{8 \sqrt{7}}&\dfrac{-\sqrt{5}}{\sqrt{21}}&\dfrac{5}{8 \sqrt{3}}&\dfrac{-\sqrt{3}}{8 \sqrt{7}}\\

\sl3H&&&&&&&&&&0^\bullet\phrac&&\\

\hline
\slj1P1~\slj3S1~\sl3P&\dfrac{\sqrt{7}}{2 \sqrt{10}}&&&&&&&&&&&\\

\sl5P&\dfrac{-\sqrt{7}}{2 \sqrt{30}}&&&\dfrac{-1}{2}&&&\dfrac{-1}{\sqrt{3}}&&&&&\\

\sl1F&&&&\dfrac{-\sqrt{5}}{2 \sqrt{21}}&\dfrac{1}{2 \sqrt{3}}&\dfrac{-\sqrt{3}}{2 \sqrt{7}}&\dfrac{-\sqrt{5}}{3 \sqrt{7}}&\dfrac{1}{12}&\dfrac{3}{4 \sqrt{7}}&&&\\

\sl3F&\dfrac{-\sqrt{7}}{3 \sqrt{10}}&\dfrac{\sqrt{2}}{3}&0^\diamond&\dfrac{-\sqrt{5}}{\sqrt{42}}&\dfrac{1}{4 \sqrt{6}}&\dfrac{3 \sqrt{3}}{4 \sqrt{14}}&\dfrac{-\sqrt{10}}{3 \sqrt{7}}&\dfrac{1}{24 \sqrt{2}}&\dfrac{-9}{8 \sqrt{14}}&0^\diamond&\dfrac{3}{8 \sqrt{2}}&\dfrac{-5 \sqrt{7}}{24 \sqrt{2}}\\

\sl5F&\dfrac{-\sqrt{7}}{3 \sqrt{5}}&\dfrac{-1}{3}&0^\diamond&\dfrac{-1}{\sqrt{14}}&\dfrac{-\sqrt{5}}{4 \sqrt{2}}&\dfrac{-\sqrt{5}}{4 \sqrt{14}}&\dfrac{-\sqrt{2}}{\sqrt{21}}&\dfrac{-\sqrt{5}}{8 \sqrt{6}}&\dfrac{\sqrt{15}}{8 \sqrt{14}}&0^\diamond&\dfrac{\sqrt{15}}{8 \sqrt{2}}&\dfrac{\sqrt{35}}{8 \sqrt{6}}\\

\sl3H&&&&&&&&&&\dfrac{\sqrt{7}}{6 \sqrt{2}}&&\\

\sl5H&&&&\dfrac{-1}{2}&&&\dfrac{\sqrt{3}}{4}&&&\dfrac{\sqrt{7}}{4 \sqrt{3}}&&\\

\hline
\slj3P1~\slj3S1~\sl3P&\dfrac{\sqrt{7}}{4 \sqrt{5}}&&&&&&&&&&&\\

\sl5P&\dfrac{-\sqrt{7}}{4 \sqrt{15}}&&&\dfrac{-1}{2 \sqrt{2}}&&&\dfrac{-1}{\sqrt{6}}&&&&&\\

\sl1F&&&&\dfrac{\sqrt{5}}{\sqrt{42}}&\dfrac{-1}{\sqrt{6}}&\dfrac{\sqrt{3}}{\sqrt{14}}&0^\dag&0^\dag&0^\dag&&&\\

\sl3F&\dfrac{4}{3 \sqrt{35}}&\dfrac{-1}{6}&\dfrac{3}{2 \sqrt{7}}&\dfrac{\sqrt{5}}{2 \sqrt{21}}&\dfrac{-1}{8 \sqrt{3}}&\dfrac{-3 \sqrt{3}}{8 \sqrt{7}}&\dfrac{\sqrt{5}}{6 \sqrt{7}}&\dfrac{-23}{48}&\dfrac{15}{16 \sqrt{7}}&\dfrac{\sqrt{5}}{2 \sqrt{7}}&\dfrac{-5}{16}&\dfrac{37}{48 \sqrt{7}}\\

\sl5F&\dfrac{1}{\sqrt{70}}&\dfrac{-1}{2 \sqrt{2}}&\dfrac{-3}{2 \sqrt{14}}&\dfrac{-1}{2 \sqrt{7}}&\dfrac{-\sqrt{5}}{8}&\dfrac{-\sqrt{5}}{8 \sqrt{7}}&\dfrac{\sqrt{3}}{2 \sqrt{7}}&\dfrac{-\sqrt{15}}{16}&\dfrac{-5 \sqrt{15}}{16 \sqrt{7}}&\dfrac{5}{2 \sqrt{21}}&\dfrac{\sqrt{5}}{16 \sqrt{3}}&\dfrac{-11 \sqrt{5}}{16 \sqrt{21}}\\

\sl3H&&&&&&&&&&\dfrac{\sqrt{7}}{12}&&\\

\sl5H&&&&\dfrac{-1}{2 \sqrt{2}}&&&\dfrac{\sqrt{3}}{4 \sqrt{2}}&&&\dfrac{\sqrt{7}}{4 \sqrt{6}}&&\\

\hline
\slj3P2~\slj3S1~\sl3P&\dfrac{\sqrt{21}}{20}&&&&&&&&&&&\\

\sl5P&\dfrac{\sqrt{7}}{12 \sqrt{5}}&&&\dfrac{-\sqrt{3}}{2 \sqrt{2}}&&&\dfrac{1}{3 \sqrt{2}}&&&&&\\

\sl7P&\dfrac{-1}{15}&&&0^\dag&&&\dfrac{-1}{3}&&&-1&&\\

\sl3F&\dfrac{17}{15 \sqrt{21}}&\dfrac{1}{6 \sqrt{15}}&\dfrac{-\sqrt{3}}{2 \sqrt{35}}&\dfrac{-5}{6 \sqrt{7}}&\dfrac{\sqrt{5}}{24}&\dfrac{3 \sqrt{5}}{8 \sqrt{7}}&\dfrac{-1}{6 \sqrt{21}}&\dfrac{-5 \sqrt{5}}{48 \sqrt{3}}&\dfrac{-\sqrt{15}}{16 \sqrt{7}}&\dfrac{-1}{2 \sqrt{21}}&\dfrac{\sqrt{5}}{16 \sqrt{3}}&\dfrac{5 \sqrt{5}}{16 \sqrt{21}}\\

\sl5F&\dfrac{-1}{\sqrt{210}}&\dfrac{1}{2 \sqrt{6}}&\dfrac{\sqrt{3}}{2 \sqrt{14}}&\dfrac{-\sqrt{3}}{2 \sqrt{7}}&\dfrac{-\sqrt{15}}{8}&\dfrac{-\sqrt{15}}{8 \sqrt{7}}&\dfrac{-1}{2 \sqrt{7}}&\dfrac{\sqrt{5}}{16}&\dfrac{5 \sqrt{5}}{16 \sqrt{7}}&\dfrac{-5}{6 \sqrt{7}}&\dfrac{-\sqrt{5}}{48}&\dfrac{11 \sqrt{5}}{48 \sqrt{7}}\\

\sl7F&\dfrac{-2 \sqrt{2}}{5 \sqrt{7}}&\dfrac{-1}{\sqrt{10}}&\dfrac{-1}{\sqrt{70}}&0^\dag&0^\dag&0^\dag&\dfrac{-1}{\sqrt{7}}&\dfrac{-\sqrt{5}}{4}&\dfrac{-\sqrt{5}}{4 \sqrt{7}}&\dfrac{-\sqrt{11}}{3 \sqrt{7}}&\dfrac{-\sqrt{55}}{12}&\dfrac{-\sqrt{55}}{12 \sqrt{7}}\\

\sl3H&&&&&&&&&&\dfrac{\sqrt{7}}{4 \sqrt{15}}&&\\

\sl5H&&&&\dfrac{-\sqrt{3}}{2 \sqrt{2}}&&&\dfrac{-1}{4 \sqrt{2}}&&&\dfrac{-\sqrt{7}}{12 \sqrt{2}}&&\\

\sl7H&-1&&&0^\dag&&&\dfrac{-1}{\sqrt{2}}&&&\dfrac{-\sqrt{13}}{3 \sqrt{10}}&&\\

\hline

\end{tabularx}
\caption{The $\xi$ coefficients for $\uF\to\uP\,\uS$.}
\label{tab:fps}\end{table*}
\begin{table}\begin{tabularx}{\columnwidth}{|>{$}r<{$}|>{$}X<{$}|>{$}X<{$}|}

\hline

-&\slj1S0~&\slj3S1~\\
\hline
\slj3D1~\slj1S0~\sl3P&\dfrac{-1}{2}&\dfrac{1}{2 \sqrt{6}}\\

\hline
\slj1D2~\slj1S0~\sl5P&&\dfrac{1}{2 \sqrt{3}}\\

\sl5F&&\dfrac{1}{2 \sqrt{3}}\\

\hline
\slj3D2~\slj1S0~\sl5P&&\dfrac{-1}{2 \sqrt{2}}\\

\sl5F&&\dfrac{1}{3 \sqrt{2}}\\

\hline
\slj3D3~\slj1S0~\sl7F&\dfrac{-1}{2}&\dfrac{-1}{3}\\

\hline

\end{tabularx}
\caption{The $\xi$ coefficients for $\uS\to\uD\,\uS$.}
\label{tab:sds}
\end{table}

\begin{table*}\begin{tabularx}{\textwidth}{|>{$}r<{$}|>{$}X<{$}>{$}X<{$}>{$}X<{$}|>{$}X<{$}>{$}X<{$}>{$}X<{$}|>{$}X<{$}>{$}X<{$}>{$}X<{$}|>{$}X<{$}>{$}X<{$}>{$}X<{$}|}

\hline

-&\slj3P0~&&&\slj1P1~&&&\slj3P1~&&&\slj3P2~&&\\

\hline
\slj3D1~\slj1S0~\sl3S&&&&\dfrac{-1}{2}&&&\dfrac{-1}{2 \sqrt{2}}&&&&&\\

\sl3D&&&&\dfrac{-1}{2 \sqrt{5}}&\dfrac{3}{4 \sqrt{5}}&\dfrac{-\sqrt{7}}{4 \sqrt{5}}&\dfrac{1}{\sqrt{10}}&\dfrac{-3}{4 \sqrt{10}}&\dfrac{-\sqrt{7}}{4 \sqrt{10}}&0^\diamond&\dfrac{-3}{20 \sqrt{2}}&\dfrac{3 \sqrt{7}}{20 \sqrt{2}}\\

\hline
\slj1D2~\slj1S0~\sl5S&&&&&&&&&&\dfrac{\sqrt{3}}{2 \sqrt{5}}&&\\

\sl5D&\dfrac{\sqrt{3}}{2}&0^\diamond&0^\diamond&0^\star&0^\star&0^\star&0^\diamond&\dfrac{-1}{2}&0^\diamond&0^\diamond&0^\diamond&\dfrac{\sqrt{3}}{2 \sqrt{5}}\\

\sl5G&&&&&&&&&&\dfrac{\sqrt{3}}{2 \sqrt{5}}&&\\

\hline
\slj3D2~\slj1S0~\sl5S&&&&&&&&&&\dfrac{-3}{2 \sqrt{10}}&&\\

\sl5D&0^\diamond&\dfrac{1}{\sqrt{2}}&0^\diamond&\dfrac{1}{2 \sqrt{3}}&\dfrac{-1}{4 \sqrt{3}}&\dfrac{-\sqrt{7}}{4 \sqrt{3}}&\dfrac{-1}{\sqrt{6}}&\dfrac{1}{4 \sqrt{6}}&\dfrac{-\sqrt{7}}{4 \sqrt{6}}&0^\diamond&\dfrac{\sqrt{7}}{4 \sqrt{10}}&\dfrac{-3}{4 \sqrt{10}}\\

\sl5G&&&&&&&&&&\dfrac{1}{\sqrt{10}}&&\\

\hline
\slj3D3~\slj1S0~\sl7D&&&&\dfrac{-\sqrt{7}}{2 \sqrt{15}}&\dfrac{-\sqrt{7}}{2 \sqrt{15}}&\dfrac{-1}{2 \sqrt{15}}&\dfrac{\sqrt{7}}{\sqrt{30}}&\dfrac{\sqrt{7}}{2 \sqrt{30}}&\dfrac{-1}{2 \sqrt{30}}&0^\diamond&\dfrac{-\sqrt{7}}{10}&\dfrac{-3}{10}\\

\sl7G&&&&\dfrac{-1}{2}&&&\dfrac{-1}{2 \sqrt{2}}&&&\dfrac{-1}{2 \sqrt{2}}&&\\

\hline

\end{tabularx}
\caption{The $\xi$ coefficients for $\uP\to\uD\,\uS$.}
\label{tab:pds}
\end{table*}

\begin{table*}\begin{tabularx}{\textwidth}{|>{$}r<{$}|>{$}X<{$}>{$}X<{$}>{$}X<{$}|>{$}X<{$}>{$}X<{$}>{$}X<{$}|>{$}X<{$}>{$}X<{$}>{$}X<{$}|>{$}X<{$}>{$}X<{$}>{$}X<{$}|}

\hline

-&\slj3D1~&&&\slj1D2~&&&\slj3D2~&&&\slj3D3~&&\\

\hline
\slj3D1~\slj1S0~\sl3P&\dfrac{-\sqrt{5}}{4 \sqrt{6}}&\dfrac{3}{4 \sqrt{2}}&0^\diamond&\dfrac{-1}{20}&\dfrac{\sqrt{3}}{4 \sqrt{5}}&\dfrac{-\sqrt{21}}{10}&\dfrac{\sqrt{3}}{20\sqrt{2}}&\dfrac{-1}{4 \sqrt{10}}&\dfrac{-\sqrt{7}}{5 \sqrt{2}}&&&\\

\sl3F&&&&\dfrac{-3}{10}&\dfrac{1}{\sqrt{10}}&\dfrac{-\sqrt{3}}{5 \sqrt{2}}&\dfrac{3 \sqrt{3}}{10 \sqrt{2}}&\dfrac{-1}{2 \sqrt{15}}&\dfrac{-1}{5}&0^\diamond&\dfrac{-1}{\sqrt{42}}&\dfrac{\sqrt{5}}{2 \sqrt{14}}\\

\hline
\slj1D2~\slj1S0~\sl5P&\dfrac{\sqrt{5}}{2 \sqrt{3}}&0^\diamond&0^\diamond&0^\star&0^\star&0^\star&0^\diamond&\dfrac{-1}{2}&0^\diamond&0^\diamond&0^\diamond&\dfrac{\sqrt{5}}{2 \sqrt{7}}\\

\sl5F&\dfrac{\sqrt{5}}{2 \sqrt{3}}&0^\diamond&0^\diamond&0^\star&0^\star&0^\star&0^\diamond&\dfrac{-1}{2}&0^\diamond&0^\diamond&0^\diamond&\dfrac{\sqrt{5}}{2 \sqrt{7}}\\

\sl5H&&&&&&&&&&\dfrac{\sqrt{5}}{2 \sqrt{7}}&&\\

\hline
\slj3D2~\slj1S0~\sl5P&\dfrac{\sqrt{5}}{4 \sqrt{2}}&\dfrac{\sqrt{3}}{4 \sqrt{2}}&0^\diamond&\dfrac{\sqrt{3}}{4 \sqrt{5}}&\dfrac{-5}{12}&\dfrac{-\sqrt{7}}{6 \sqrt{5}}&\dfrac{-3}{4 \sqrt{10}}&\dfrac{5}{12 \sqrt{6}}&\dfrac{-\sqrt{7}}{3 \sqrt{30}}&0^\diamond&\dfrac{1}{3 \sqrt{6}}&\dfrac{-2 \sqrt{10}}{3 \sqrt{21}}\\

\sl5F&\dfrac{-\sqrt{5}}{6 \sqrt{2}}&\dfrac{1}{2}&0^\diamond&\dfrac{1}{\sqrt{10}}&0^\ddag&\dfrac{-\sqrt{3}}{2 \sqrt{5}}&\dfrac{-\sqrt{3}}{2 \sqrt{5}}&0^\ddag&\dfrac{-1}{\sqrt{10}}&0^\diamond&\dfrac{1}{\sqrt{14}}&\dfrac{-\sqrt{5}}{2 \sqrt{42}}\\

\sl5H&&&&&&&&&&\dfrac{\sqrt{5}}{\sqrt{42}}&&\\

\hline
\slj3D3~\slj1S0~\sl7P&&&&\dfrac{-\sqrt{21}}{10}&\dfrac{-\sqrt{7}}{6 \sqrt{5}}&\dfrac{-1}{30}&\dfrac{3 \sqrt{7}}{10 \sqrt{2}}&\dfrac{\sqrt{7}}{6 \sqrt{30}}&\dfrac{-1}{15 \sqrt{6}}&0^\diamond&\dfrac{-2}{3 \sqrt{3}}&\dfrac{-\sqrt{5}}{3 \sqrt{21}}\\

\sl7F&\dfrac{\sqrt{5}}{6}&\dfrac{1}{2 \sqrt{2}}&0^\diamond&\dfrac{-\sqrt{3}}{5 \sqrt{2}}&\dfrac{-\sqrt{3}}{2 \sqrt{5}}&\dfrac{-1}{5}&\dfrac{3}{10}&\dfrac{1}{2 \sqrt{10}}&\dfrac{-\sqrt{2}}{5 \sqrt{3}}&0^\diamond&\dfrac{-1}{\sqrt{14}}&\dfrac{-\sqrt{5}}{\sqrt{42}}\\

\sl7H&&&&\dfrac{-1}{2}&&&\dfrac{-1}{\sqrt{6}}&&&\dfrac{-\sqrt{5}}{\sqrt{42}}&&\\

\hline

\end{tabularx}
\caption{The $\xi$ coefficients for $\uD\to\uD\,\uS$.}
\label{tab:dds}
\end{table*}

\begin{table*}\begin{tabularx}{\textwidth}{|>{$}r<{$}|>{$}X<{$}>{$}X<{$}>{$}X<{$}|>{$}X<{$}>{$}X<{$}>{$}X<{$}|>{$}X<{$}>{$}X<{$}>{$}X<{$}|>{$}X<{$}>{$}X<{$}>{$}X<{$}|}

\hline

-&\slj3F2~&&&\slj1F3~&&&\slj3F3~&&&\slj3F4~&&\\

\hline
\slj3D1~\slj1S0~\sl3D&\dfrac{-7}{10 \sqrt{6}}&\dfrac{\sqrt{14}}{5 \sqrt{3}}&0^\diamond&\dfrac{-1}{2 \sqrt{35}}&\dfrac{1}{2 \sqrt{5}}&\dfrac{-3 \sqrt{3}}{2 \sqrt{35}}&\dfrac{1}{\sqrt{105}}&\dfrac{-1}{4 \sqrt{15}}&\dfrac{-9}{4 \sqrt{35}}&&&\\

\sl3G&&&&\dfrac{-\sqrt{3}}{2 \sqrt{7}}&\dfrac{\sqrt{3}}{4 \sqrt{2}}&\dfrac{-\sqrt{11}}{4 \sqrt{14}}&\dfrac{1}{\sqrt{7}}&\dfrac{-1}{8 \sqrt{2}}&\dfrac{-\sqrt{33}}{8 \sqrt{14}}&0^\diamond&\dfrac{-\sqrt{21}}{8 \sqrt{10}}&\dfrac{7 \sqrt{11}}{24 \sqrt{10}}\\

\hline
\slj1D2~\slj1S0~\sl5S&\dfrac{\sqrt{7}}{2 \sqrt{5}}&&&&&&&&&&&\\

\sl5D&\dfrac{\sqrt{7}}{2 \sqrt{5}}&0^\diamond&0^\diamond&0^\star&0^\star&0^\star&0^\diamond&\dfrac{-1}{2}&0^\diamond&0^\diamond&0^\diamond&\dfrac{\sqrt{7}}{6}\\

\sl5G&\dfrac{\sqrt{7}}{2 \sqrt{5}}&0^\diamond&0^\diamond&0^\star&0^\star&0^\star&0^\diamond&\dfrac{-1}{2}&0^\diamond&0^\diamond&0^\diamond&\dfrac{\sqrt{7}}{6}\\

\sl5I&&&&&&&&&&\dfrac{\sqrt{7}}{6}&&\\

\hline
\slj3D2~\slj1S0~\sl5S&\dfrac{\sqrt{7}}{\sqrt{30}}&&&&&&&&&&&\\

\sl5D&\dfrac{\sqrt{7}}{2 \sqrt{30}}&\dfrac{\sqrt{2}}{\sqrt{15}}&0^\diamond&\dfrac{1}{\sqrt{14}}&\dfrac{-1}{2 \sqrt{2}}&\dfrac{-\sqrt{3}}{2 \sqrt{14}}&\dfrac{-\sqrt{2}}{\sqrt{21}}&\dfrac{1}{4 \sqrt{6}}&\dfrac{-3}{4 \sqrt{14}}&0^\diamond&\dfrac{1}{4 \sqrt{2}}&\dfrac{-5 \sqrt{7}}{12 \sqrt{6}}\\

\sl5G&\dfrac{-\sqrt{14}}{3 \sqrt{15}}&\dfrac{\sqrt{5}}{3 \sqrt{3}}&0^\diamond&\dfrac{5}{6 \sqrt{7}}&\dfrac{1}{12 \sqrt{2}}&\dfrac{-\sqrt{33}}{4 \sqrt{14}}&\dfrac{-5}{3 \sqrt{21}}&\dfrac{-1}{24 \sqrt{6}}&\dfrac{-3 \sqrt{11}}{8 \sqrt{14}}&0^\diamond&\dfrac{\sqrt{11}}{8 \sqrt{2}}&\dfrac{-\sqrt{7}}{8 \sqrt{6}}\\

\sl5I&&&&&&&&&&\dfrac{\sqrt{7}}{3 \sqrt{6}}&&\\

\hline
\slj3D3~\slj1S0~\sl7S&&&&\dfrac{-1}{2}&&&\dfrac{1}{\sqrt{3}}&&&&&\\

\sl7D&\dfrac{\sqrt{7}}{5 \sqrt{3}}&\dfrac{1}{5 \sqrt{3}}&0^\diamond&\dfrac{-\sqrt{6}}{\sqrt{35}}&\dfrac{-\sqrt{3}}{2 \sqrt{10}}&\dfrac{-1}{2 \sqrt{70}}&\dfrac{2 \sqrt{2}}{\sqrt{35}}&\dfrac{1}{4 \sqrt{10}}&\dfrac{-\sqrt{3}}{4 \sqrt{70}}&0^\diamond&\dfrac{-\sqrt{5}}{4 \sqrt{2}}&\dfrac{-\sqrt{35}}{12 \sqrt{6}}\\

\sl7G&\dfrac{\sqrt{7}}{3 \sqrt{6}}&\dfrac{2}{3 \sqrt{3}}&0^\diamond&\dfrac{-\sqrt{11}}{6 \sqrt{7}}&\dfrac{-\sqrt{11}}{6 \sqrt{2}}&\dfrac{-\sqrt{3}}{2 \sqrt{14}}&\dfrac{\sqrt{11}}{3 \sqrt{21}}&\dfrac{\sqrt{11}}{12 \sqrt{6}}&\dfrac{-3}{4 \sqrt{14}}&0^\diamond&\dfrac{-\sqrt{11}}{4 \sqrt{10}}&\dfrac{-\sqrt{21}}{4 \sqrt{10}}\\

\sl7I&&&&\dfrac{-1}{2}&&&\dfrac{-\sqrt{3}}{4}&&&\dfrac{-7}{12 \sqrt{3}}&&\\

\hline

\end{tabularx}
\caption{The $\xi$ coefficients for $\uF\to\uD\,\uS$.}
\label{tab:fds}
\end{table*}
\begin{table}\begin{tabularx}{\columnwidth}{|>{$}r<{$}|>{$}X<{$}|>{$}X<{$}|>{$}X<{$}|>{$}X<{$}|}

\hline

-&\slj3P0~&\slj1P1~&\slj3P1~&\slj3P2~\\

\hline
\slj1S0~\slj1S0~\sl1P&&0^\star&\dfrac{-1}{2}&\\

\hline
\slj3S1~\slj1S0~\sl3P&\dfrac{1}{\sqrt{2}}&\dfrac{-1}{2}&\dfrac{1}{2 \sqrt{2}}&\dfrac{-1}{2 \sqrt{2}}\\

\sl3F&&&&0^\triangle\phrac\\

\hline
\slj1S0~\slj3S1~\sl3P&\dfrac{-1}{\sqrt{2}}&\dfrac{-1}{2}&\dfrac{-1}{2 \sqrt{2}}&\dfrac{1}{2 \sqrt{2}}\\

\sl3F&&&&0^\triangle\phrac\\

\hline
\slj3S1~\slj3S1~\sl1P&&0^\dag&\dfrac{-1}{2 \sqrt{3}}&\\

\sl3P&0^\dag&\dfrac{-1}{\sqrt{2}}&0^\dag&0^\dag\\

\sl5P&&0^\dag&\dfrac{-\sqrt{5}}{2 \sqrt{3}}&\dfrac{-\sqrt{3}}{2}\\

\sl3F&&&&0^\triangle\phrac\\

\sl5F&&0^\triangle&0^\triangle&0^\triangle\phrac\\

\hline

\end{tabularx}
\caption{Abnormal parity $\xi$ coefficients for $\uP\to\uS\,\uS$.}
\label{tab:pss:abnormal}
\end{table}

\begin{table}\begin{tabularx}{\columnwidth}{|>{$}r<{$}|>{$}X<{$}>{$}X<{$}|>{$}X<{$}>{$}X<{$}|>{$}X<{$}>{$}X<{$}|>{$}X<{$}>{$}X<{$}|}

\hline

-&\slj3P0~&&\slj1P1~&&\slj3P1~&&\slj3P2~&\\

\hline
\slj3P0~\slj1S0~\sl1S&\dfrac{1}{\sqrt{2}}&&&&&&&\\

\sl1D&&&&&&&\dfrac{-1}{2 \sqrt{10}}&\dfrac{\sqrt{3}}{2 \sqrt{10}}\\

\hline
\slj1P1~\slj1S0~\sl3S&&&0^\star&&\dfrac{-1}{2}&&&\\

\sl3D&&&0^\star&0^\star&\dfrac{-1}{2}&0^\diamond&0^\diamond&\dfrac{\sqrt{3}}{2 \sqrt{5}}\\

\hline
\slj3P1~\slj1S0~\sl3S&&&\dfrac{-1}{2}&&\dfrac{1}{2 \sqrt{2}}&&&\\

\sl3D&&&\dfrac{1}{4}&\dfrac{-\sqrt{3}}{4}&\dfrac{-1}{4 \sqrt{2}}&\dfrac{-\sqrt{3}}{4 \sqrt{2}}&\dfrac{3}{4 \sqrt{10}}&\dfrac{-\sqrt{3}}{4 \sqrt{10}}\\

\hline
\slj3P2~\slj1S0~\sl5S&&&&&&&\dfrac{-1}{2 \sqrt{2}}&\\

\sl5D&\dfrac{1}{\sqrt{2}}&0^\diamond&\dfrac{-\sqrt{3}}{4}&\dfrac{-1}{4}&\dfrac{\sqrt{3}}{4 \sqrt{2}}&\dfrac{-1}{4 \sqrt{2}}&\dfrac{-\sqrt{7}}{4 \sqrt{10}}&\dfrac{-\sqrt{21}}{4 \sqrt{10}}\\

\hline
\slj3P2~\slj1S0~\sl5G&&&&&&&0^\triangle\phrac&\\

\hline
\slj3P0~\slj3S1~\sl3S&&&\dfrac{-1}{\sqrt{6}}&&\dfrac{-1}{\sqrt{3}}&&&\\

\sl3D&&&\dfrac{1}{2 \sqrt{6}}&\dfrac{-1}{2 \sqrt{2}}&\dfrac{-1}{4 \sqrt{3}}&\dfrac{1}{4}&\dfrac{-\sqrt{3}}{4 \sqrt{5}}&\dfrac{3}{4 \sqrt{5}}\\

\hline
\slj1P1~\slj3S1~\sl1S&\dfrac{-1}{\sqrt{2}}&&&&&&&\\

\sl3S&&&\dfrac{1}{2}&&\dfrac{1}{2 \sqrt{2}}&&&\\

\sl5S&&&&&&&\dfrac{1}{2 \sqrt{2}}&\\

\sl1D&&&&&&&\dfrac{1}{2 \sqrt{10}}&\dfrac{-\sqrt{3}}{2 \sqrt{10}}\\

\sl3D&&&\dfrac{-1}{4}&\dfrac{\sqrt{3}}{4}&\dfrac{-1}{4 \sqrt{2}}&\dfrac{-\sqrt{3}}{4 \sqrt{2}}&\dfrac{3}{4 \sqrt{10}}&\dfrac{-\sqrt{3}}{4 \sqrt{10}}\\

\sl5D&\dfrac{-1}{\sqrt{2}}&0^\diamond&\dfrac{-\sqrt{3}}{4}&\dfrac{-1}{4}&\dfrac{-\sqrt{3}}{4 \sqrt{2}}&\dfrac{1}{4 \sqrt{2}}&\dfrac{\sqrt{7}}{4 \sqrt{10}}&\dfrac{\sqrt{21}}{4 \sqrt{10}}\\

\sl5G&&&&&&&0^\triangle\phrac&\\

\hline
\slj3P1~\slj3S1~\sl1S&0^\dag&&&&&&&\\

\sl3S&&&\dfrac{-1}{2 \sqrt{2}}&&\dfrac{1}{4}&&&\\

\sl5S&&&&&&&\dfrac{-3}{4}&\\

\sl1D&&&&&&&0^\dag&0^\dag\phrac \\

\sl3D&&&\dfrac{1}{4 \sqrt{2}}&\dfrac{-\sqrt{3}}{4 \sqrt{2}}&\dfrac{-1}{8}&\dfrac{-\sqrt{3}}{8}&\dfrac{3}{8 \sqrt{5}}&\dfrac{-\sqrt{3}}{8 \sqrt{5}}\\

\sl5D&0^\diamond&\dfrac{-\sqrt{3}}{2}&\dfrac{-\sqrt{3}}{4 \sqrt{2}}&\dfrac{-1}{4 \sqrt{2}}&\dfrac{\sqrt{3}}{8}&\dfrac{-5}{8}&\dfrac{3 \sqrt{7}}{8 \sqrt{5}}&\dfrac{-\sqrt{21}}{8 \sqrt{5}}\\

\sl5G&&&&&&&0^\triangle\phrac&\\

\hline
\slj3P2~\slj3S1~\sl3S&&&\dfrac{\sqrt{5}}{2 \sqrt{6}}&&\dfrac{-\sqrt{5}}{4 \sqrt{3}}&&&\\

\sl5S&&&&&&&\dfrac{\sqrt{3}}{4}&\\

\sl3D&&&\dfrac{-\sqrt{5}}{4 \sqrt{6}}&\dfrac{\sqrt{5}}{4 \sqrt{2}}&\dfrac{-7}{8 \sqrt{15}}&\dfrac{1}{8 \sqrt{5}}&\dfrac{-\sqrt{3}}{40}&\dfrac{9}{40}\\

\sl5D&0^\diamond&\dfrac{1}{2}&\dfrac{-3}{4 \sqrt{2}}&\dfrac{-\sqrt{3}}{4 \sqrt{2}}&\dfrac{-1}{8}&\dfrac{5}{8 \sqrt{3}}&\dfrac{-\sqrt{21}}{8 \sqrt{5}}&\dfrac{\sqrt{7}}{8 \sqrt{5}}\\

\sl7D&&&0^\dag&0^\dag&\dfrac{-\sqrt{7}}{2 \sqrt{5}}&\dfrac{-\sqrt{7}}{2 \sqrt{15}}&\dfrac{-\sqrt{21}}{5 \sqrt{2}}&\dfrac{-\sqrt{7}}{5 \sqrt{2}}\\

\sl5G&&&&&&&0^\triangle\phrac&\\

\sl7G&&&0^\triangle&&0^\triangle&&0^\triangle\phrac&\\

\hline

\end{tabularx}
\caption{Abnormal parity $\xi$ coefficients for $\uP\to\uP\,\uS$.}
\label{tab:pps:abnormal}
\end{table}

\begin{table}\begin{tabularx}{\columnwidth}{|>{$}r<{$}|>{$}X<{$}>{$}X<{$}|>{$}X<{$}>{$}X<{$}|>{$}X<{$}>{$}X<{$}|>{$}X<{$}>{$}X<{$}|}

\hline

-&\slj3P0~&&\slj1P1~&&\slj3P1~&&\slj3P2~&\\

\hline
\slj3D1~\slj1S0~\sl3P&\dfrac{1}{\sqrt{2}}&0^\diamond&\dfrac{1}{4}&\dfrac{-\sqrt{3}}{4}&\dfrac{-1}{4 \sqrt{2}}&\dfrac{-\sqrt{3}}{4 \sqrt{2}}&\dfrac{-1}{20 \sqrt{2}}&\dfrac{3 \sqrt{3}}{20 \sqrt{2}}\\

\sl3F&&&&&&&\dfrac{-3}{10 \sqrt{2}}&\dfrac{3}{10}\\

\hline
\slj1D2~\slj1S0~\sl5P&&&0^\star&0^\star&\dfrac{-1}{2}&0^\diamond&0^\diamond&\dfrac{\sqrt{3}}{2 \sqrt{5}}\\

\sl5F&&&0^\star&0^\star&\dfrac{-1}{2}&0^\diamond&0^\diamond&\dfrac{\sqrt{3}}{2 \sqrt{5}}\\

\hline
\slj3D2~\slj1S0~\sl5P&&&\dfrac{-\sqrt{3}}{4}&\dfrac{-1}{4}&\dfrac{\sqrt{3}}{4 \sqrt{2}}&\dfrac{-1}{4 \sqrt{2}}&\dfrac{\sqrt{3}}{4 \sqrt{10}}&\dfrac{-\sqrt{5}}{4 \sqrt{2}}\\

\sl5F&&&\dfrac{1}{2 \sqrt{3}}&\dfrac{-1}{\sqrt{6}}&\dfrac{-1}{2 \sqrt{6}}&\dfrac{-1}{2 \sqrt{3}}&\dfrac{1}{2 \sqrt{5}}&0^\ddag\\

\hline
\slj3D3~\slj1S0~\sl7P&&&&&&&\dfrac{-\sqrt{21}}{10 \sqrt{2}}&\dfrac{-\sqrt{7}}{10 \sqrt{2}}\\

\sl7F&\dfrac{1}{\sqrt{2}}&0^\diamond&\dfrac{-1}{\sqrt{6}}&\dfrac{-1}{2 \sqrt{3}}&\dfrac{1}{2 \sqrt{3}}&\dfrac{-1}{2 \sqrt{6}}&\dfrac{-\sqrt{3}}{10}&\dfrac{-3 \sqrt{3}}{10 \sqrt{2}}\\

\sl7H&&&&&&&0^\triangle\phrac&\\

\hline

\end{tabularx}
\caption{Abnormal parity $\xi$ coefficients for $\uP\to\uD\,\uS$.}
\label{tab:pds:abnormal}
\end{table}
\endgroup

Entries in the tables are channels which are allowed by angular momentum. An explicit zero therefore corresponds to a selection rule which arises from the assumed $\spino\cdot\spato$ structure. Each zero in the tables is marked according to whether it is due to the spin-singlet selection rule ($^\star$), the spin-triplet selection rule ($^\dag$), the spatial-vector selection rule ($^\triangle$), due to violation of triangular relations in the spin and spatial degrees of freedom ($^\diamond$), a cancellation ($^\bullet$) or an accident zero in a 6-$j$ coefficient ($^\ddag$). For details, see Section \ref{zeroes}.

\bibliography{tjb}

\end{document}